\pdfoutput=1
\documentclass[twocolumn]{aastex63}

\newcommand{\RN}[1]{\uppercase\expandafter{\romannumeral #1\relax}}

\submitjournal{ApJS}

\graphicspath{{./}}

\begin{document}

\title{A catalogue of 323 cataclysmic variables from LAMOST DR6}

\correspondingauthor{Wenyuan Cui}
\email{wenyuancui@126.com,cuiwenyuan@hebtu.edu.cn}

\author{Yongkang Sun\footnote[1]{The first author}}
\affiliation{College of Physics, Hebei Normal University, Shijiazhuang 050024, China}

\author{Zhenghao Cheng\footnote[1]{The first author}}
\affiliation{College of Physics, Hebei Normal University, Shijiazhuang 050024, China}

\author{Shuo Ye}
\affiliation{Software College, Hebei Normal University, Shijiazhuang 050024, China}

\author{Ruobin Ding}
\affiliation{College of Physics, Hebei Normal University, Shijiazhuang 050024, China}

\author{Yijiang Peng}
\affiliation{College of Physics, Hebei Normal University, Shijiazhuang 050024, China}

\author{Jiawen Zhang}
\affiliation{College of Physics, Hebei Normal University, Shijiazhuang 050024, China}

\author{Zhenyan Huo}
\affiliation{College of Physics, Hebei Normal University, Shijiazhuang 050024, China}

\author[0000-0003-1359-9908]{Wenyuan Cui}
\affiliation{College of Physics, Hebei Normal University, Shijiazhuang 050024, China}

\author{Xiaofeng Wang}
\affiliation{Physics Department, Tsinghua University, Beijing 100084, China}
\affiliation{Beijing Planetarium, Beijing Academy of Sciences and Technology, Beijing 100044, China}

\author{Jianrong Shi}
\affiliation{School of Astronomy and Space Science, University of Chinese Academy of Sciences, Beijing 100049, China}
\affiliation{Key Laboratory of Optical Astronomy, National Astronomical Observatories, Chinese Academy of Sciences, Beijing 100012, China}

\author{Jie Lin}
\affiliation{Physics Department, Tsinghua University, Beijing 100084, China}

\author{Chengyuan Wu}
\affiliation{Physics Department, Tsinghua University, Beijing 100084, China}

\author{Linlin Li}
\affiliation{College of Physics, Hebei Normal University, Shijiazhuang 050024, China}

\author{Shuai Feng}
\affiliation{College of Physics, Hebei Normal University, Shijiazhuang 050024, China}

\author{Yang Yu}
\affiliation{College of Physics, Hebei Normal University, Shijiazhuang 050024, China}

\author{Xiaoran Ma}
\affiliation{Physics Department, Tsinghua University, Beijing 100084, China}

\author{Xin Li}
\affiliation{Beijing Planetarium, Beijing Academy of Sciences and Technology, Beijing 100044, China}

\author{Cheng Liu}
\affiliation{Beijing Planetarium, Beijing Academy of Sciences and Technology, Beijing 100044, China}

\author{Ziping Zhang}
\affiliation{Beijing Planetarium, Beijing Academy of Sciences and Technology, Beijing 100044, China}

\author{Zhenzhen Shao}
\affiliation{Beijing Planetarium, Beijing Academy of Sciences and Technology, Beijing 100044, China}

\begin{abstract}

In this work, we present a catalogue of cataclysmic variables (CVs) identified from the Sixth Data Release (DR6) of the Large Sky Area Multi-Object Fiber Spectroscopic Telescope (LAMOST). To single out the CV spectra, we introduce a novel machine-learning algorithm called UMAP to screen out a total of 169,509 H$\alpha$-emission spectra, and obtain a classification accuracy of the algorithm of over 99.6$\%$ from the cross-validation set.  We then apply the template matching program PyHammer v2.0 to the LAMOST spectra to obtain the optimal spectral type with metallicity, which help us identify the chromospherically active stars and potential binary stars from the 169,509 spectra. After visually inspecting all the spectra, we identify 323 CV candidates from the LAMOST database, among  them 52 objects are new. We further discuss the new CV candidates in subtypes based on their spectral features, including five DN subtype during outbursts, five NL subtype and four magnetic CVs (three AM Her type and one IP type). We also find two CVs that have been previously identified by photometry, and confirm their previous classification by the LAMOST spectra.

\end{abstract}


\section{Introduction} \label{sec:intro}
Cataclysmic variable stars (CVs) are semi-detached binary systems with short orbital periods ranging from shorter than one hour to longer than six hours \citep{warner2003cataclysmic, zorotovic2016detached}. They typically consist of a white dwarf star and a late-type main-sequence companion. With the loss of angular momentum, the mass transfer is continuously or intermittently activated through the inner Lagrangian point of the Roche lobe, usually leading to the formation of an accretion disc. 

The CVs can be divided into several types: classical nova (CN), recurrent nova (RN), dwarf nova (DN), nova-like variables (NL), polars (or AM Her type), and intermediate polars (IPs, including DQ Her type) \citep{warner2003cataclysmic}. The latter two are magnetic CVs. Note that in polars, the magnetic field is typically stronger than 20 MG that an accretion disc can not form. Classical novae usually have only one observed eruption by definition. If a classical nova is observed with a second eruption, it will be classified as a recurrent nova. Dwarf novae usually have frequent and regular bursts as a result of instability of the accretion disk \citep{morales2002spectral}. U Gem stars are the prototype of DN subtype of CVs which have repeated normal outbursts. The SU UMa subtype of DNs presents superhumps in addition to normal outbursts and the Z Cam subtype of DNs are characteristic of standstills in the light curve \citep{szkody1987photometry, hellier2001echo}. Nova-like objects (NLs) are recognized as novae between eruptions and do not have a recorded outburst. Some NLs have been observed to occasionally change from a high state with higher brightness to a low state with lower brightness \citep{la1994observations, king1998low,10.1093/mnras/stv1244}. In the AM Her systems, the magnetic field of the white dwarf is strong enough for both stars to be locked into the same rotation period. In this system, the mass flow is controlled by the magnetic field and falls directly toward the surface of the white dwarf. Correspondingly, the accretion disk of IPs is truncated at the radius of the magnetosphere. Sometimes these two types of magnetic CVs are referred to jointly as one subtype of nova-like variables \citep{coppejans2015novalike}. 

As a special kind of interacting binary systems, CVs have complex physical processes and unique observation properties in various wave bands and show luminosity variations of different time scales and amplitudes. They are important samples that help us to understand the evolution of binary stars and even the explosion mechanism of type Ia supernovae (SNe Ia) \citep{hachisu2001recurrent,anupama_2011}. 
The acquisition and analysis of more CV samples are helpful to the refining of the theoretical models of CVs.

So far, numerous works have been conducted to search for CVs. \citet{downes2001} published a catalogue including spectra, which is a revision and supplement to the first two editions (Edition 1 - \citet{downes1993catalog} and Edition 2 - \citet{downes1997catalog}). This catalogue consists of 1034 CVs and 194 non-CV objects. 
Using the Sloan Digital Sky Survey (SDSS) I/II, \citet {szkody2011} published a summary table of 285 CVs with the SDSS spectra. \citet {jiang2013data} identified CVs from the low-resolution spectra of LAMOST by using the support vector machine technique (SVM) combined with principal component analysis (PCA). They identified 10 cataclysmic variables, of which two are new CV candidates. \citet {breedt20141000} provided spectroscopic identification of 85 systems fainter than g$\geq$19 mag and analyzed the photometric properties of the 1043 CV candidates from the Catalina Real-time Transient Survey (CRTS). \citet {coppejans2016statistical} published an outburst catalogue which contains a wide variety of observational properties for 722 dwarf nova-type CVs and 309 CVs of other types from the CRTS. By cross-matching the published catalogues with LAMOST DR3, \citet {han2018cataclysmic} collected 48 known CVs, and they found three new CVs using the method adopted by \citet {jiang2013data}. \citet {szkody2020cataclysmic} published 329 objects as known or candidate CVs during the first year of testing and operation of the Zwicky Transient Facility. \citet {hou2020spectroscopically} used the Bagging and the Random Forest algorithms to search for CVs from LAMOST DR5, and they identified 245 cataclysmic variables, of which 52 CV candidates are new.

In this work, we present a catalogue of 323 CV candidates identified from LAMOST DR6, of which 52 objects are new. The structure of this paper is as follows. Section 2 describes the method and process to identify the CV candidates according to the spectral features. Section 3 presents the results. Section 4 is the discussion of the spectral properties and subtype classification of the LAMOST CV candidates. Section 5 is the summary.

\section{Data} \label{sec:data}
The Large Sky Area Multi-Object Fiber Spectroscopic Telescope (LAMOST) is a 4-meter quasi-meridian reflecting Schmidt telescope \citep{zhao2012lamost, cui2012large} that is operated by the National Astronomical Observatories, Chinese Academy of Sciences. With a field of view of 20 deg$^2$ and 4000 fibers at its focal plane, LAMOST can obtain up to 4000 spectra in one exposure and it has been the most efficient telescope for spectral acquisition in the world. Up to July 2018, LAMOST has completed its pilot survey and the first six years of regular survey which was initiated on September 2012.
After the six-year-survey, the whole data (DR6) were released in 2018, containing 9,911,337 low-resolution spectra. They are flux(relatively)- and wavelength-calibrated, sky-subtracted and have a resolution of R $\sim$ 1800 at 5500\AA\ and a wavelength coverage of 3700\AA\ $\leq$ $\lambda$ $\leq$ 9000\AA. Each spectrum is spliced by the blue part and the red part. The integration time is 600, 1500 or 1800 seconds. All the low-resolution spectra were generally classified into different types through the pipeline including 9,231,057 stellar spectra, 177,270 galaxy spectra, 62,168 quasar spectra, and 440,842 spectra marked as unknown. 
The massive spectral database obtained by LAMOST is a rich resource for us to identify different kinds of peculiar stars. As CVs are one kind of valuable objects, searching them from the massive database is of much importance. And a larger CV catalogue is greatly helpful for follow-up studies.

\section{Method} \label{sec:method}
We first introduce a machine learning algorithm called UMAP for screening out the spectra with prominent H$\alpha$ emission features, which leads to a total of 169,509 spectra. The sample data used for supervised machine learning comes from 10 published CV or X-ray binary catalogues which are: \citet{downes1993catalog}; \citet{downes1997catalog}; \citet{downes2001}; \citet{szkody2011}; \citet{jiang2013data}; \citet{breedt20141000}; \citet{coppejans2016statistical}; \citet{han2018cataclysmic}; \citet{ritter2003catalogue}; \citet{hou2020spectroscopically}. The PyHammer v2.0 program is introduced to fit the LAMOST low-resolution spectra to help the manual inspection. There are two rounds in our manual inspection. In the first round, those spectra of late-type main-sequence stars with enhanced chromospheric activities are removed. In the second round, we identify the early-type emission-line stars, pre-main-sequence, and other types of non-CV emission-line objects. For the left, we finally picked out the potential CV candidates.
 
\subsection{H$\alpha$ classification based on machine learning}
As the H$\alpha$ is common to be an emission line for CV spectra, we first screen out the spectra with H$\alpha$ emission features from the whole LAMOST DR6 database.

\subsubsection{Training sample} \label{section:sample}
\begin{figure}[ht!]
	\plotone{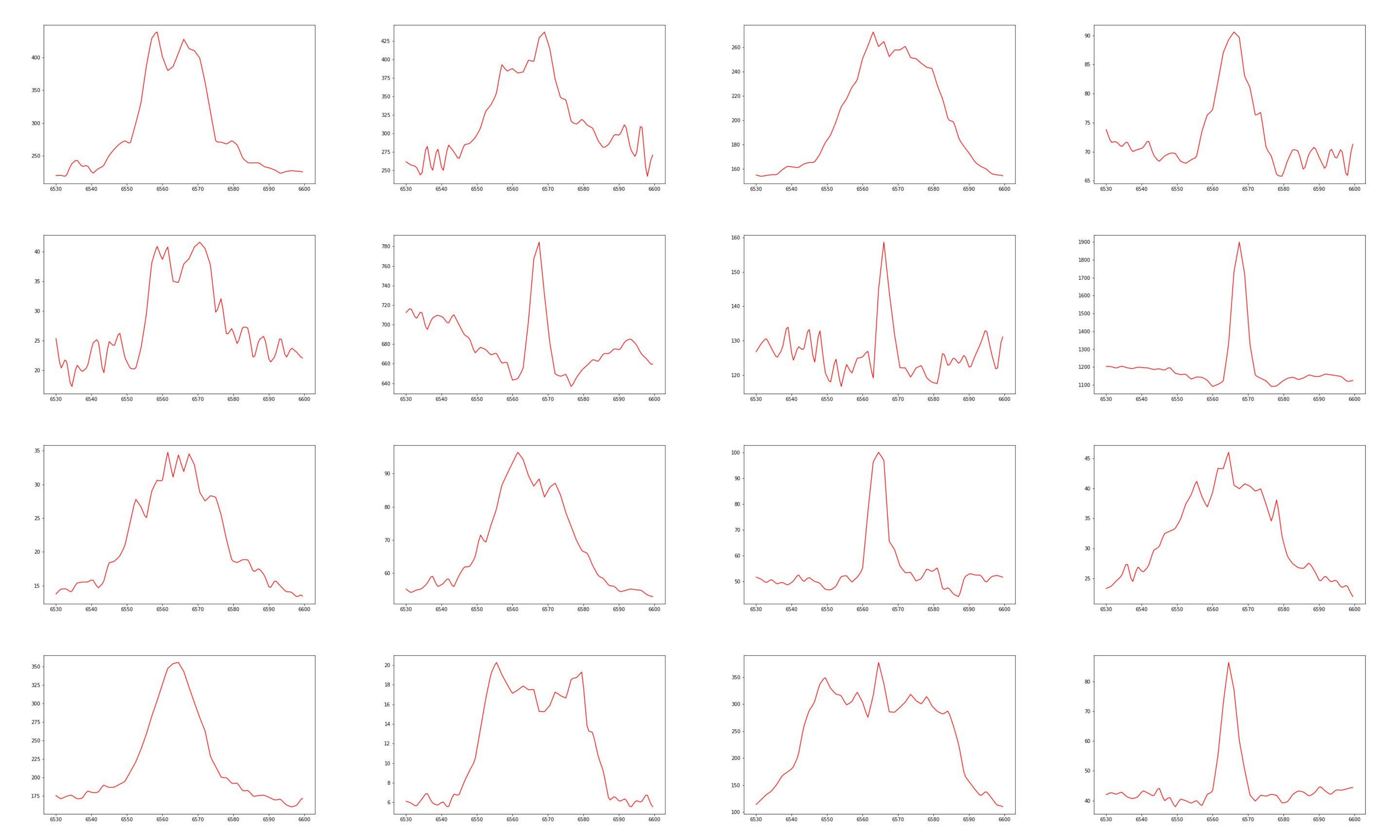}
	\caption{These are 16 examples of 392 sample spectra that are used for machine-learning training. The wavelength segment 6530-6600\,\AA\ associated with the H$\alpha$ emission line is extracted out and used to screen for the same H$\alpha$ emission line features. \label{fig:sample}}
\end{figure}

To obtain initial samples for supervised machine learning, we first cross-match the above mentioned 10 published catalogues with the LAMOST DR6 database. 
We manually examine each of the spectra in the cross-matching result and remove those that do not show H$\alpha$ emission features. Thus, we obtaine 392 specrta with H$\alpha$ emission features. 
Note that not all these spectra are actually CVs, as long as they show H$\alpha$ emission features.
The wavelength range 6530-6600\,\AA\ of these spectra are extracted out as an input to the machine learning program (see Figure \ref{fig:sample}.)  In order to avoid loss of CVs during outbursts, those spectra that have H$\alpha$ wide absorption with emission cores are included in the sample. The double-peaked H$\alpha$ profile is also included in the sample. 

Then we randomly select 1000 spectra in the whole LAMOST DR6 database. We manually check the H$\alpha$ line features, retaining those without typical H$\alpha$ emission features, and obtain 973 spectra that can be used as controls in the supervised machine learning.

\subsubsection{UMAP with k-NN algorithm}
Uniform Manifold Approximation and Projection (UMAP) is a novel machine-learning algorithm for dimension reduction proposed by \citet{mcinnes2018umap}, which is based on the mathematical framework of Riemannian geometry and algebraic topology. UMAP tends to preserve more of the global structure in dimension reduction while many other algorithms like PCA (Principal Components Analysis) seek to preserve the distance structure within the data. UMAP is also capable of preserving the local fine structure while still retaining much of the large-scale global structure like how PCA does. Thus, UMAP has natural advantages for dimension reduction of nonlinear data. 
Through our tests, the PCA algorithm often has the problem that the classification boundary in parameter space is not clear in comparison with the result of the UMAP algorithm, and may cause large contamination. 
Moreover, UMAP also shows superior run time performance for high-dimensional massive data compared to PCA.
We therefore select UMAP as the dimension reduction algorithm for the high-dimensional spectral data. 
K-Nearest Neighbour (k-NN) is a common machine learning classification algorithm. The k-NN classifier is utilized to classify the data after the dimension reduction.

\begin{figure}
	\plotone{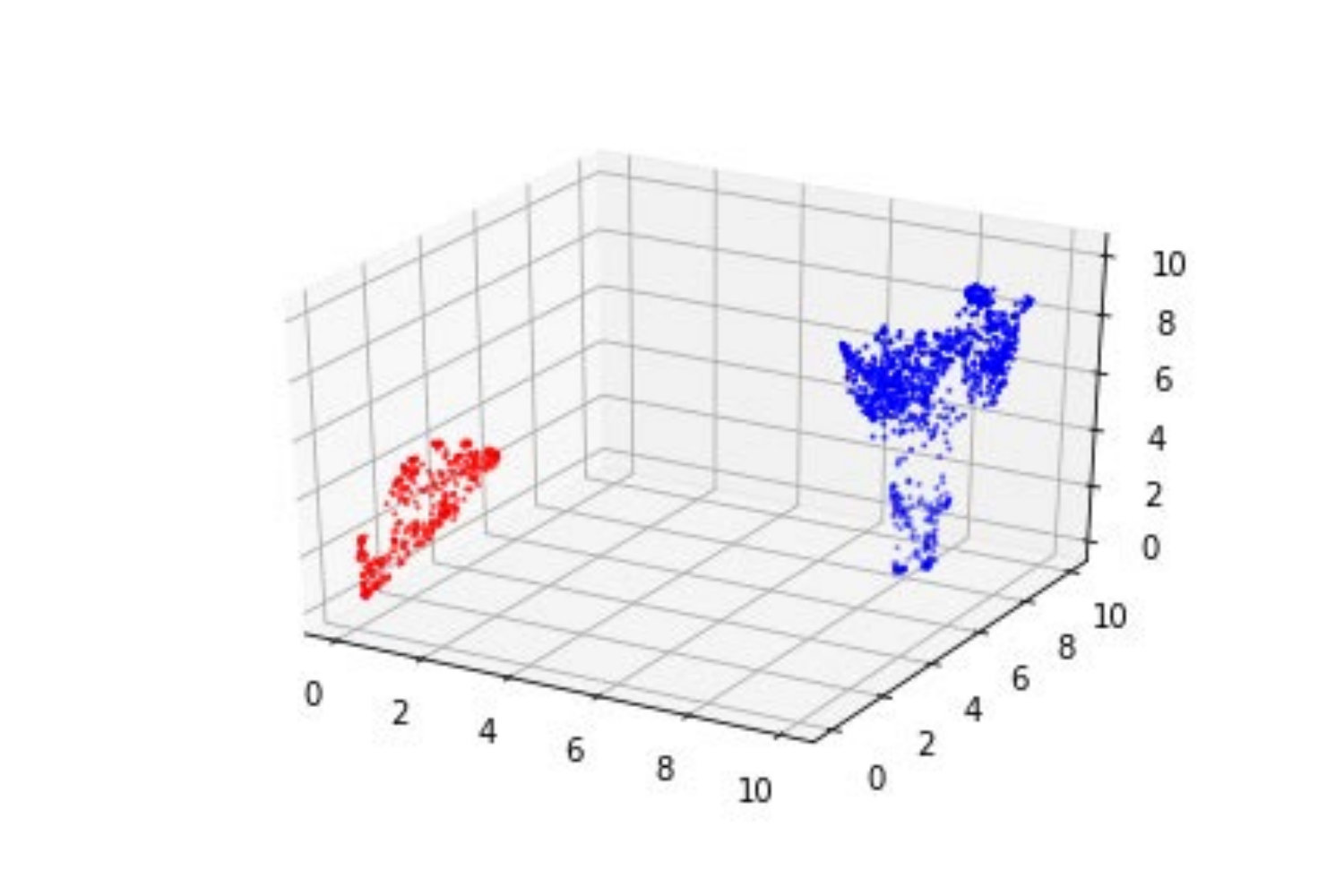}
	\caption{After dimension reduction by UMAP to 3 dimensions, 392 H$\alpha$-emission spectra and 973 randomly selected spectra without H$\alpha$ emission can be separated clearly.  \label{fig:separation}}
\end{figure}

After training with the selected samples and testing several groups of parameters in the UMAP with k-NN algorithm, 392 sample spectra with H$\alpha$ emission features and 973 without H$\alpha$ emission features can be clearly separated in the three-dimensional space, as shown in Figure \ref{fig:separation}. The same algorithm is applied to all spectra in the LAMOST DR6 database, and we obtain a total of 169,509 H$\alpha$-emission spectra.
After we conduct a five-fold cross-validation \citep{shao1993linear}, we obtain an classification accuracy of over 99.6$\%$ in the cross-validation set.

\subsection{PyHammer v2.0}
\begin{figure}
	\plotone{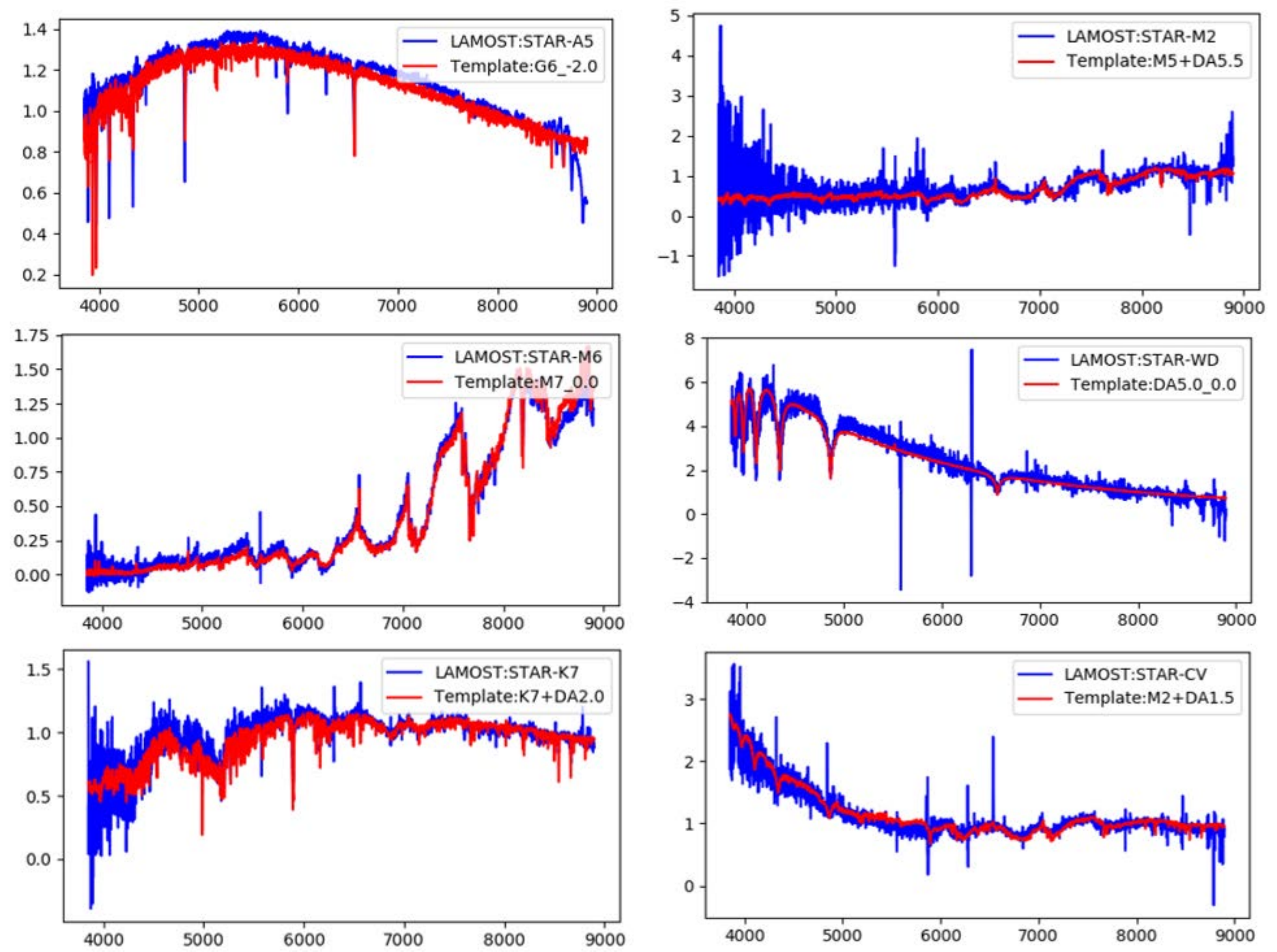}
	\caption{Examples of spectral fitting results obtained by using the Pyhammer v2.0 program. We use such spectra to identify the continuum and screen out the late type main sequence stars affected by chromosphere activity. \label{fig:fitting}}
\end{figure}

The PyHammer v2.0 program was published by \citet{roulston2020classifying} to give a more accurate classification of stellar spectra. We introduce this program to obtain the optimal fitted spectra as a reference for our manual inspection. We have modified the code to fit the LAMOST low-resolution spectra while still use the original templates. In the first round of manual inspection, those spectra that can be fitted well to the G/K/M-type templates considered as normal main-sequence stars with enhanced chromospheric activities, are removed. The spectra of this kind make up about half of the total 169,509 H$\alpha$-emission spectra. 
After removing the chromospherically active stars from the first round, we go through the rest of the spectra to finally pick out CV candidates. Figure \ref{fig:fitting} shows some examples of the spectral fitting using PyHammer v2.0.

\subsection{Manual inspection}
\subsubsection{CV features} \label{subsec:CVfeatures}
Typical CV spectra show a high-temperature continuum superimposed with strong emission lines, which are mainly Balmer and He\,{\small\RN{1}}/He\,{\small\RN{2}} lines. An important feature is that the line width of Balmer and helium emission lines appear to be significantly broad in most CV spectra, i.e. 70\,\AA\ for H$\alpha$. Thus the existence of uniformly broad Balmer and He\,{\small\RN{1}}/He\,{\small\RN{2}} emission lines is a general criterion to identify CVs. The CV spectra usually do not have a steep decrement towards the higher order of the Balmer series as they can be seen up to H8 or higher. The strongest Balmer line may usually be H$\alpha$ to H$\delta$.
The line width of the helium emission lines is usually close to that of the Balmer lines. The double-peaked line profiles present in the Balmer and He emission lines indicate the existence of a rotating accretion disk. According to \citet{knigge2000self}, when the disk is self-occulted, the double-peaked emission lines tend to be narrower and straighter, which may be the reason of the straight double-peaked profiles in some CV spectra. In some cases, the Ca\,{\small\RN{2}} H/K and Ca\,{\small\RN{2}} triple lines also show broad emission, and the superposition of Ca\,{\small\RN{2}} H and H$\epsilon$ lines make H$\epsilon$ appear stronger (close to or beyond H$\delta$) than it originally is. Moreover, when a spectrum is dominated by the companion's contribution (often a M-type spectrum), the blue continuum would show an excess compared to the single M-type star. However, it is possible that at certain orbital phases the emission lines are obscured due to the eclipses, so that the spectrum does not show any CV feature. For very rare cases such as AM CVn systems, they do not have hydrogen lines due to the hydrogen-deficient nature.

For those spectra observed during outbursts, the Balmer lines show absorption through optically thick materials. The central emission cores can be absent at outburst peak and become larger if the spectrum is slightly past outburst. The central emission cores often present in all visible Balmer series, and the H$\alpha$ emission core is usually strong enough to maintain higher than the continuum (this can be called a core emission). The He\,{\small\RN{1}} lines are also affected by wide absorption, while in many cases He\,{\small\RN{2}} $\lambda$4686 is still an emission line. During outbursts, He\,{\small\RN{1}} $\lambda$6678 behaves similarly to H$\alpha$ which usually shows emission and higher than the continuum. In some rare cases, the absorption lines are very strong, which cause the emission core of He\,{\small\RN{1}} $\lambda$6678 to be lower than the continuum. 

Although in some cases the Balmer emission lines can be quite narrow, asymmetric broadening components can be found at the line wings that meet our empirical criteria of 24\,\AA, thus the spectra can be classified as CV candidates. Some of these cases belong to the AM Her-type CVs, for which the TiO absorption bands from the companion can be identified in low states. As mentioned in \citet{warner2003cataclysmic}, the spectral lines in AM Her-type CVs can display a combination of a broad `base' component and narrow components due to a combination of contributions of emission in the stream and pole regions. We find that the known AM Her-type CVs can be identified by the complex line structures and the profile differences between Balmer lines. The emission lines of AM Her-type CVs are more complicated than those of IPs. Strong He\,{\small\RN{2}} $\lambda$4686 lines are common in NLs and magnetic CVs. For AM Her-type CVs, He\,{\small\RN{2}} $\lambda$4686 are often stronger than or comparable to H$\beta$. However, in the spectra of some known AM Her-type CVs, He\,{\small\RN{2}} lines including He\,{\small\RN{2}} $\lambda$4686 are very weak or absent. Accordingly, He\,{\small\RN{2}} lines are significant in the spectra of the known IPs (the strength ratio of He\,{\small\RN{2}} $\lambda$4686/H$\beta$ $\ge$ 0.5, or He\,{\small\RN{2}} $\lambda$4686 is close to but sightly weaker than H$\beta$). 
In NLs, the relative strength of C\,{\small\RN{3}}/N\,{\small\RN{3}} $\lambda$4560 to He\,{\small\RN{2}} $\lambda$4686 emission lines is generally greater than that in magnetic CVs, and in some cases, C\,{\small\RN{3}}/N\,{\small\RN{3}} $\lambda$4560 emission lines are as strong as He\,{\small\RN{2}} $\lambda$4686. 
Also, the emission lines of NLs can be narrower during low states than in higher states \citep{10.1093/mnras/staa612}. As C\,{\small\RN{2}} emission lines appear commonly in NLs but not in DNs, the presence of C\,{\small\RN{2}} lines can be an important feature to distinguish between the NL and the DN subclass.

More spectral line features of different subtypes of CV from the LAMOST observations have been described in \citet{hou2020spectroscopically}. 
For the manual inspection, the empirical criteria we chose to identify the potencial CVs are listed below:

\begin{itemize}
	\item{The apparent line width of H$\alpha$ is greater than or equal to 24\,\AA},
	\item{He\,{\small\RN{1}} $\lambda$6678 emission line presents in the spectra and it is not too narrow (less than 14\,\AA) compared to H$\alpha$; or other He emission lines present},
	\item{Balmer series or He lines show double-peaked profiles},
	\item{Balmer series show wide absorption with emission cores and the emission cores can be seen at the Balmer series higher than H$\gamma$}.
\end{itemize}

Satisfying any of the above criteria will allow us to mark it out in the manual inspection process. For low signal-to-noise ratio (SNR) spectra, when limited Balmer series or only the H$\alpha$ emission line can be identified visually, they will be included as a CV candidate.

\subsubsection{Inspection process}
Each of the 169,509 spectra obtained from the machine learning program has been inspected manually in two rounds. In the first round, we remove the chromospherically active main-sequence stars (which can be fitted well to the G/K/M-type templates) by visually checking the whole spectra. In this process, we compare the LAMOST spectra with the fitted template by PyHammer v2.0 program to check whether there is an excess at the blue end. The remaining spectra, including those well-fitted to the O/B/A/F-type templates and binary star templates, as well as those not well-fitted to the templates (many CVs fall into this category), are retained for the second round. In the second round, we visually check the detailed spectral features carefully and single out those spectra that have CV features. 
Near the galactic plane, stellar spectra may be affected by the H\,{\small\RN{2}} regions. By examining the presence of He absorption lines, we can remove the contaminations of the Oe/Be stars. By judging the line width of H$\alpha$, we can remove the H\,{\small\RN{2}} regions.

\section{Results} \label{sec:results}
Finally, we obtained 475 spectra for 323 objects in total from LAMOST DR6. After checking the SIMBAD database, we find 271 known CVs or CV candidates, and 52 objects are new.

\subsection{Distribution}
\begin{figure}
 	\plotone{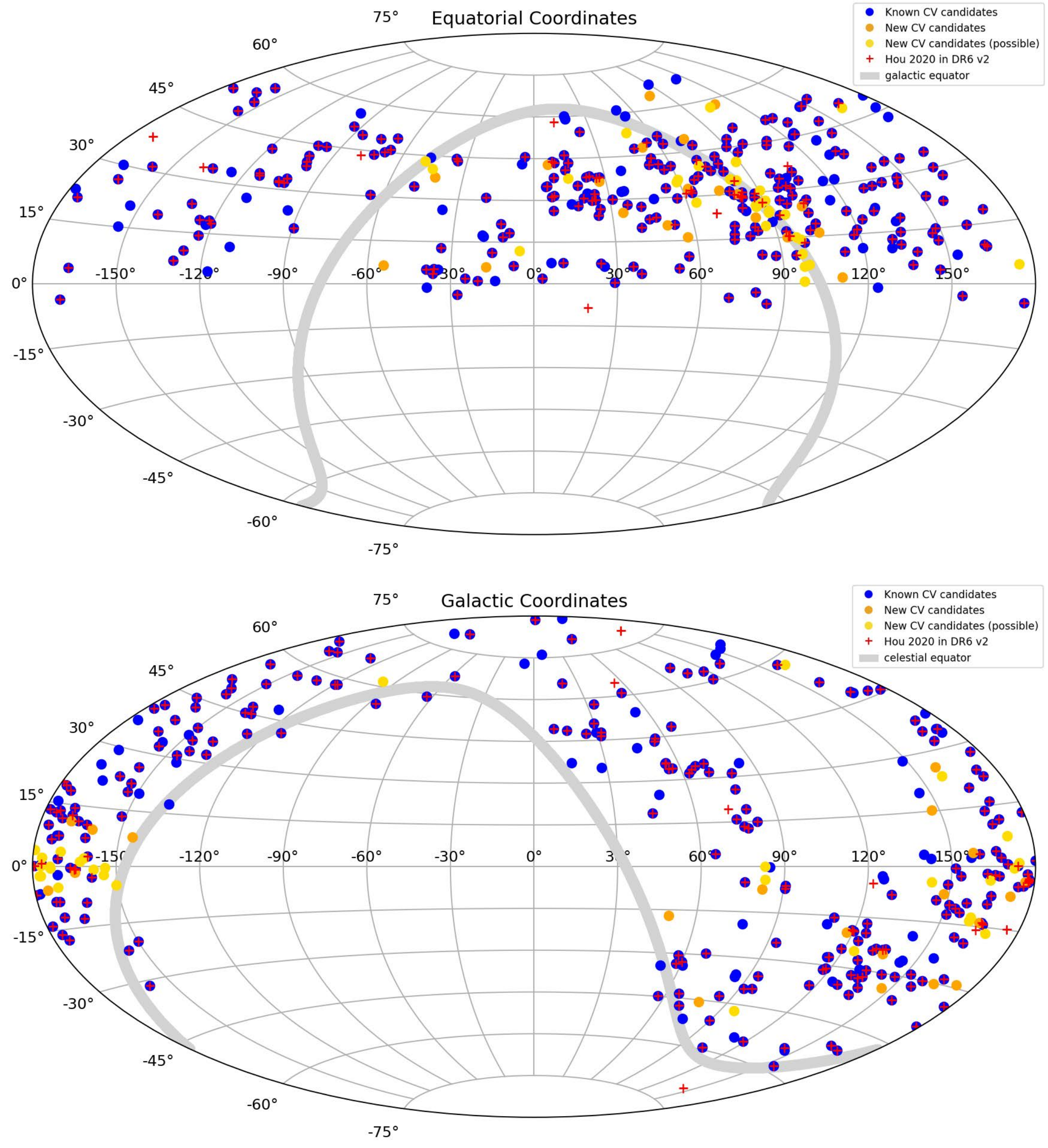}
 	\caption{Spacial distributions of our results drawn in equatorial coordinates and galactic coordinates respectively. The blue spots represent 271 known CV objects from our analysis of LAMOST DR6 v2. The orange spots are 20 confirmed new CV candidates in our results and the yellow spots are 32 possible new CV candidates. The red cross markers indicate the 224 objects which are from cross-matching the result of \citet{hou2020spectroscopically} with LAMOST DR6 v2. \label{fig:spacedistribution}} 
 \end{figure}

\begin{figure}
 	\plotone{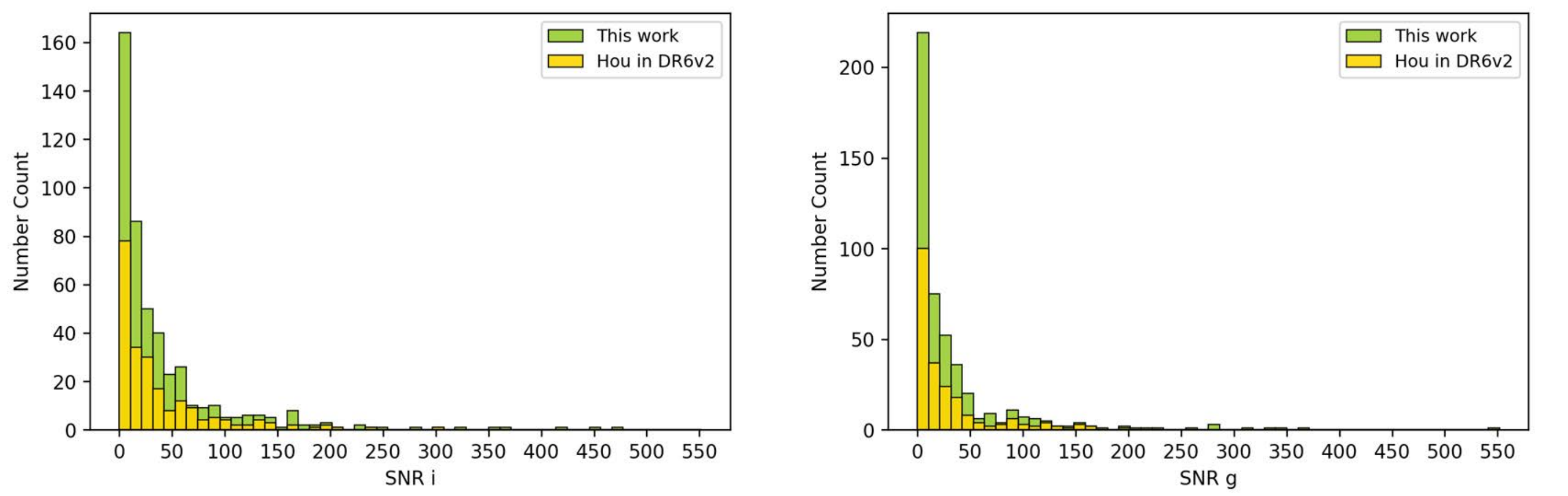}
 	\caption{SNR distributions of g and i band. The green bars represent all 475 spectra in our result. The yellow bars represent the 220 spectra which are from cross-matching the result of \citet{hou2020spectroscopically} with LAMOST DR6 v2. \label{fig:snrdistribution}}
 \end{figure}

 The spatial distribution of the identified CVs is shown in Figure \ref{fig:spacedistribution}. The upper diagram is in equatorial coordinate, while the lower diagram is in galactic coordinate. For the 224 objects indentified by \citet{hou2020spectroscopically} from LAMOST DR5, our results cover 215 of them, and nine of them are not recovered in our results. In Section \ref{sec:hou}, we will specifically discuss why these nine objects are not included in our results. From Figure \ref{fig:spacedistribution}, we can see that the known CVs in LAMOST DR6 v2 are distributed evenly in space. The new CV candidates are more abundant near the galactic plane; however, they are especially rare in the northern galactic latitudes.
 Figure \ref{fig:snrdistribution} shows the SNR distributions in $g$ and $i$ band for the LAMOST data. We can see that with our method, we have found more CV candidates with lower SNRs than those of \citet{hou2020spectroscopically} in LAMOST DR6.
 
\subsection{Cross-match with Hou et al. 2020} \label{sec:hou}
\begin{figure}
	\gridline{\fig{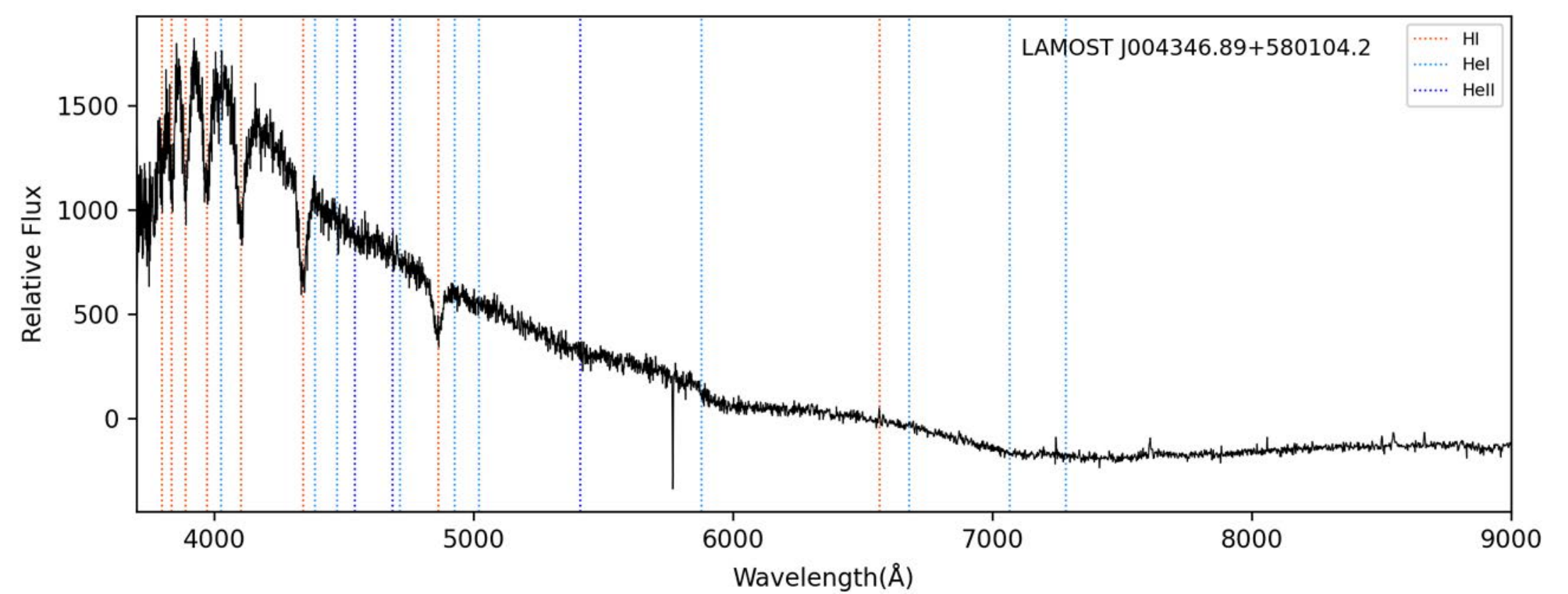}{0.5\textwidth}{}
		\fig{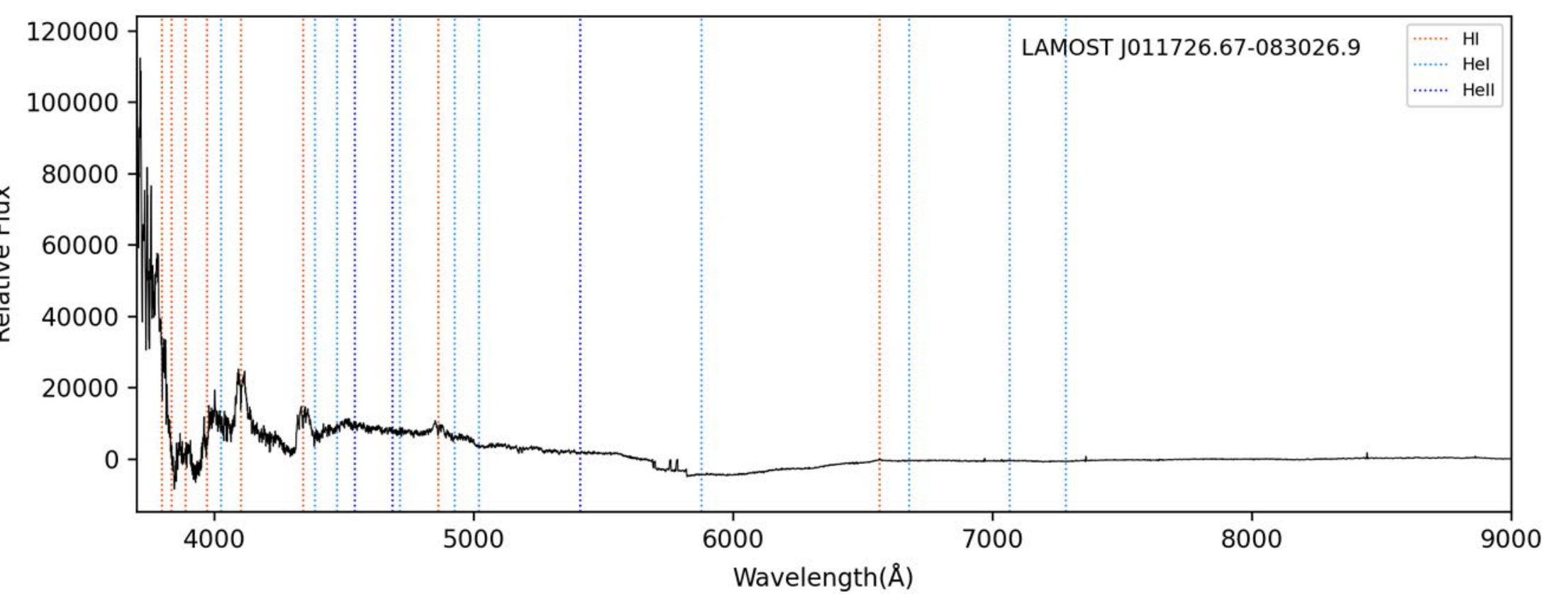}{0.5\textwidth}{}}
	\gridline{\fig{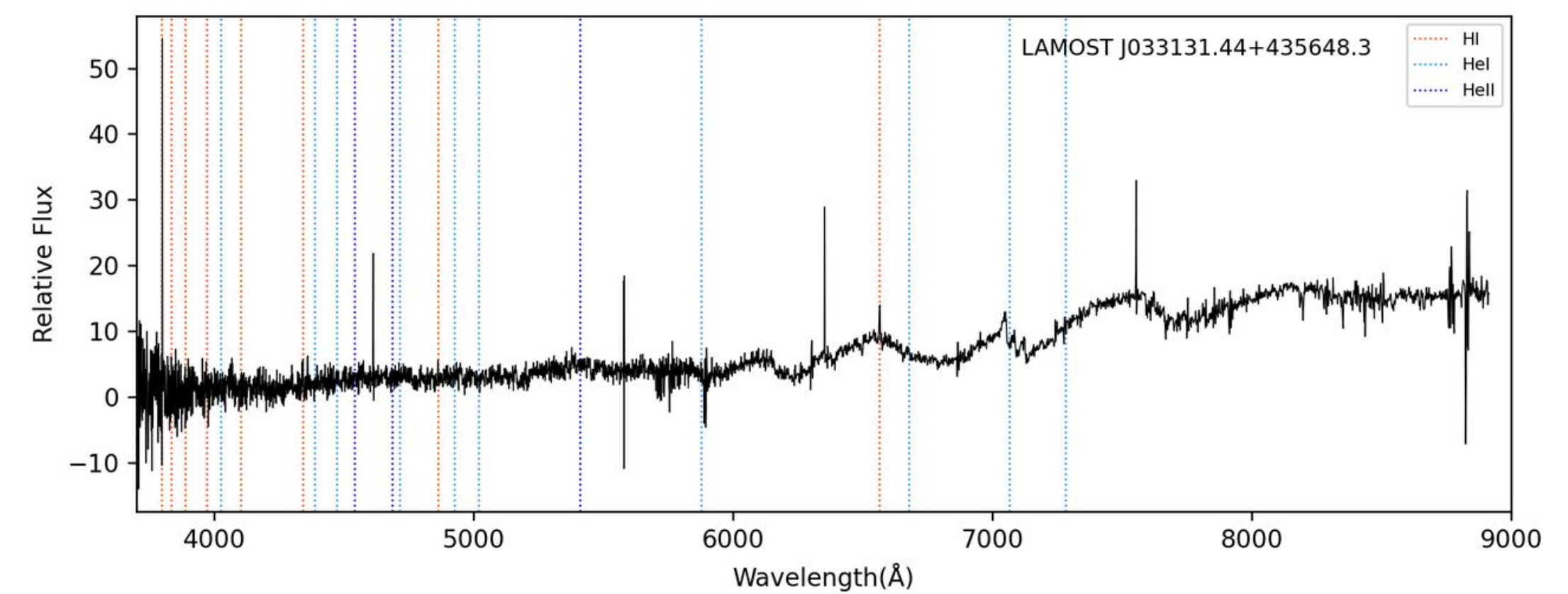}{0.5\textwidth}{}
		\fig{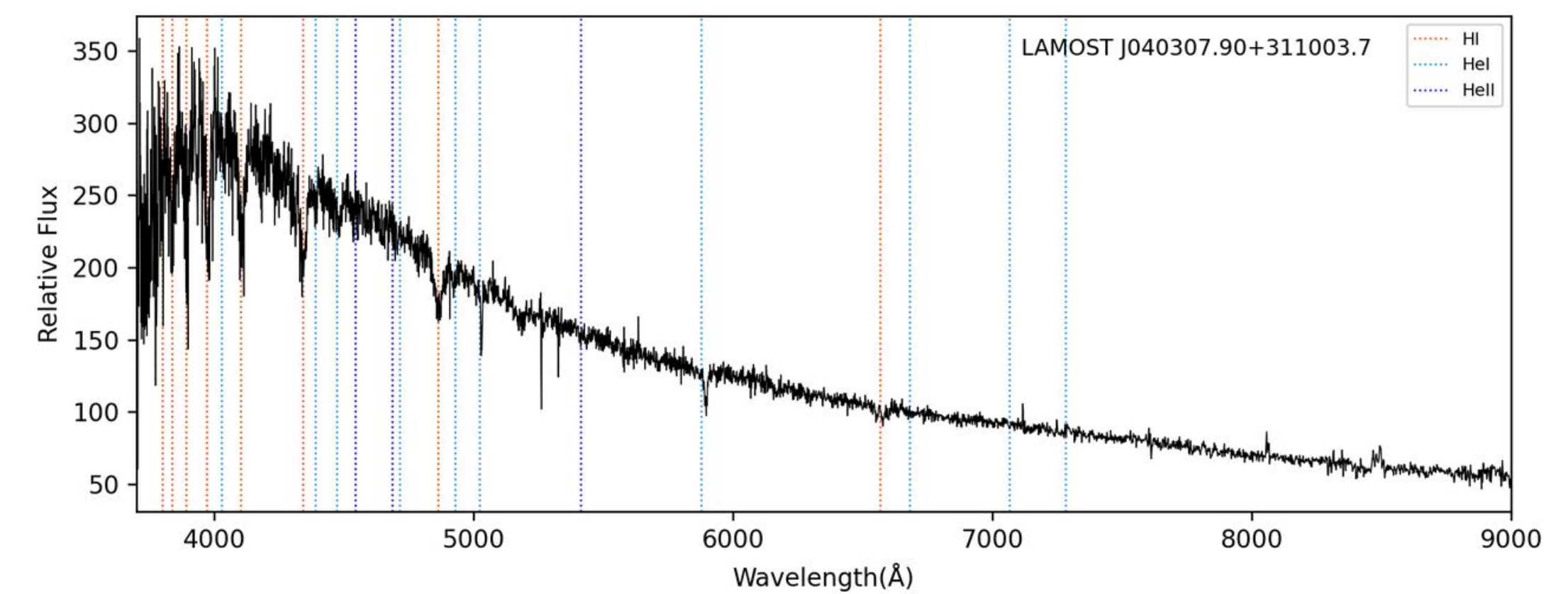}{0.5\textwidth}{}}
	\gridline{\fig{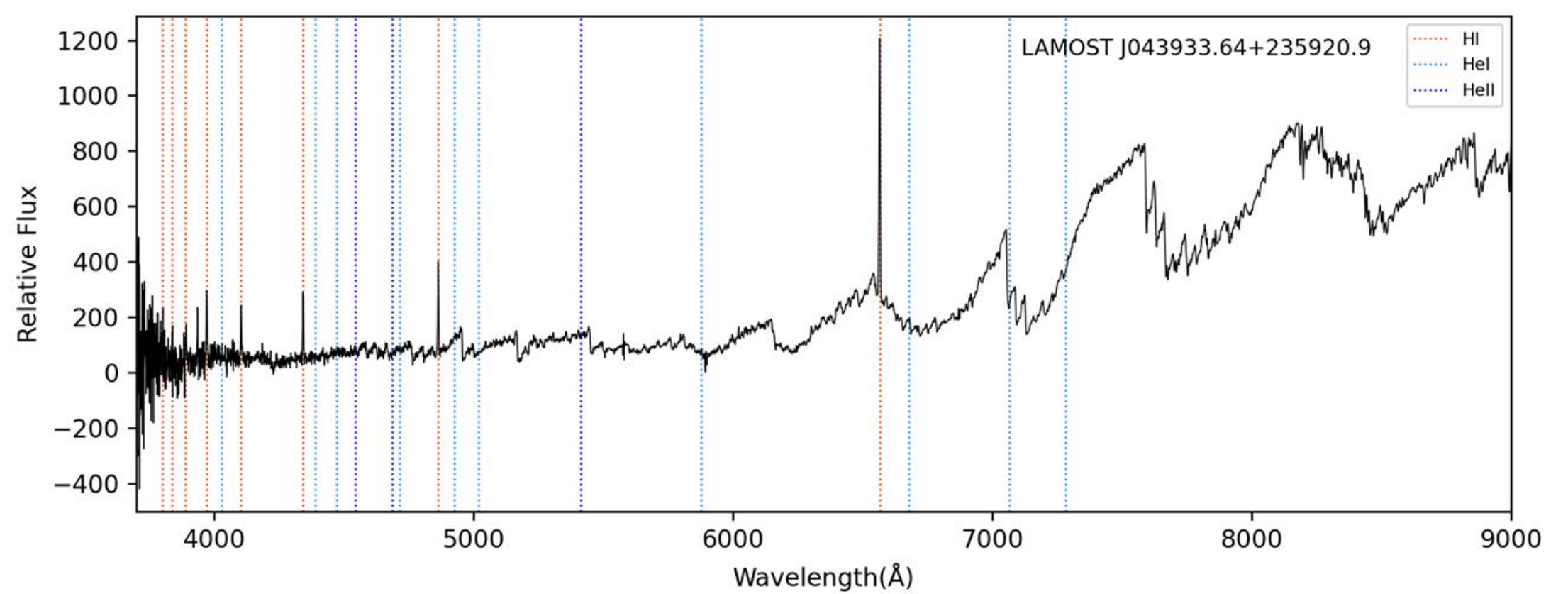}{0.5\textwidth}{}
		\fig{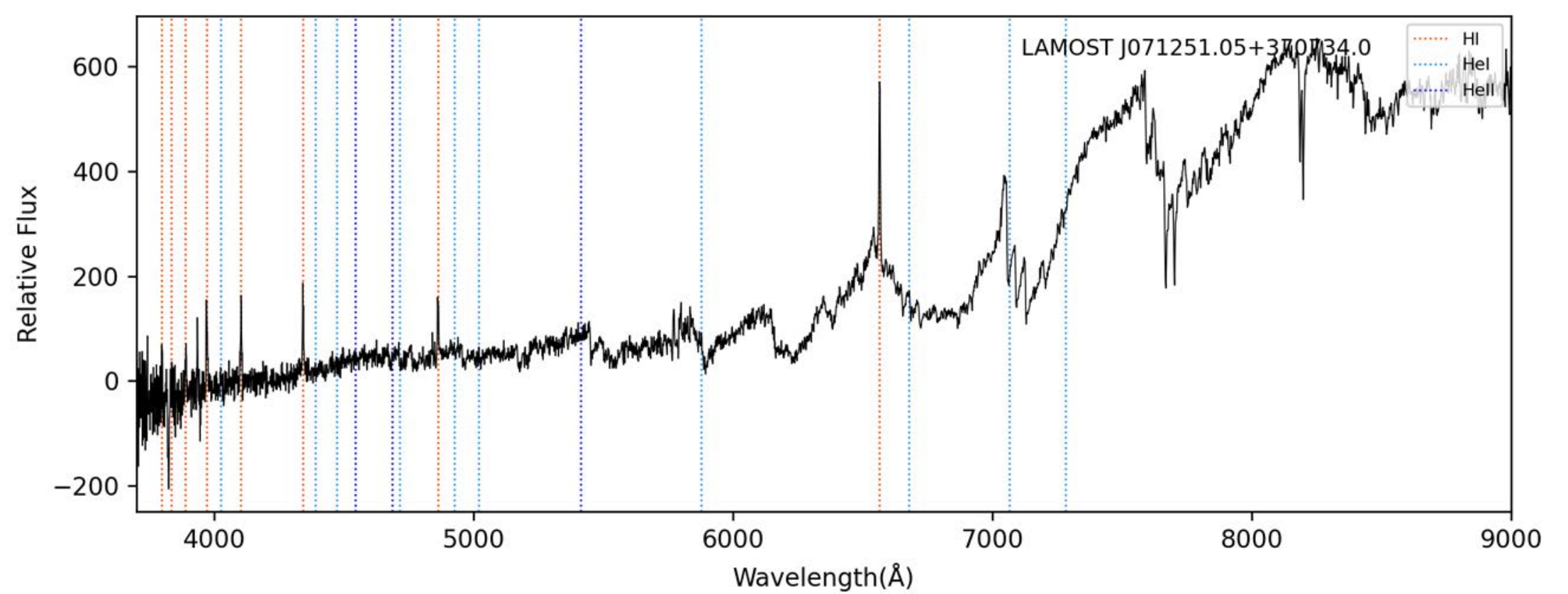}{0.5\textwidth}{}}
	\gridline{\fig{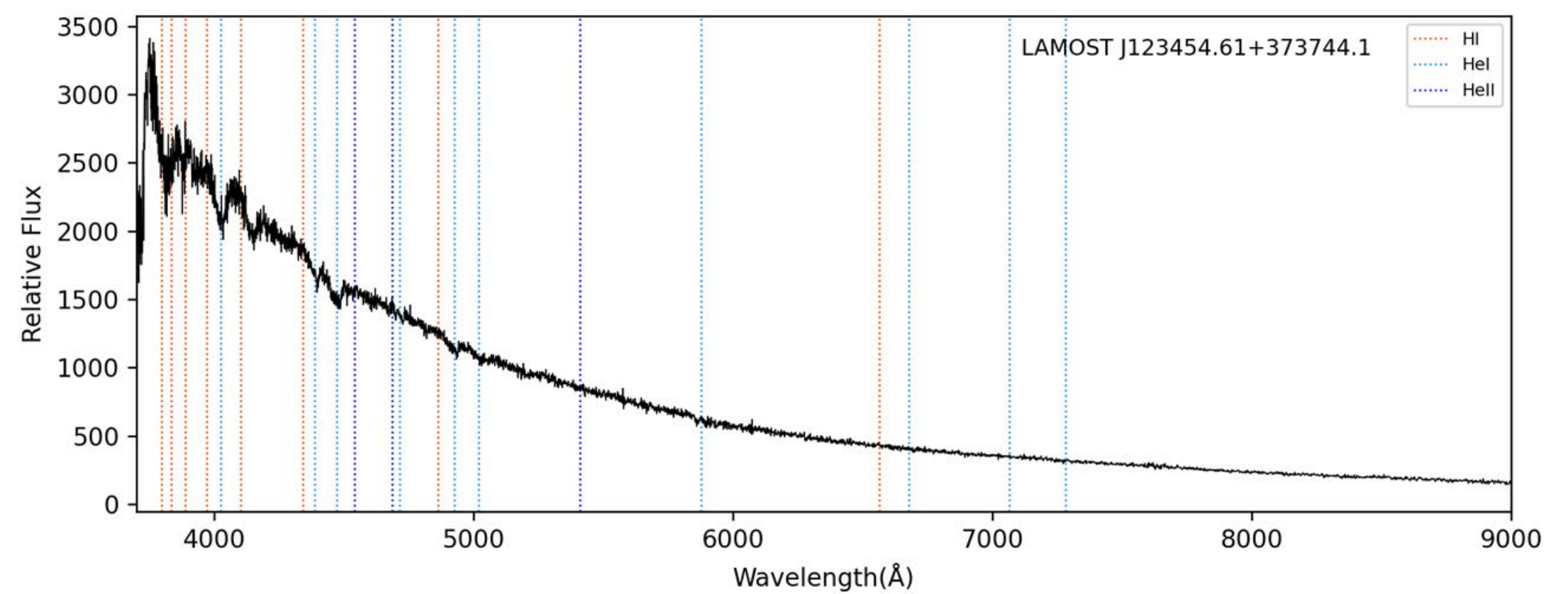}{0.5\textwidth}{}
		\fig{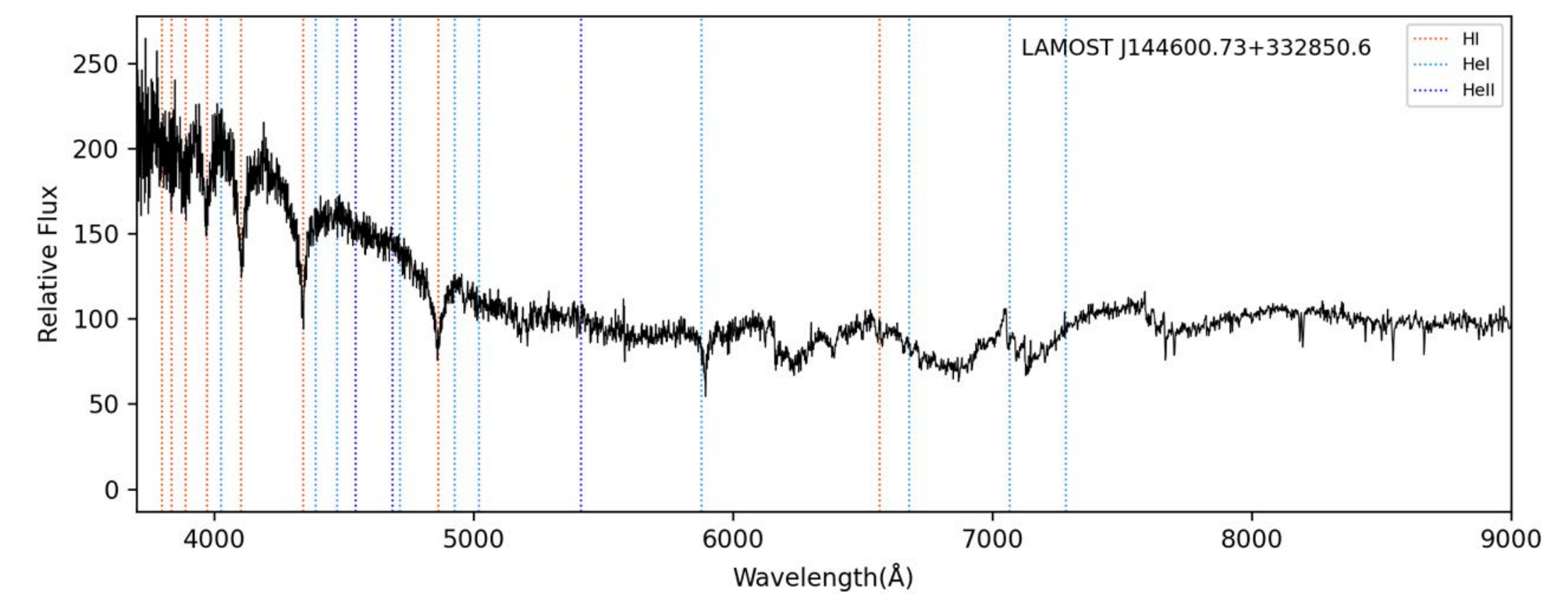}{0.5\textwidth}{}}
	\gridline{\fig{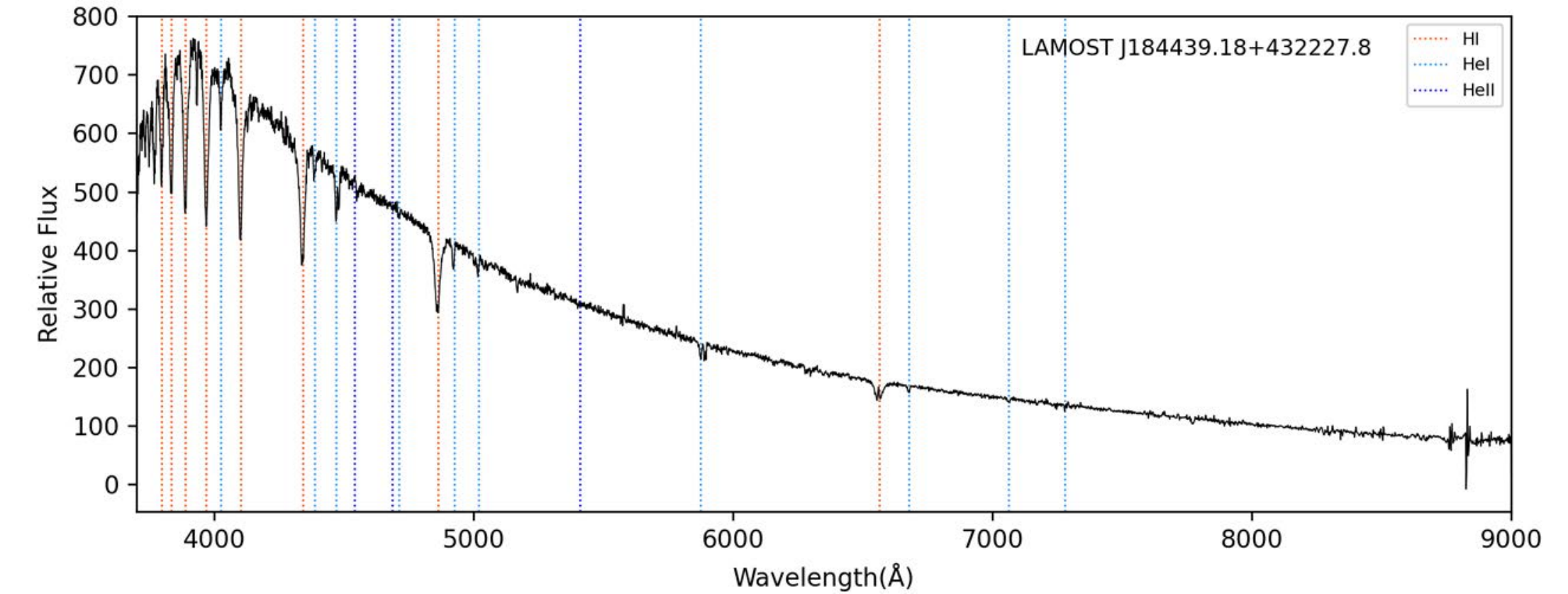}{0.5\textwidth}{}}
	\caption{Spectra of nine objects from the catalogue of \citet{hou2020spectroscopically} but are not found in our result. The wavelength range is from 3700\,\AA\ to 9000\,\AA.   \label{fig:9}}
\end{figure}

\citet{hou2020spectroscopically} have presented 245 CV candidates from LAMOST DR5 and 224 objects are found in LAMOST DR6 v2 by cross-matching the coordinates. Out of these 224 objects, we have 215 objects in common with \citet{hou2020spectroscopically} and nine objects are not included in our results (see Figure \ref{fig:9}). We give the reasons why we lost these nine objects.

The spectrum of J0117 is affected by the bad quality. The red part is flat and does not have any feature, while the blue part shows an abnormally violent change. Therefore, in our manual inspection, we consider this spectrum unreliable and don't include it in our results. 

J1234 is the prototype star and is a representative of the AM CVn-type systems with hydrogen-deficient. Due to the lack of H$\alpha$ emission, our results do not include this type objects. The two spectra of this object do not show visible hydrogen lines on the high-temperature continuum, and He {\small\RN{1}} lines on the blue part show broad absorption with partial filling at the center. 

J0403 has two spectra in which the Balmer lines show deep absorption with emission cores, and He {\small\RN{1}} lines show weak emission cores. Because the H$\alpha$ emission core does not exceed the continuum, the UMAP algorithm failed to screen it out, but it can be a CV candidate as the Balmer emission cores exist up to H8 and He emission lines are marginally visible. 

The spectrum of J1446 shows a component of a white dwarf on the blue part and a M-type companion on the red part, and it is fitted to the M3+DA2.5 template by the PyHammer program. Because its H$\alpha$ line shows an emission core that does not exceed the continuum, the UMAP algorithm fails to identify it. 

J1844 is a known dwarf nova, showing strong absorption with emission cores for both Balmer and He {\small\RN{1}} lines. Its H$\alpha$ line appears to have multiple components, and the emission core at the center is below the continuum so the UMAP algorithm fails to identify it. 

The Balmer lines in the spectrum of J0043 show strong absorption. H$\beta$ and higher members of Balmer series are partially filled up by emission components, while the H$\alpha$ is fully filled in so the UMAP algorithm fails to identify it as a CV. The He emission lines are barely visible.

J0331 is a known dwarf nova. The only spectrum shows very weak Balmer emission lines and the He lines are almost invisible. As the spectrum can be fitted to the M4 main-sequence template and does not show an excess on the blue end, we cannot identify it as a CV from this spectrum.

One of the spectra of J0439 can be fitted with the M6 template. The narrow Balmer emission lines may come from chromospheric activities. The other two spectra of this object show an indistinctive excess on the blue part, and they lack He lines, so we do not include the object in the manual inspection process. But it can still be a possible CV candidate as it presents enhanced Balmer emission lines and excess on the blue part compared with the template.

The spectrum of J0712 does not show an excess on the blue end relative to the M5 template, so we do not include it in the manual inspection process. But the enhanced Balmer emission and absorption center at higher Balmer lines still make it a possible CV candidate.

In conclusion, constrained by our machine learning samples, the UMAP algorithm has some limitations to perfectly classify the spectra that have strong wide absorption in H$\alpha$ line with weak emission cores. 
If the spectra are dominated by the main-sequence star and they don't show an excess on the blue part or any obvious emission lines, we can not identify it as a CV in the manual inspection process. This phenomenon may be due to high inclination and specific orbital phases, which is relatively rare for CVs. In addition, a few spectra are affected by bad quality or very low SNRs. We considered them not trustworthy in the manual inspection and hence do not include them.

\subsection{Cross-match with other catalogues}
We examine the cases which were previously classified as CV objects but are not included in our results from analysis of the LAMOST spectra. 
We cross-match the catalogs from the literature listed in section \ref{section:sample} with LAMOST DR6. From the cross-matching result, a total of 230 objects are classified as CVs or CV candidates in the SIMBAD database, in which 193 objects are found by us. For the rest 37 objects, three objects show CV features in their LAMOST spectra, and are discribed as follows.
J1234, the prototype star of the hydrogen-deficient AM CVn system, and the dwarf nova J1844 have been explained in Section \ref{sec:hou}. J0713 (ASASSN-14mv) is also an AM CVn-type object \citep{denisenko2014master}, which shows strong He {\small\RN{1}} and He {\small\RN{2}} emission lines but doesn't show hydrogen lines in its spectrum. 
The other 34 objects do not show features that could be identified as CVs in their LAMOST spectra, though they have been identified by spectroscopy or photometry. The likely reason is that they are too faint or in a fainter state that LAMOST can not obtain relatively high SNR spectra. Also, it is possible that the emission regions are obscured, or they were incorrectly identified as CVs in the catalogue (e.g. 2MASS J01095921+2801244 in \citet{downes2001}). 

\subsection{New CV Candidates}	
In our result, 52 objects are newly identified as CV candidates. Among them, 20 objects that can be classified as CVs as they show broad Balmer and He emission lines, are listed in Table \ref{confirmed}. The other 32 objects, which can be classified as possible CV candidates, are listed in Table \ref{possible}, as they lack some of the distinct features of CVs, or the spectra may be affected by abnormal circumstances, or they have a lower SNR. The 21 spectra of 20 confirmed CVs and 52 spectra of 32 possible CV candidates are plotted in Figures \ref{fig:newcv_confirmed} and \ref{fig:newcv_possible} in the Appendix, respectively.

\subsection{Catalogue Description}
The 20 new confirmed CVs are presented in Table \ref{confirmed}, while the 32 new possible CV candidates are shown in Table \ref{possible}. In Table \ref{catalogue}, we present the 271 previously known ones.
The first two columns of each table are the LAMOST obsid (a unique spectra ID) and designation. The 16 spectra from LAMOST DR6 v1 which no longer exist in LAMOST DR6 v2 are marked by an asterisk before their obsid number. Column (3)-(5) are the MJD, right ascension and declination, respectively. 
Columns (6)-(8) are the CV subtype identified by us, \citet{hou2020spectroscopically}, and from SIMBAD, respectively.  We mark all the new CV candidates with a single asterisk and label the 52 spectra of 32 new possible CV candidates as `CV?' in column (6). If a source has more than one spectrum but those spectra do not show CV characteristics, we leave the row blank in column (6).
Columns (9) and (10) are the median magnitude and the magnitude difference between the maximum and minimum taken from the ZTF DR4 data, respectively. The ZTF data are obtained by best cross-matching coordinates within 2 arcsec. 

\begin{longrotatetable}
	\begin{deluxetable*}{rlrrrlrrrr}
		\tablecaption{New confirmed CVs from LAMOST DR6 \label{confirmed}}
		\tablewidth{700pt}
		\tabletypesize{\scriptsize}
		\tablehead{
			\colhead{obsid} & \colhead{designation} & \colhead{mjd} & 
			\colhead{ra} & \colhead{dec} &
			\colhead{subtype$^a$} & \colhead{subtype$^b$} & \colhead{subtype$^c$} &
			\colhead{medianmag$^d$} & \colhead{magdifference$^e$} \\
			\colhead{} & \colhead{(LAMOST)} & \colhead{} & 
			\colhead{} & \colhead{} &
			\colhead{(this work)} & \colhead{(Hou2020)} & \colhead{(SIMBAD)} &
			\colhead{(ZTF)} & \colhead{(ZTF)}
		} 
		\startdata
	472302069 & J002356.46+424707.8 & 57687 & 5.985285 & 42.7855 & *DN\_burst &       & none  &       &  \\
	354712003 & J014841.06+363350.0 & 57281 & 27.17111 & 36.56389 & *DN\_burst &       & none  & 15.68463 & 0.52242 \\
	371007239 & J014841.06+363350.0 & 57309 & 27.17111 & 36.56389 & *DN\_burst &       & none  &       &  \\
	600008212 & J021633.89+251346.5 & 58067 & 34.14122 & 25.22961 & *MCV(polar) &       &       & 18.80127 & 1.86958 \\
	353714066 & J032034.51+203607.2 & 57278 & 50.14383 & 20.602 & *MCV(polar) &       & none  &       &  \\
	587803184 & J032641.99+475344.6 & 58018 & 51.67497 & 47.89573 & *MCV(polar) &       &       & 18.19204 & 2.89637 \\
	605413195 & J034713.84+161108.2 & 58074 & 56.80769 & 16.18563 & *CV   &       & none  & 16.91274 & 4.93641 \\
	203507028 & J040901.83+323955.6 & 56660 & 62.25766 & 32.66547 & *NL   &       &       & 14.66591 & 0.60534 \\
	631412182 & J045452.12+491709.2 & 58135 & 73.71717 & 49.28589 & *NL   &       &       & 17.20238 & 5.18567 \\
	642708209 & J045745.94+312517.2 & 58156 & 74.44143 & 31.42146 & *NL   &       &       & 17.7648 & 0.92789 \\
	593404034 & J051949.51+642921.6 & 58045 & 79.95633 & 64.48935 & *NL   &       & none  & 17.29306 & 1.69211 \\
	19714085 & J052346.16+295034.5 & 55917 & 80.94235 & 29.84292 & *CV   &       &       & 17.48023 & 0.13482 \\
	198710203 & J053532.64+220006.8 & 56652 & 83.88601 & 22.0019 & *CV   &       &       & 19.52661 & 1.4481 \\
	400115034 & J061249.27+150642.1 & 57391 & 93.20533 & 15.11172 & *DN\_burst &       & none  & 17.73853 & 0.33707 \\
	108714229 & J065237.19+243622.1 & 56309 & 103.155 & 24.60617 & *CV   &       & none  & 15.70214 & 1.40907 \\
	422415102 & J070133.82+161010.3 & 57427 & 105.3909 & 16.16954 & *DN\_burst &       &       & 18.35857 & 4.36762 \\
	631513222 & J071553.88+581606.4 & 58135 & 108.9745 & 58.26846 & *MCV(IP) &       &       & 16.17882 & 1.2921 \\
	600414109 & J072304.77+020437.4 & 58067 & 110.7699 & 2.077058 & *NL   &       & none  & 15.48614 & 0.38905 \\
	587208191 & J202300.37+063142.4 & 58016 & 305.7516 & 6.528447 & *NL?  &       & none  &       &  \\
	593810181 & J211249.93+374225.8 & 58046 & 318.2081 & 37.70719 & *DN\_burst &       &       & 16.50493 & 0.41542 \\
	173307033 & J225017.79+060645.8 & 56598 & 342.5741 & 6.112739 & *CV   &       & none  & 16.19793 & 0.38934 \\
		\enddata
		\tablecomments{
			\\ $^a$ If a source has more than one spectrum but the spectrum has no visible CV characteristics, we leave the row blank. If the row is the spectrum of a new CV candidate, we mark it with a single asterisk in this column. For new confirmed CV candidates, we particularly give the subtype inferred from the spectrum. Two poorly known objects discussed in section \ref{sec:discussion2} are marked by double asterisks.
			\\ $^b$ The column is the CV subtype for the corresponding LAMOST DR6 spectra given by \citet{hou2020spectroscopically}.  
			\\ $^c$ The column is the CV subtype from SIMBAD, or from the literature subtype given in the catalogue of \citet{hou2020spectroscopically}.  If the source exists in SIMBAD but is given a type other than CV, it is marked as `none'. 
			\\ $^d$ This is the median magnitude taken from the ZTF DR4. The ZTF data is obtained by best matching within 2 arcsec with our result.
			\\ $^e$ This is the difference between the maximum and minimum magnitude taken from the ZTF DR4. }
	\end{deluxetable*}
\end{longrotatetable}

\begin{longrotatetable}
	\begin{deluxetable*}{rlrrrrrrrr}
		\tablecaption{New possible CV candidates from LAMOST DR6 \label{possible}}
		\tablewidth{700pt}
		\tabletypesize{\scriptsize}
		\tablehead{
			\colhead{obsid} & \colhead{designation} & \colhead{mjd} & 
			\colhead{ra} & \colhead{dec} &
			\colhead{subtype$^a$} & \colhead{subtype$^b$} & \colhead{subtype$^c$} &
			\colhead{medianmag$^d$} & \colhead{magdifference$^e$} \\
			\colhead{} & \colhead{(LAMOST)} & \colhead{} & 
			\colhead{} & \colhead{} &
			\colhead{(this work)} & \colhead{(Hou2020)} & \colhead{(SIMBAD)} &
			\colhead{(ZTF)} & \colhead{(ZTF)}
		} 
		\startdata
		    18508092 & J005651.51+375620.4 & 55914 & 14.21465 & 37.93903 & *CV?  &       &       & 19.83481 & 0.52011 \\
		16115150 & J031108.31+530719.3 & 55910 & 47.78466 & 53.12206 & *CV?  &       & none  &       &  \\
		181509193 & J035603.23+351450.5 & 56617 & 59.01347 & 35.24738 & *CV?  &       &       & 14.89156 & 1.63763 \\
		212414002 & J035603.23+351450.5 & 56684 & 59.01347 & 35.24738 & *CV?  &       &       &       &  \\
		182009193 & J035603.23+351450.6 & 56617 & 59.01347 & 35.24741 & *CV?  &       &       &       &  \\
		194714002 & J035603.23+351450.6 & 56645 & 59.01347 & 35.24741 &       &       &       &       &  \\
		194716052 & J035933.23+360323.1 & 56645 & 59.88847 & 36.05643 & *CV?  &       &       & 14.9497 & 0.10124 \\
		181912155 & J035933.23+360323.1 & 56617 & 59.88847 & 36.05643 &       &       &       &       &  \\
		182012155 & J035933.23+360323.1 & 56617 & 59.88847 & 36.05643 &       &       &       &       &  \\
		194616052 & J035933.23+360323.1 & 56645 & 59.88847 & 36.05643 &       &       &       &       &  \\
		194816052 & J035933.23+360323.1 & 56645 & 59.88847 & 36.05643 &       &       &       &       &  \\
		17506020 & J041407.98+280120.3 & 55913 & 63.53326 & 28.02231 & *CV?  &       &       & 15.74186 & 0.0913 \\
		119012144 & J041407.98+280120.3 & 56331 & 63.53326 & 28.02231 &       &       &       &       &  \\
		214612144 & J041407.98+280120.4 & 56687 & 63.53325 & 28.02234 &       &       &       &       &  \\
		214712144 & J041407.98+280120.4 & 56687 & 63.53325 & 28.02234 &       &       &       &       &  \\
		39906248 & J044644.51+400141.0 & 55967 & 71.68547 & 40.02806 & *CV?  &       &       & 14.14061 & 0.05539 \\
		68812240 & J052443.05+350123.3 & 56219 & 81.17938 & 35.02314 & *CV?  &       &       & 18.15303 & 1.49098 \\
		269314147 & J053416.07+342316.1 & 56981 & 83.567 & 34.38782 & *CV?  &       &       &       &  \\
		132413240 & J053416.08+342316.1 & 56358 & 83.567 & 34.38781 & *CV?  &       &       & 15.34793 & 0.1759 \\
		300813240 & J053416.08+342316.1 & 57042 & 83.567 & 34.38781 & *CV?  &       &       &       &  \\
		16202089 & J054311.80+254445.8 & 55910 & 85.79917 & 25.74608 & *CV?  &       &       & 14.88737 & 0.11771 \\
		81412213 & J054311.80+254445.8 & 56254 & 85.79917 & 25.74608 & *CV?  &       &       & 14.88737 & 0.11771 \\
		406715055 & J054345.28+252601.0 & 57400 & 85.93869 & 25.43363 & *CV?  &       &       & 17.47745 & 0.10737 \\
		108002166 & J054627.34+190011.9 & 56308 & 86.61396 & 19.00333 & *CV?  &       &       & 13.49713 & 0.09616 \\
		81116232 & J055303.79+262659.4 & 56253 & 88.26582 & 26.44984 & *CV?  &       &       & 20.81788 & 1.02977 \\
		124105039 & J055444.33+401846.5 & 56343 & 88.68471 & 40.31292 & *CV?  &       &       & 14.74165 & 0.1076 \\
		83211001 & J055444.33+401846.5 & 56256 & 88.68471 & 40.31292 &       &       &       &       &  \\
		124405039 & J055455.84+401952.3 & 56343 & 88.73269 & 40.33121 & *CV?  &       &       & 14.94291 & 0.19556 \\
		83111007 & J055455.84+401952.3 & 56256 & 88.73269 & 40.33121 &       &       &       &       &  \\
		116308162 & J055834.38+232004.9 & 56321 & 89.64327 & 23.33472 & *CV?  &       &       & 15.45488 & 0.11825 \\
		116308163 & J055924.27+233707.6 & 56321 & 89.85116 & 23.61878 & *CV?  &       &       & 14.19076 & 0.12284 \\
		81108248 & J055924.73+234341.1 & 56253 & 89.85307 & 23.72811 & *CV?  &       &       & 19.08045 & 1.30034 \\
		11107016 & J060116.87+271327.1 & 55898 & 90.32029 & 27.22422 & *CV?  &       &       & 18.20835 & 0.12946 \\
		127912024 & J060214.93+303536.8 & 56349 & 90.56224 & 30.59356 & *CV?  &       &       & 14.32692 & 0.10813 \\
		485106105 & J062104.00+221554.7 & 57720 & 95.2667 & 22.26522 & *CV?  &       &       &       &  \\
		486306105 & J062104.00+221554.7 & 57723 & 95.2667 & 22.26522 & *CV?  &       &       &       &  \\
		486806105 & J062104.00+221554.7 & 57724 & 95.2667 & 22.26522 & *CV?  &       &       &       &  \\
		497206105 & J062104.00+221554.7 & 57739 & 95.2667 & 22.26522 & *CV?  &       &       &       &  \\
		506606105 & J062104.00+221554.7 & 57718 & 95.2667 & 22.26522 & *CV?  &       &       &       &  \\
		508806105 & J062104.00+221554.7 & 57757 & 95.2667 & 22.26522 & *CV?  &       &       &       &  \\
		509806105 & J062104.00+221554.7 & 57758 & 95.2667 & 22.26522 & *CV?  &       &       &       &  \\
		510706105 & J062104.00+221554.7 & 57759 & 95.2667 & 22.26522 & *CV?  &       &       &       &  \\
		603206105 & J062104.00+221554.7 & 58076 & 95.2667 & 22.26522 & *CV?  &       &       &       &  \\
		606206105 & J062104.00+221554.7 & 58075 & 95.2667 & 22.26522 & *CV?  &       &       & 14.57116 & 0.03718 \\
		615306105 & J062104.00+221554.7 & 58098 & 95.2667 & 22.26522 & *CV?  &       &       &       &  \\
		617806105 & J062104.00+221554.7 & 58104 & 95.2667 & 22.26522 & *CV?  &       &       &       &  \\
		620206105 & J062104.00+221554.7 & 58108 & 95.2667 & 22.26522 & *CV?  &       &       &       &  \\
		631106105 & J062104.00+221554.7 & 58134 & 95.2667 & 22.26522 & *CV?  &       &       &       &  \\
		644906105 & J062104.00+221554.7 & 58161 & 95.2667 & 22.26522 & *CV?  &       &       &       &  \\
		483206105 & J062104.00+221554.7 & 57715 & 95.2667 & 22.26522 &       &       &       &       &  \\
		490506105 & J062104.00+221554.7 & 57727 & 95.2667 & 22.26522 &       &       &       &       &  \\
		501106105 & J062104.00+221554.7 & 57748 & 95.2667 & 22.26522 &       &       &       &       &  \\
		525406105 & J062104.00+221554.7 & 57781 & 95.2667 & 22.26522 &       &       &       &       &  \\
		529006105 & J062104.00+221554.7 & 57787 & 95.2667 & 22.26522 &       &       &       &       &  \\
		529606105 & J062104.00+221554.7 & 57788 & 95.2667 & 22.26522 &       &       &       &       &  \\
		280405205 & J062549.60+145347.4 & 57006 & 96.45668 & 14.89651 & *CV?  &       &       & 16.60911 & 0.40176 \\
		101201021 & J062811.33+004654.5 & 56295 & 97.04721 & 0.781809 & *CV?  &       &       & 14.90546 & 0.08848 \\
		419307028 & J062842.43+093209.2 & 57422 & 97.17679 & 9.535889 & *CV?  &       &       & 16.76195 & 1.77769 \\
		44314014 & J063018.87+053831.3 & 55983 & 97.57867 & 5.642036 & *CV?  &       &       & 13.31386 & 0.10978 \\
		442712204 & J063018.88+053831.4 & 57468 & 97.57867 & 5.642059 &       &       &       &       &  \\
		406804155 & J063209.32+140334.5 & 57400 & 98.03886 & 14.05961 & *CV?  &       &       & 18.28807 & 3.54584 \\
		44515210 & J063722.97+060755.8 & 55983 & 99.34572 & 6.132189 & *CV?  &       &       &       &  \\
		111003180 & J065815.75+575142.9 & 56316 & 104.5656 & 57.86194 & *CV?  &       &       & 16.01802 & 0.18338 \\
		110903180 & J065815.75+575142.9 & 56316 & 104.5656 & 57.86194 &       &       &       &       &  \\
		217303180 & J065815.75+575142.9 & 56699 & 104.5656 & 57.86194 &       &       &       &       &  \\
		629705190 & J104751.79+483503.7 & 58130 & 161.9658 & 48.58438 & *CV?  &       & none  & 18.48886 & 0.34744 \\
		525913199 & J113835.67+044454.7 & 57782 & 174.6486 & 4.748538 & *CV?  &       & none  & 19.32805 & 3.88678 \\
		587910071 & J204502.35+430520.6 & 58018 & 311.2598 & 43.08906 & *CV?  &       & none  & 15.39373 & 1.23171 \\
		594305076 & J210306.10+402306.3 & 58050 & 315.7754 & 40.38511 & *CV?  &       &       & 17.1679 & 0.09375 \\
		355016173 & J233900.38+115707.2 & 57283 & 354.7516 & 11.95201 & *CV?  &       & none  & 18.19033 & 0.27358 \\
		\enddata
		\tablecomments{\\ (This table is available in its entirety in machine-readable form.)	}
	\end{deluxetable*}
\end{longrotatetable}

\begin{longrotatetable}
	\begin{deluxetable*}{rlrrrlllrrll}
		\tablecaption{Previously known CV candidates from LAMOST DR6 (examples of 15 rows) \label{catalogue}}
		\tablewidth{700pt}
		\tabletypesize{\scriptsize}
		\tablehead{
			\colhead{obsid} & \colhead{designation} & \colhead{mjd} & 
			\colhead{ra} & \colhead{dec} &
			\colhead{subtype$^a$} & \colhead{subtype$^b$} & \colhead{subtype$^c$} &
			\colhead{medianmag$^d$} & \colhead{magdifference$^e$}\\
			\colhead{} & \colhead{(LAMOST)} & \colhead{} & 
			\colhead{} & \colhead{} &
			\colhead{(this work)} & \colhead{(Hou2020)} & \colhead{(SIMBAD)} &
			\colhead{(ZTF)} & \colhead{(ZTF)}
		} 
		\startdata
		266703059 & J001204.51+020129.8 & 56977 & 3.018792 & 2.024947 & CV    & \multicolumn{1}{l}{CV} & none  & 16.5676 & 0.42809 \\
		592201085 & J001856.93+345444.2 & 58039 & 4.737242 & 34.91228 & CV    &       & CV    & 17.27818 & 0.7825 \\
		592301085 & J001856.93+345444.2 & 58039 & 4.737242 & 34.91228 & CV    &       & CV    & 17.27818 & 0.7825 \\
		614101085 & J001856.93+345444.2 & 58095 & 4.737242 & 34.91228 & CV    &       & CV    & 17.27818 & 0.7825 \\
		75902147 & J002148.44+350451.1 & 56231 & 5.45187 & 35.08089 & CV    &       & CV    & 14.51631 & 0.88123 \\
		284616096 & J002148.44+350451.1 & 57014 & 5.45187 & 35.08089 & CV    & \multicolumn{1}{l}{CV} & CV    & 14.51631 & 0.88123 \\
		592307039 & J002148.44+350451.2 & 58039 & 5.451867 & 35.08089 & CV    &       & CV    & 14.51631 & 0.88123 \\
		614107039 & J002148.44+350451.2 & 58095 & 5.451867 & 35.08089 & CV    &       & CV    & 14.51631 & 0.88123 \\
		76002147 & J002148.45+350451.1 & 56231 & 5.451892 & 35.08088 & CV    &       & CV    & 14.51631 & 0.88123 \\
		81816096 & J002148.45+350451.2 & 56254 & 5.451877 & 35.08091 & CV    &       & CV    & 14.43212 & 0.90129 \\
		81916096 & J002148.45+350451.2 & 56254 & 5.451877 & 35.08091 & CV    &       & CV    & 14.43212 & 0.90129 \\
		\multicolumn{1}{l}{*174106175} & J002500.18+073349.3 & 56601 & 6.250776 & 7.563698 & CV    & \multicolumn{1}{l}{CV} & CV    & 19.15958 & 3.25442 \\
		483901014 & J002842.53+311819.4 & 57716 & 7.177215 & 31.3054 & CV    & \multicolumn{1}{l}{CV} &       & 18.73505 & 2.02 \\
		82609065 & J003005.80+261726.3 & 56255 & 7.524205 & 26.29067 & CV    & \multicolumn{1}{l}{NL-candidate} & NL    & 15.2266 & 5.27938 \\
		90111206 & J003135.88+434905.3 & 56265 & 7.899519 & 43.81814 & CV    &       & Nova  & 15.67982 & 0.20626 \\
		\nodata & \nodata & \nodata & \nodata & \nodata & \nodata & \nodata & \nodata & \nodata & \nodata \\
		\enddata
		\tablecomments{
			\\ (This table is available in its entirety in machine-readable form.)	}
	\end{deluxetable*}
\end{longrotatetable}

\begin{figure}
	\plotone{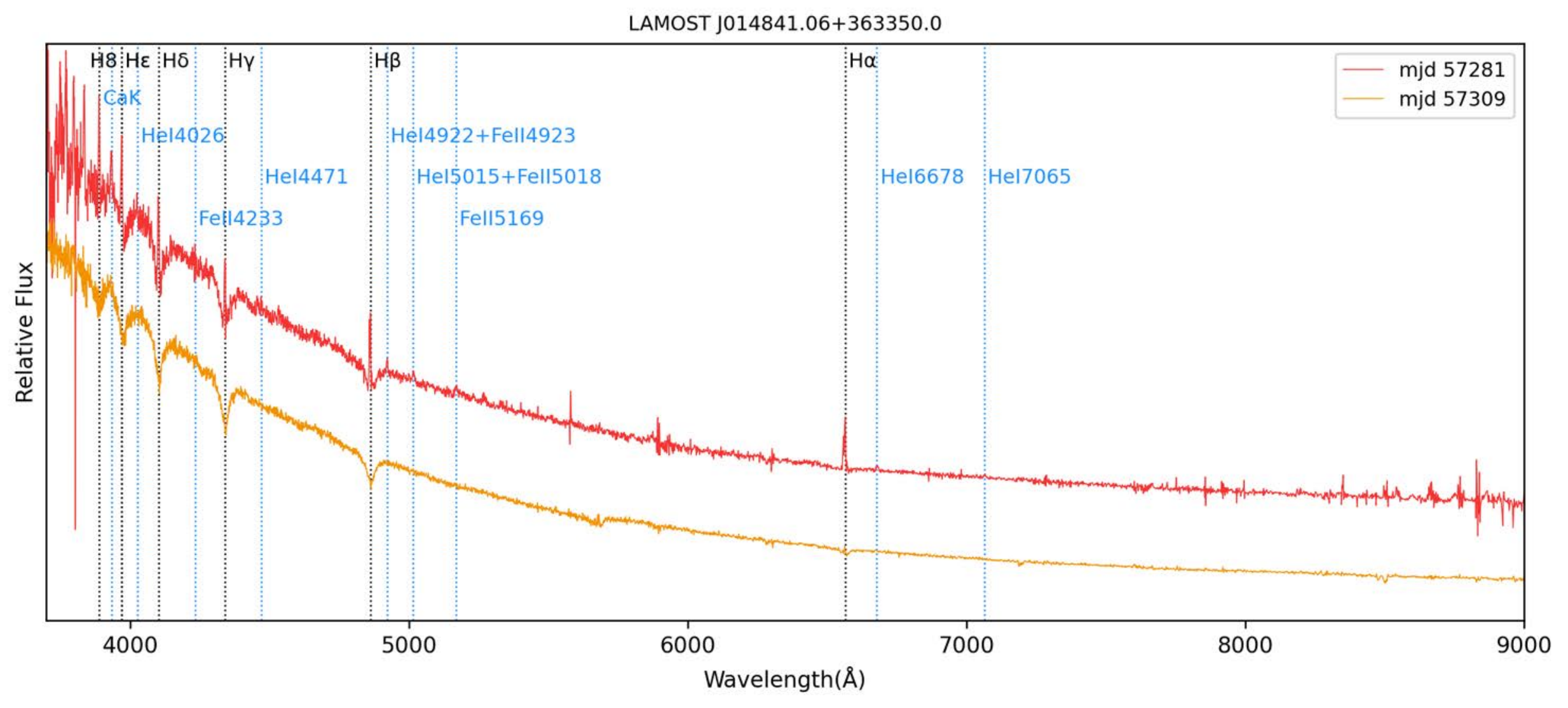}
	\caption{Two spectra of J0148 with MJDs 57821 and 57309, respectively, show variation. The wavelength range is from 3700\,\AA\ to 9000\,\AA. The upper red line represents the spectrum of MJD 57281, while the lower orange line for MJD 57309. The Balmer lines are indicated with dotted black lines, and the He\,{\footnotesize\RN{1}}, He\,{\footnotesize\RN{2}} and Fe\,{\footnotesize\RN{2}} lines are indicated with dotted blue lines. \label{fig:J014841}}
\end{figure}

\begin{figure}
	\plotone{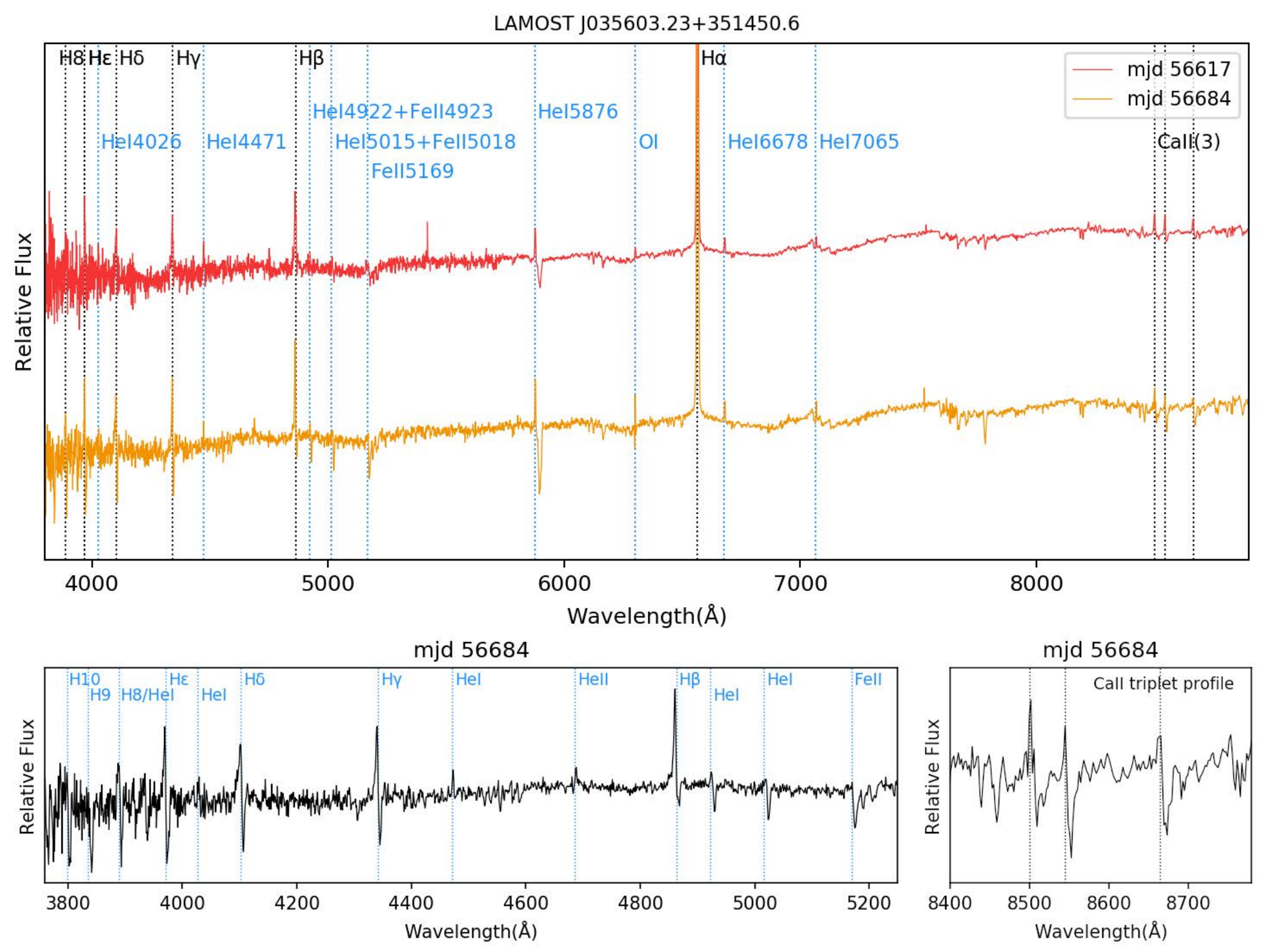} 
	\caption{Upper panel: Two spectra of J0356, taken on MJDs 56617 and 56684, respectively. The upper red line represents the spectrum of MJD 56617, while the lower orange line for MJD 56684. The Balmer lines are indicated with dotted black lines, and the He\,{\footnotesize\RN{1}}, He\,{\footnotesize\RN{2}} and Fe\,{\footnotesize\RN{2}} lines are indicated with dotted blue lines. The wavelength range is from 3800\,\AA\ to 8900\,\AA. 	
	Lower panel: the blue part and the Ca {\footnotesize\RN{2}} triplet region of the spectrum of MJD 56684 show the anti-P Cyg profile. \label{fig:J035603}}
\end{figure}

\begin{figure}
	\plotone{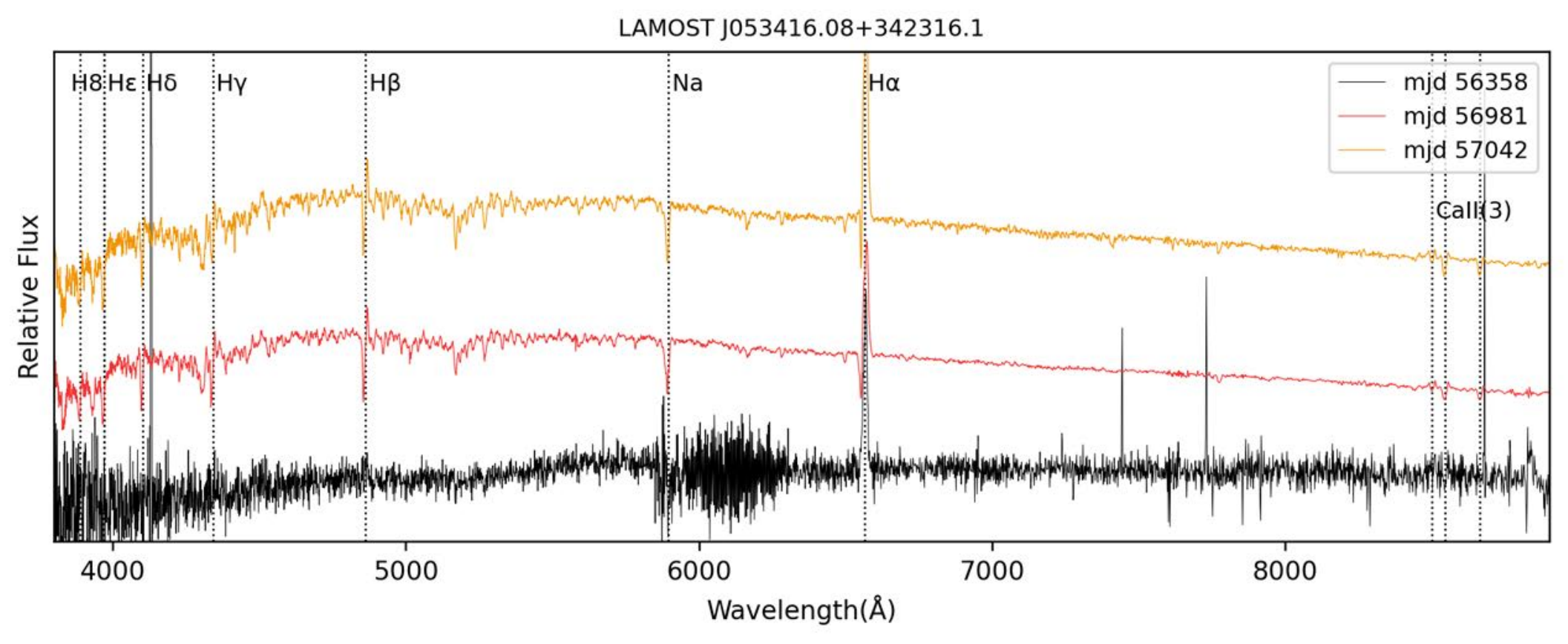} 
	\caption{Three spectra of J0534+3423 are taken at MJDs 56358, 56981 and 57042, respectively. The black, red and orange lines represent the spectrum taken at MJDs 57042, 56981 and 57042, respectively. The lines of Balmer series, Na $\lambda$5895.6 and Ca {\footnotesize\RN{2}} triplets are indicated with dotted black lines. The wavelength range is from 3800\,\AA\ to 8900\,\AA. \label{fig:J053416}}
\end{figure}
\section{Discussion} \label{sec:discussion}

\subsection{Objects of New CV Candidates} \label{sec:discussion1}
Among the new CV candidates, we find some of them showing interesting features and discuss them in the following subsections according to their subtypes.

\subsubsection{DN in outburst}
J0148, J0612, J0701, J0023, and J2112 reveal typically spectral features of DNs undergoing outbursts, and the wide absorption is visible in each Balmer line in the spectra of these objects. We note that the Balmer lines for J0148, J0612, J0701, and J2112 are narrow, for example, the H$\alpha$ is less than 24\,\AA, while J0023 shows slightly broad emission components (the H$\alpha$ emission core is 33\,\AA). Asymmetric central absorption also presents at the Balmer emission cores in all the spectra.

There are two spectra of J0148 (Figure \ref{fig:newcv_confirmed} on page 19). The emission of the Balmer series is significant in the middle of the wide absorption lines, while He\,{\footnotesize\RN{1}} lines are in emission and does not exhibit wide absorption features. Thus, we can identify it as a DN in outburst. The other spectrum mainly shows wide Balmer absorption lines. The variations of the two spectra observed at different times are shown in Figure \ref{fig:J014841}. The spectrum of MJD 57281 presents central emission cores of Balmer lines, while H$\alpha$ and He\,{\small\RN{1}} $\lambda$6678 are almost pure emission. In the spectrum of MJD 57309, the Balmer lines become nearly all absorption, while the H$\alpha$ still presents an obvious emission core. Although the multiple emission components with asymmetric features are not obvious, they can be seen in the absorption lines of Balmer series higher than H$\alpha$. This means that the emission components have a steeper decrement than those of the absorption. Besides, He\,{\small\RN{1}} $\lambda$6678 also shows very weak emission. For these features above, the spectrum of MJD 57309 is likely to be taken during the outburst maximum \citep{warner2003cataclysmic}.

The spectrum of J0612 (Figure \ref{fig:newcv_confirmed} on page 20) shows narrow asymmetric core emissions with weak wide absorption in the Balmer series, and the He\,{\small\RN{1}} $\lambda$4471, $\lambda$5876, $\lambda$6678, $\lambda$7065 lines are also in emission. 

In the spectrum of J0701 (Figure \ref{fig:newcv_confirmed} on page 20), the absorption lines are strong while the emission cores are weak for the higher order Balmer series., and the He\,{\small\RN{1}} $\lambda$6678 line is affected by the absorption.
 
The spectrum of J0023 (Figure \ref{fig:newcv_confirmed} on page 19) shows strong wide Balmer absorption with broad emission cores, and there seems to be two emission components of the Balmer lines. The He\,{\small\RN{2}} $\lambda$4686 is a strong emission line, and the C\,{\small\RN{3}}/N\,{\small\RN{3}} blend at 4650\,\AA\ is also in emission, which may indicate the presence of a hot emission zone in the disk.
 
The spectrum of J2112 (Figure \ref{fig:newcv_confirmed} on page 20) shows weak wide absorption of Balmer lines, while the weak He\,{\small\RN{1}} $\lambda$6678 line presents emission. The C\,{\small\RN{3}}/N\,{\small\RN{3}} blend at 4650\,\AA\ is in emission, while He\,{\small\RN{2}} is absent. Note that Fe\,{\small\RN{2}} lines are present in all of the spectra above.

\subsubsection{Nova-like type}
J0409, J0519, J0723, J0454, and J0457 mainly show strong C\,{\small\RN{3}}/N\,{\small\RN{3}} emission features and C\,{\small\RN{2}} lines, resembling the NL subtype of CVs. The common central absorption or double-peaked profile in the emission lines suggests the presence of disks.

\begin{figure}
	\gridline{\fig{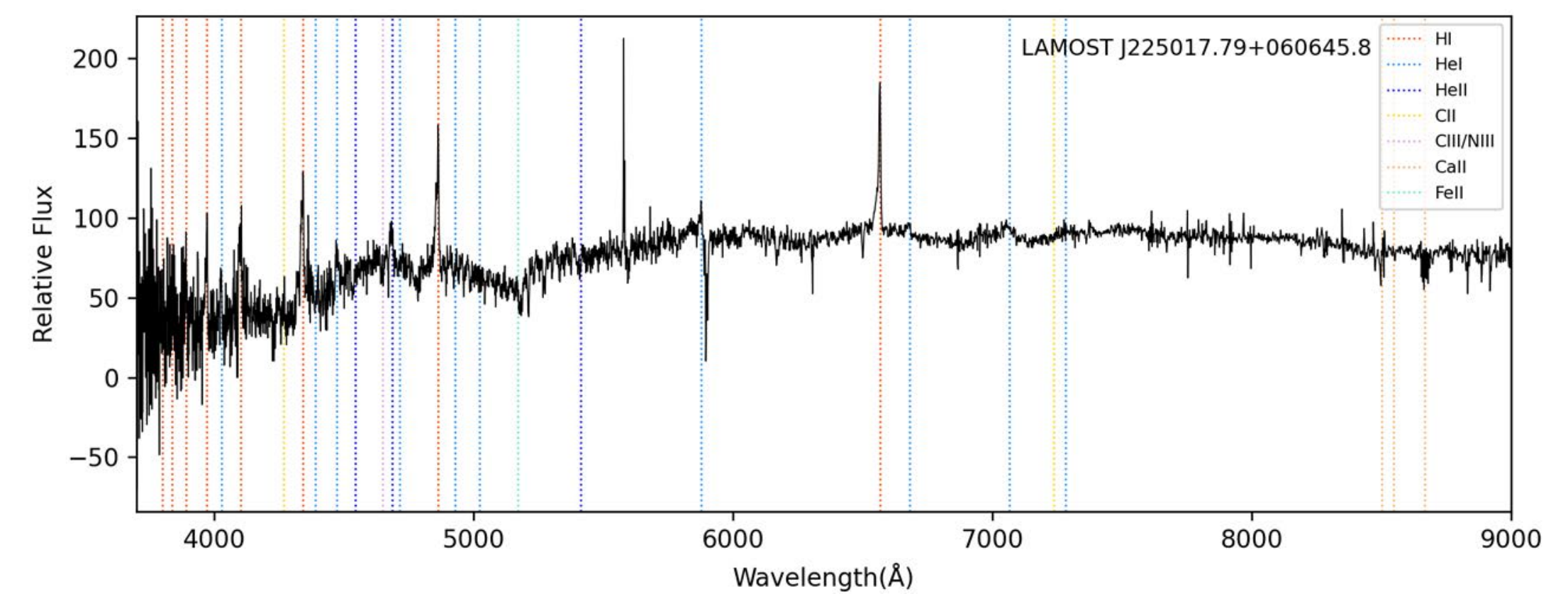}{0.5\textwidth}{}}
	\caption{21 spectra of 20 new confirmed CV candidates in LAMOST DR6 (ordered by increasing RA). The wavelength range is from 3700\,\AA\ to 9000\,\AA. Cosmic ray hits are marked by short red slashes. \label{fig:newcv_confirmed}}
\end{figure}

J0409 (Figure \ref{fig:newcv_confirmed} on page 19) presents peculiar features of a high-temperature continuum along with obvious H$\alpha$, H$\beta$, He\,{\small\RN{1}}, He\,{\small\RN{2}}, C\,{\small\RN{3}}/N\,{\small\RN{3}}, and C\,{\small\RN{2}} emission lines. The Balmer decrement is quite strong, and the asymmetric double-peaked profiles present. The He\,{\small\RN{2}} $\lambda$4686 line and C\,{\small\RN{3}}/N\,{\small\RN{3}} blend are extremely strong, which are much higher than the apparent strength of H$\alpha$ and H$\beta$. The strength of C\,{\small\RN{3}}/N\,{\small\RN{3}} is comparable to that of He {\small\RN{2}} $\lambda$4686. The Ca\,{\small\RN{2}} H/K lines show single narrow absorption, which could be due to interstellar absorption. Interestingly, we find that J2023, which is also one of the newly discovered CV candidates, has very similar features, while the prominent difference is that in J2023 the He\,{\small\RN{2}} $\lambda$4686 is absent, for which it can only be classified as a possible NL. 

J0723 (Figure \ref{fig:newcv_confirmed} on page 20) shows strong Balmer emission lines but weak He\,{\small\RN{1}} lines. The Balmer lines show asymmetric shapes and the strongest Balmer line is H$\alpha$. For H$\gamma$, an absorption center is present which may indicate the existence of an accretion disk. The He\,{\small\RN{2}} $\lambda$4686 is much weaker than H$\beta$, while the C\,{\small\RN{3}}/N\,{\small\RN{3}} blend is as strong as He\,{\small\RN{2}} $\lambda$4686. The Ca\,{\small\RN{2}} $\lambda$3934 line shows narrow absorption, while C\,{\small\RN{2}} $\lambda$7234 is in emission.

J0454 (Figure \ref{fig:newcv_confirmed} on page 19) shows very broad strong Balmer, He\,{\small\RN{1}}, He\,{\small\RN{2}}, and C\,{\small\RN{2}} emission lines. The C\,{\small\RN{3}}/N\,{\small\RN{3}} is about half the height of that of He\,{\small\RN{2}} $\lambda$4686. The broad and asymmetric double-peaked shapes in the Balmer and He lines are significant, which indicates that there is an accretion disk. 

The spectra of J0519 (Figure \ref{fig:newcv_confirmed} on page 20) and J0457 (Figure \ref{fig:newcv_confirmed} on page 19) have relatively low SNR. He\,{\small\RN{2}} $\lambda$4686 emission can be identified in both spectra. The C\,{\small\RN{3}}/N\,{\small\RN{3}} emission possibly exists in J0519, while C\,{\small\RN{2}} $\lambda$7234 emission is visible in J0457. For J0519, the central absorption is seen in the Balmer lines, which indicates there is an accretion disk. 

\subsubsection{Magnetic CVs}
Four new CV candidates, J0320, J0326, J0216, and J0715, can be classified as magnetic CVs by the strong He\,{\small\RN{2}} emission and other spectral features.

J0320 (Figure \ref{fig:newcv_confirmed} on page 19) shows strong broad Balmer emission lines, the strongest being H$\delta$, and they are all blue-shifted. He\,{\small\RN{2}} $\lambda$4686 is stronger than H$\beta$. He\,{\small\RN{1}} $\lambda$7281, $\lambda$7065, $\lambda$6678, $\lambda$5876, $\lambda$4388, $\lambda$4471 and $\lambda$4026 are in emission. The Na\,{\small\RN{1}} D line shows very strong absorption, which may be contributed by the companion star. The C\,{\small\RN{3}} $\lambda$4634 and C\,{\small\RN{4}} $\lambda$5805 lines also show emission. As He\,{\small\RN{2}} $\lambda$4686 is strong and Balmer lines split, this object could be a magnetic CV, possibly of AM Her type. 

J0326 (Figure \ref{fig:newcv_confirmed} on page 19) reveals an M-type continuum at the red part along with very strong Balmer and He\,{\small\RN{1}}/He\,{\small\RN{2}} emission lines. Both the narrow and broad emission components in the Balmer and He\,{\small\RN{2}} lines can be seen, and the strongest Balmer line is H$\delta$. C\,{\small\RN{2}} lines only show weak emissions. Deep Na\,{\small\RN{1}} D absorption line can be seen at 5895\,\AA. The strength of He\,{\small\RN{2}} $\lambda$4686 is almost the same as H$\beta$, and C\,{\small\RN{3}} is nearly absent. The spectroscopic features are consistent with that of the AM Her systems.

J0216 (Figure \ref{fig:newcv_confirmed} on page 19) shows a relatively flat continuum with strong Balmer and He\,{\small\RN{1}}/He\,{\small\RN{2}} emission lines. As the He\,{\small\RN{2}} $\lambda$4686 emission line is stronger than H$\beta$, and there are mutiple components or splits in the Balmer lines, we classify it as a magnetic CV, possibly of the AM Her type.

J0715 (Figure \ref{fig:newcv_confirmed} on page 20) shows broad Balmer emissions. The H$\alpha$ emission line presents a `wine bottle' profile. The He\,{\small\RN{2}} $\lambda$4686 line is quite strong as the strength is comparable to that of H$\beta$, and the C\,{\small\RN{3}}/N\,{\small\RN{3}} blend also show strong emission. In view of the above characteristics, we consider that it as a possible magnetic CV, specifically an IP.

\subsubsection{Binary spectra}
The spectra of J0523, J2250, J0535, J0628, J0356 and J2045 present doninent cool companion star features on the red part. For J2339 and J0311, both the companion and the white dwarf features can be seen.

The spectrum of J0523 (Figure \ref{fig:newcv_confirmed} on page 20) presents strong narrow Balmer emission lines with weak decrement to higher order. The spectrum shows prominent M-type star features on the red part, and it can be fitted with a M2-type star. There is a significant excess at the blue part. The Balmer series present an asymmetric broadening component superimposing on the strong narrow emission. He {\small\RN{1}} $\lambda$6678, $\lambda$5876, and $\lambda$4471 lines are in emission, while He {\small\RN{2}} lines are absent.

J2250 (Figure \ref{fig:newcv_confirmed} on page 21) shows a continuum that can be fitted with a late K-type star, while presents strong Balmer, He {\small\RN{1}} and He\,{\small\RN{2}} emission lines. The Balmer lines have a broad component and show obvious asymmetric structure. The narrow Balmer emission components show a decrement, while the broad components do not show obvious decrement and shift to blue with respect to the narrow components. We consider that the narrow Balmer emission may come from chromospheric activities of the companion star, while the broad one come from an accretion disk or hot spot. In addition, the deep Na\,{\small\RN{1}} D $\lambda$5895 absorption line as well as [O\,{\small\RN{1}}] $\lambda$6302 and O\,{\small\RN{1}} $\lambda$6365 absorption lines is visible. 

\begin{figure}
	\gridline{\fig{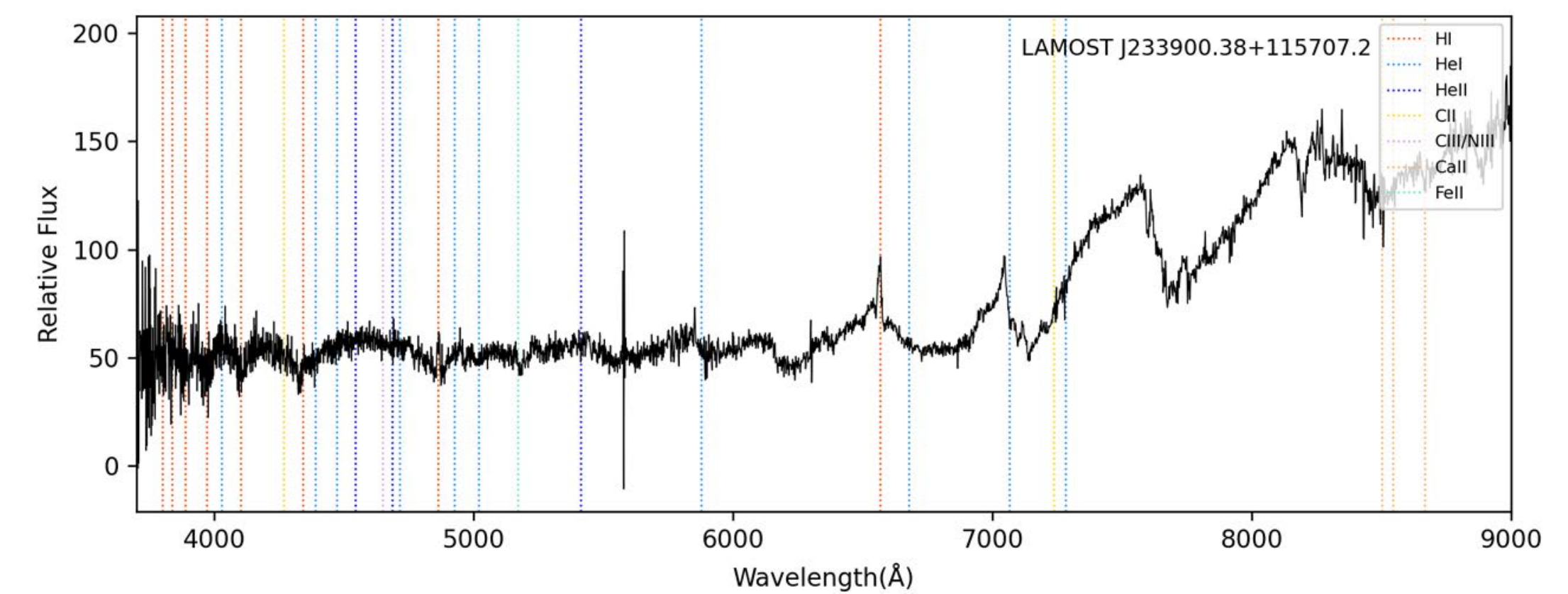}{0.5\textwidth}{}}
	\caption{52 spectra of 32 new possible CV candidates in LAMOST DR6 (ordered by increasing RA). The wavelength range is from 3700\,\AA\ to 9000\,\AA. Cosmic ray hits are marked by short red slashes. \label{fig:newcv_possible}}
\end{figure}

J0356 (Figure \ref{fig:newcv_possible} on page 22) has multiple spectra in LAMOST DR6, and shows prominent variations. We present two of the spectra of MJDs 56617 and 56684 in Figure \ref{fig:J035603}. Both spectra show the M-type companion features along with strong Balmer and He\,{\small\RN{1}}/He\,{\small\RN{2}} emission lines. Compared with the fitted template spectrum, the relative flux of the LAMOST spectra has obvious excesses at the blue part. In the lower panel of the figure, the near-ultraviolet and near-infrared regions of the spectrum of MJD 56684 are shown. In the observation of MJD 56684, the Balmer series members higher than H$\alpha$, He {\small\RN{1}} and Ca\,{\small\RN{2}} lines all show anti-P Cyg profile, which reveals the explosive nature of the object, while He\,{\small\RN{2}} $\lambda$4686 line appears to be a single emission or have a P Cyg profile. However, we can not completely rule out the possibility that it is an explosive pre-main-sequence object (like FU Orionis type stars) by the spectral features alone.

The spectra of J0535 (Figure \ref{fig:newcv_confirmed} on page 20), J0628 (Figure \ref{fig:newcv_possible} on page 26) and J2045 (Figure \ref{fig:newcv_possible} on page 26) all show features of a M-type companion. In those spectra, the Balmer emission lines are quite narrow and all show asymmetric features, and the Ca\,{\small\RN{2}} H/K lines are in strong emission. The H$\beta$ is much weaker than H$\alpha$, which is not typical for CVs. Moreover, J0535 shows He\,{\small\RN{1}} emission lines, while J0628 and J2045 show both He\,{\small\RN{1}} and He\,{\small\RN{2}} emission lines. The Ca\,{\small\RN{2}} triplets at 8500\,\AA\ present slight emission in the former two but strong in J2045. [O\,{\small\RN{1}}] $\lambda$6302 also presents weak emission in all those spectra. 

The spectrum of J2339 (Figure \ref{fig:newcv_possible} on page 27) presents both a M-type companion and a white dwarf features. In the spectra, both H$\alpha$ and H$\beta$ show strong emission with double-peaked profiles, and He\,{\small\RN{1}} $\lambda$6678 presents weak and broad emission, which may indicates that there is an accretion disk. 

J0311 (Figure \ref{fig:newcv_possible} on page 22) has been classified as a DA-type white dwarf by \citet{zhang2013white}, and they suggested it has unknown emission process. The spectrum presents strong H$\alpha$ emission core and has strong  Balmer decrement for the emission components, while the broad Balmer absorption resemble a white dwarf. It also presents strong Ca\,{\small\RN{2}} triplets but weak He\,{\small\RN{1}} emission lines. As there are mutiple emission components in the Balmer series and some are broad, we consider this object may have a CV nature.

\subsubsection{Other Objects} \label{sec:J0534}
There are 21 spectra of J0621 in LAMOST DR6, and 15 of them show H$\alpha$ emission features for which can be identified as a possible CV candidate, as shown in Figure \ref{fig:newcv_possible} on pages 24 and 25. The H$\alpha$ emission lines in its spectra present complex structure. There is a sharp flux increase at the blue end with emission features. Four spectra of this object taken before MJD 58076 show very wide absorption (the line width of H$\alpha$ beyond 130\,\AA) of Balmer and Paschen series, and the Na\,{\small\RN{1}} absorption lines are strong. Seven spectra taken at MJD 58076 and later show an unusual double hump at 4000\,\AA\ - 5000\,\AA\ for unknown reasons. As these spectra are taken at different dates, the double hump feature should be real rather than instrumental problems.

J0534 is identified as a possible CV candidate due to the broad H$\alpha$ emission feature. It has three spectra in LAMOST DR6 as shown in Figure \ref{fig:J053416}. The spectrum of MJD 56358 has a relatively low SNR, in which only symmetrical H$\alpha$ emission line can be seen. In the spectra of MJDs 56981 and 57042, a late G- to early K-type companion can be clearly seen. These two spectra present a strong P Cyg profiles for H$\alpha$ to H$\gamma$ and Ca\,{\small\RN{2}} triplets, which could indicate the presence of disk winds or jets. However, in the spectrum of MJD 56981, both the strength ratio of Sr\,{\small\RN{2}} $\lambda$4077 to Fe\,{\small\RN{1}} $\lambda$4046 and the strength of C/N $\lambda$4215 band indicate that the K-type star resembles a giant rather than a main-sequence \citep{gray2009stellar}. The lack of He emission lines also suggests that it may not be a CV. Follow up observations are required to determine the nature of this object.

\subsection{Poorly investigated objects} \label{sec:discussion2}
J0631 was investigated by \citet{2013ATel.5451....1B} using the MASTER-Tunka auto-detection system. Based on the blue color and photometric variability, it was identified as a probable dwarf nova. It has also a UV counterpart GALEX J0631. There are three spectra of this object. Although the SNRs are low for all of them, the DN properties can be found. In the spectrum of MJD 55861, the continuum reveals a high-temperature component, and the emission features of the Balmer and He lines are obvious. The Balmer core emissions are not so broad and may have mutiple components, while the week wide absorption can be seen for H$\beta$ and higher Balmer series members. Strong He\,{\small\RN{2}} $\lambda$4686 emission line also presents in the spectra. The above features are in consistant with a dwarf nova during an outburst. For the other two spectra of MJDs 55873 and 57096, the continuum is quite flat. They both present distinctive asymmetric Balmer and He\,{\small\RN{1}} $\lambda$6678 emission lines, while He\,{\small\RN{2}} $\lambda$4686 is absent in the two spectra. We thus consider both spectra were taken at the quiescent phase of a DN.

J0720 has been identified to be most likely a polar with an extreme hot spot \citep{2018ATel11626....1D}, as it shows a large amplitude of variation presenting in its light curve with no eclipses found. It has an X-ray counterpart XMMSL2 J0720. There is only one spectrum in LAMOST DR6, which shows sharp and asymmetric Balmer emission lines with no obvious continuum. He\,{\small\RN{1}} $\lambda$4471, $\lambda$6678, and $\lambda$7065 emission lines also present while He\,{\small\RN{2}} $\lambda$4686 is absent. These spectral features are consistent with those of a polar. However, the H$\beta$ shows a double-peaked profile, which might imply the presence of a residual disk.

\section{Conclusion} \label{animation}
Using the UMAP with k-NN method, we have spectroscopically identified 475 CV spectra of 323 objects from LAMOST DR6, of which 52 objects are new.
After cross-match with the DR5 CV catalogue presented by \citet{hou2020spectroscopically}, our results include 215 of their 224 objects in LAMOST DR6 v2. 
The other nine are not included as their spectra don't show obvious H$\alpha$-emission features. This is due to the limitations of the current UMAP method. 

By visually examining the CV features of the spectra, we can remove contamination of the Oe/Be stars and H {\small\RN{2}} regions, and identify more potential CV candidates that are previously difficult to single out. The new CVs include five DNs during outbursts, five NLs and four magnetic CVs (three of AM Her type and one of IP type). We also discuss objects that directly show the binary systems. In addition, we listed two objects that have been poorly investigated from the known CVs in our catalogue. Their spectral features confirm the subtype they previously classified. Nevertheless, for J0720, the H$\beta$ presents an double-peaked profile which may suggest that there is a residual disk.

Our results show that the UMAP algorithm can identify CV spectra that present an core emission at H$\alpha$, however, there are still shortcomings if the emission core is weak, the program may not pick them out. More samples are needed for training of the UMAP method to improve the classification accuracy. Our results do not include the AM CVn type, and the pre-polars which may only show broad cyclotron emission lines, because they both lack hydrogen lines. New methods should be adopted to identify these two types of CVs as they may not have any H$\alpha$ emission lines.

In the future, we will carry out follow-up photometric studies on the new CV candidates and poorly studied or interesting CV objects, obtain the orbital parameters to further study their properties.

\section*{Acknowledgements}
We would like to thank the anonymous referee for his/her very heplful feedback. 
This work is supported by the National Key Basic R\&D Program of China via 2019YFA0405500, 2019YFA0405501,
the National Natural Science Foundation of China (NSFC) under grant 11773009, 11673007, 12090040, 12090044, 11833006, 11903012, 11703038, 12033003 and 12173103, the Natural Science Foundation of Hebei Province under grants A2021205006.
W.Y.C. is also supported by the `333 talents project' of Hebei Province. The Guo Shou Jing Telescope (the Large Sky Area Multi-Object Fiber Spectroscopic Telescope, LAMOST) is a National Major Scientific Project built by the Chinese Academy of Sciences. Funding for the project has been provided by the National Development and Reform Commission. LAMOST is operated and managed by National Astronomical Observatories, Chinese Academy of Sciences. 
We acknowledge the science research grants from the China Manned Space Project with NO.CMS-CSST-2021-A08.
This research has made use of the SIMBAD database, operated at CDS, Strasbourg, France. This work has made use of data from the Zwicky Transient Facility (ZTF) data.

\bibliography{bibliography}{}
\bibliographystyle{aasjournal}
\section*{APPENDIX\\ THE LAMOST SPECTRA}
\begin{figure}
	\gridline{\fig{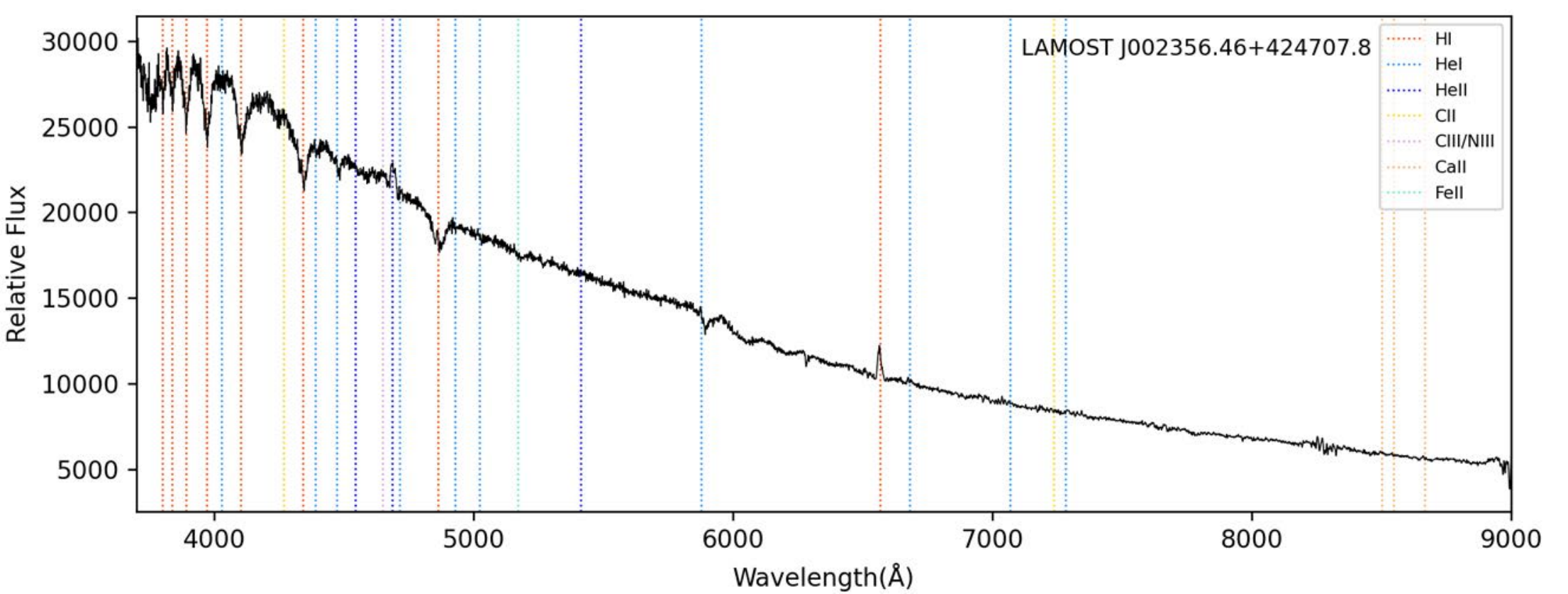}{0.5\textwidth}{}
		\fig{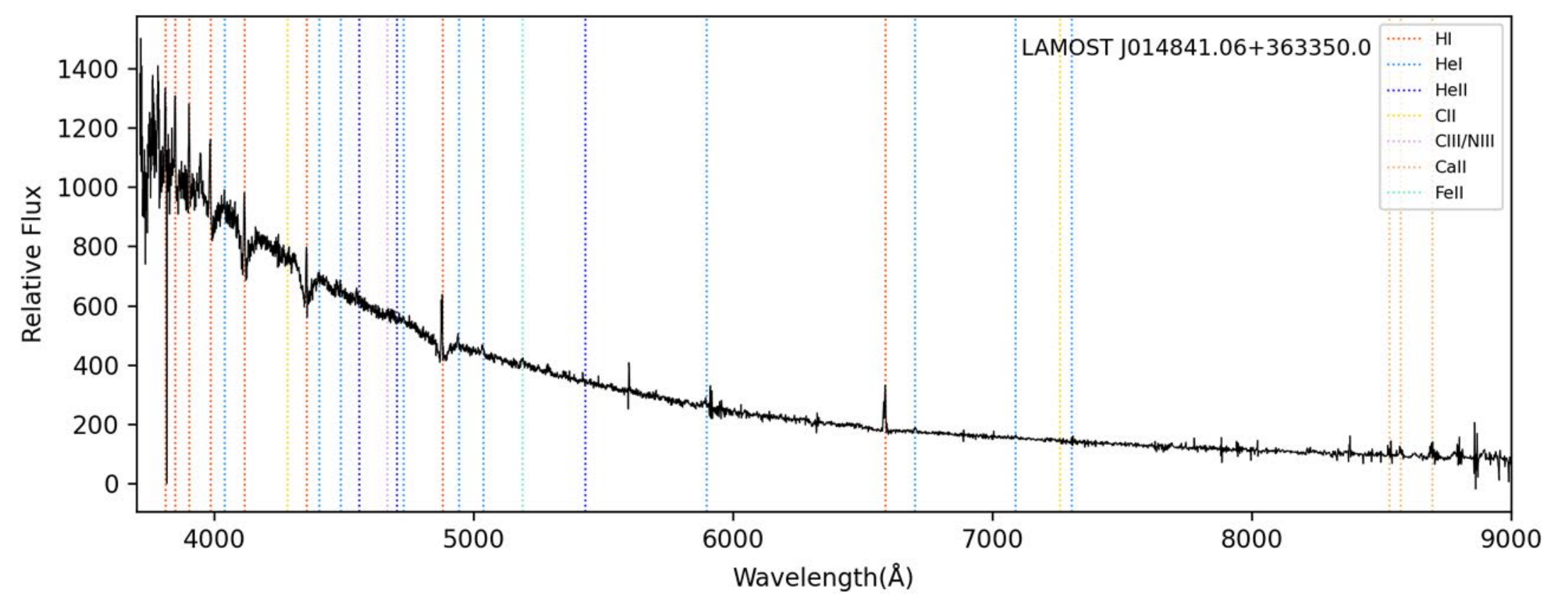}{0.5\textwidth}{}}
	\gridline{\fig{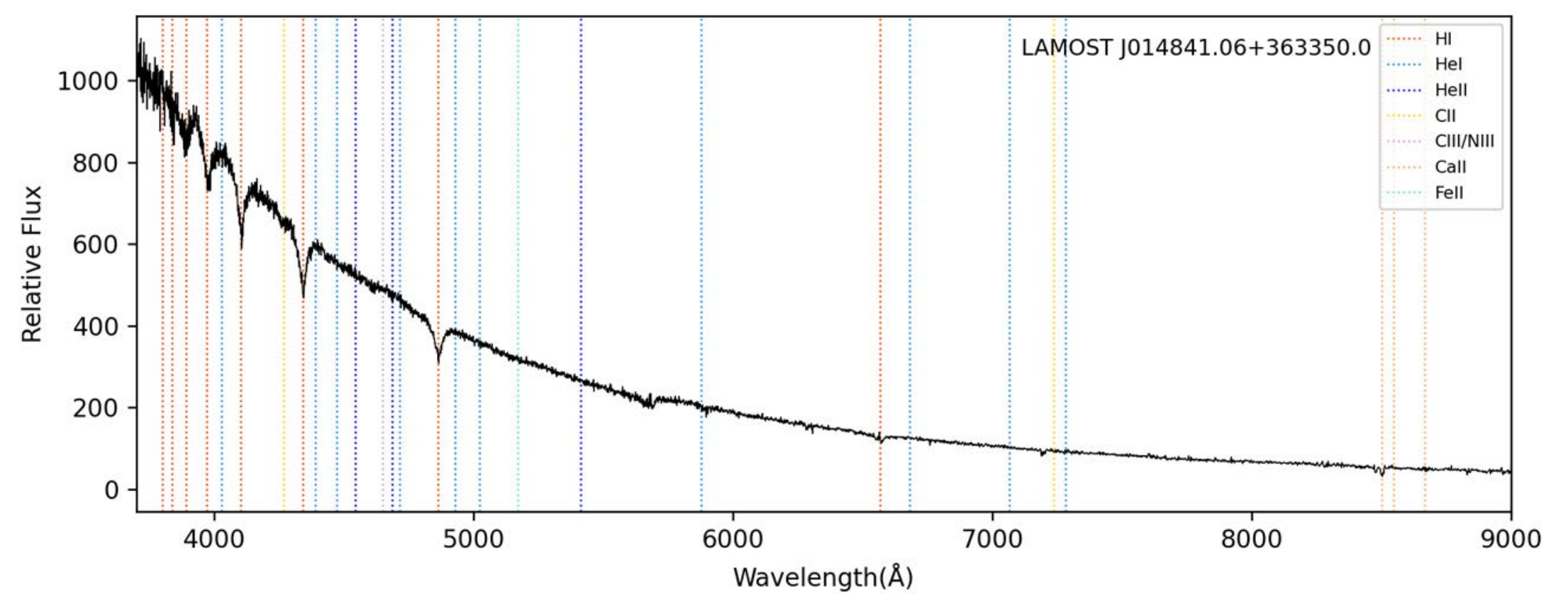}{0.5\textwidth}{}
		\fig{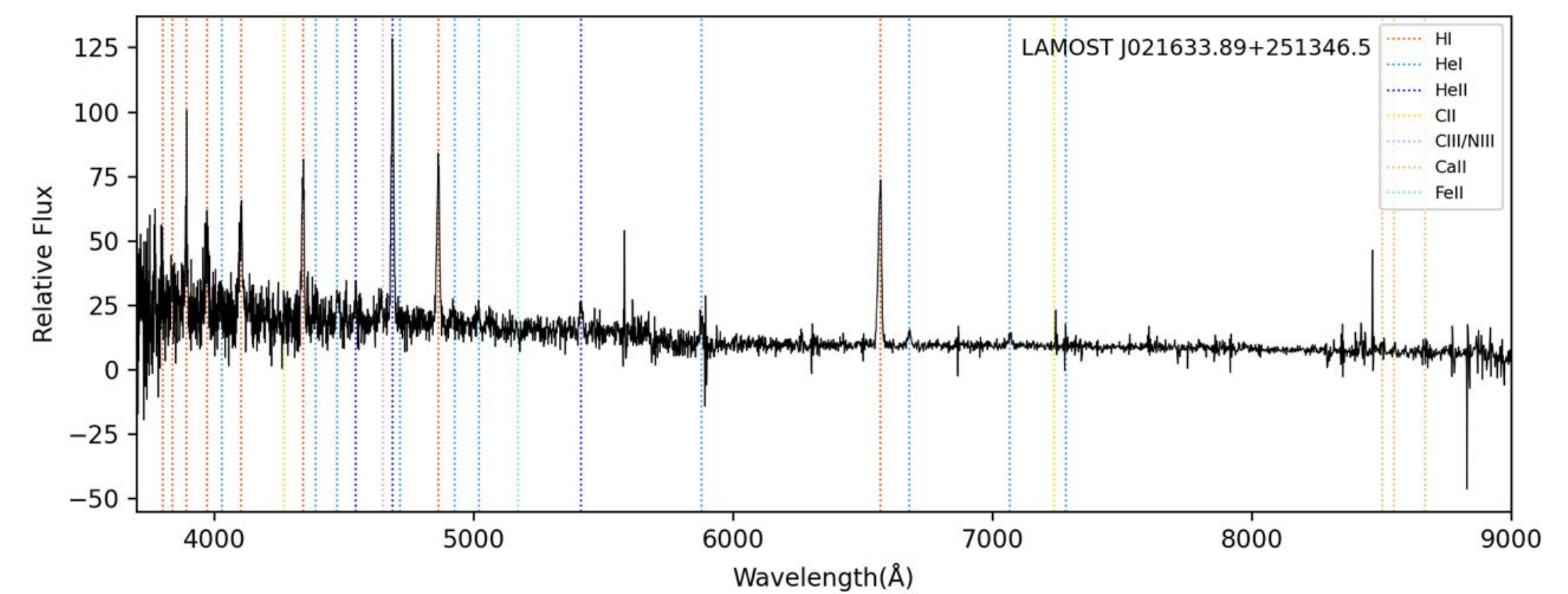}{0.5\textwidth}{}}
	\gridline{\fig{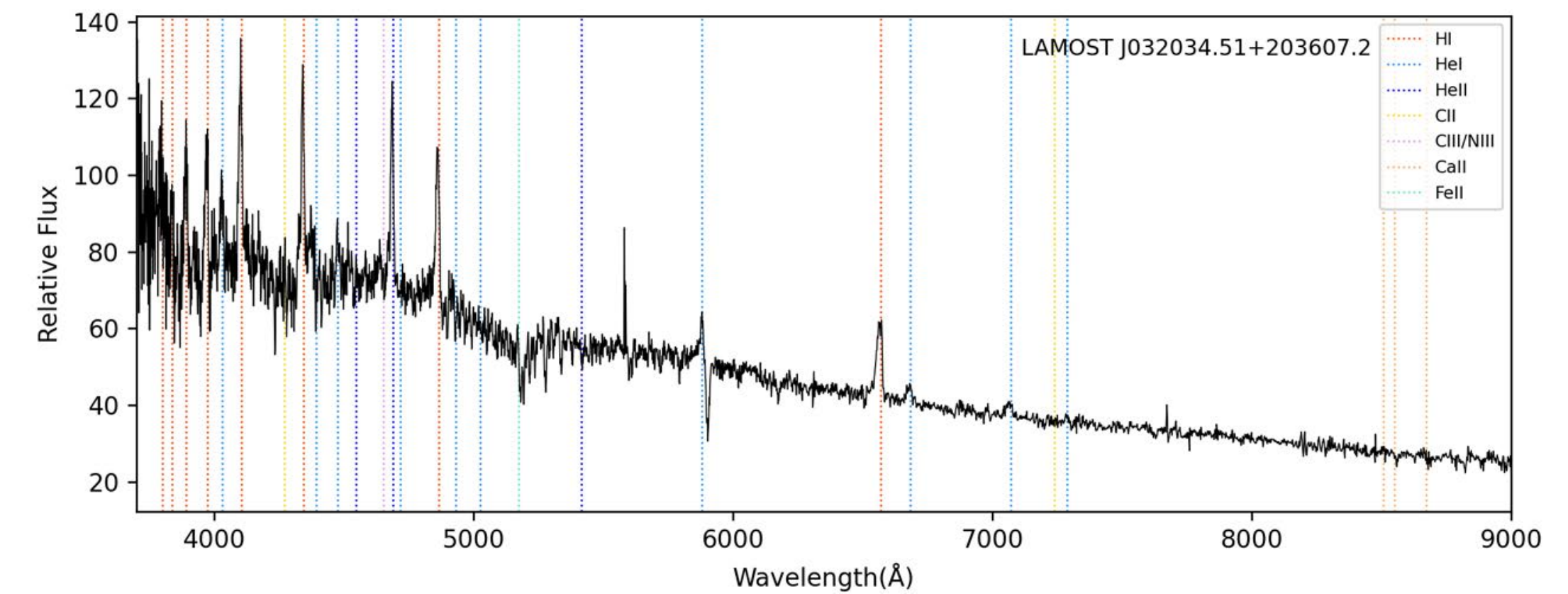}{0.5\textwidth}{}
		\fig{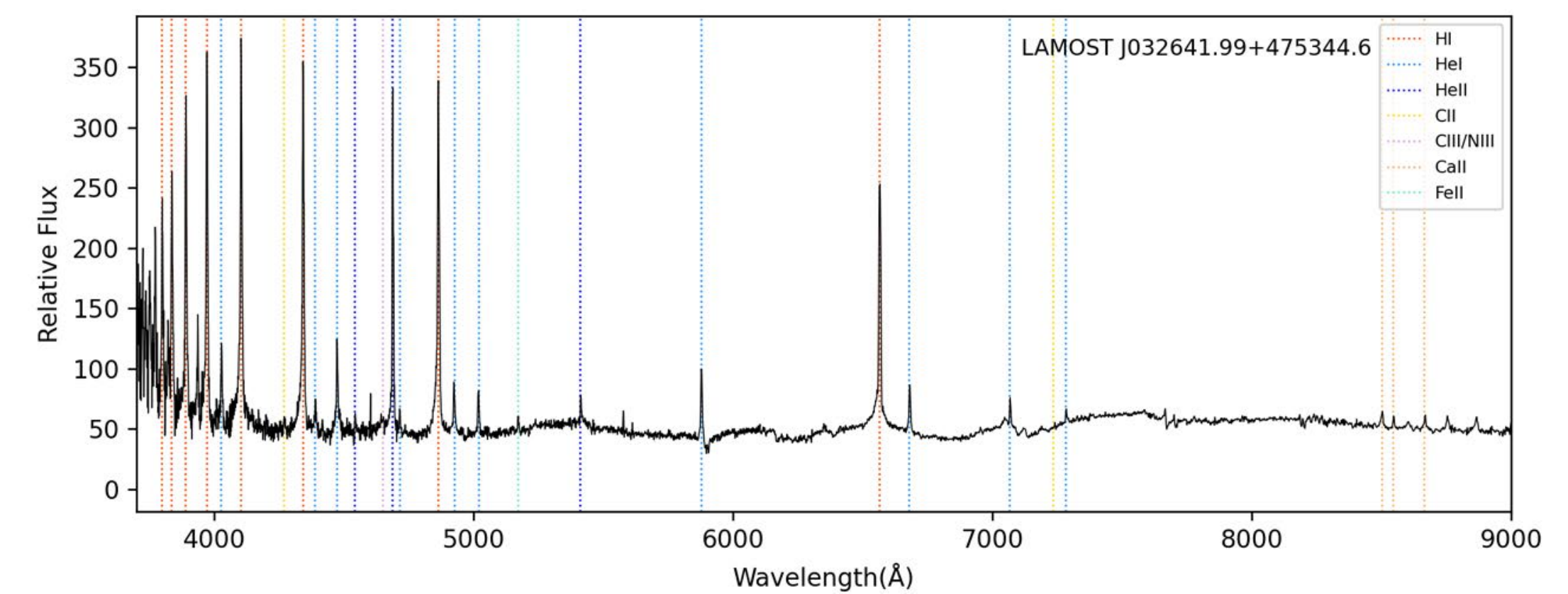}{0.5\textwidth}{}}
	\gridline{\fig{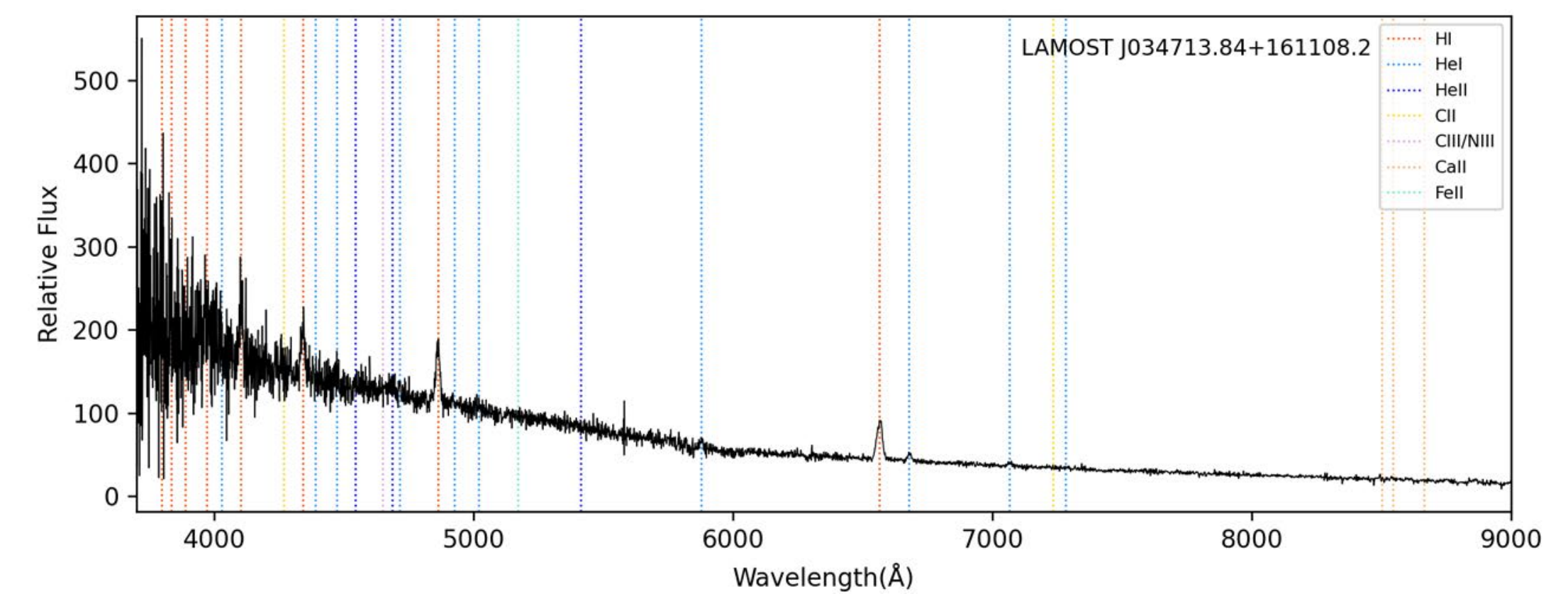}{0.5\textwidth}{}
		\fig{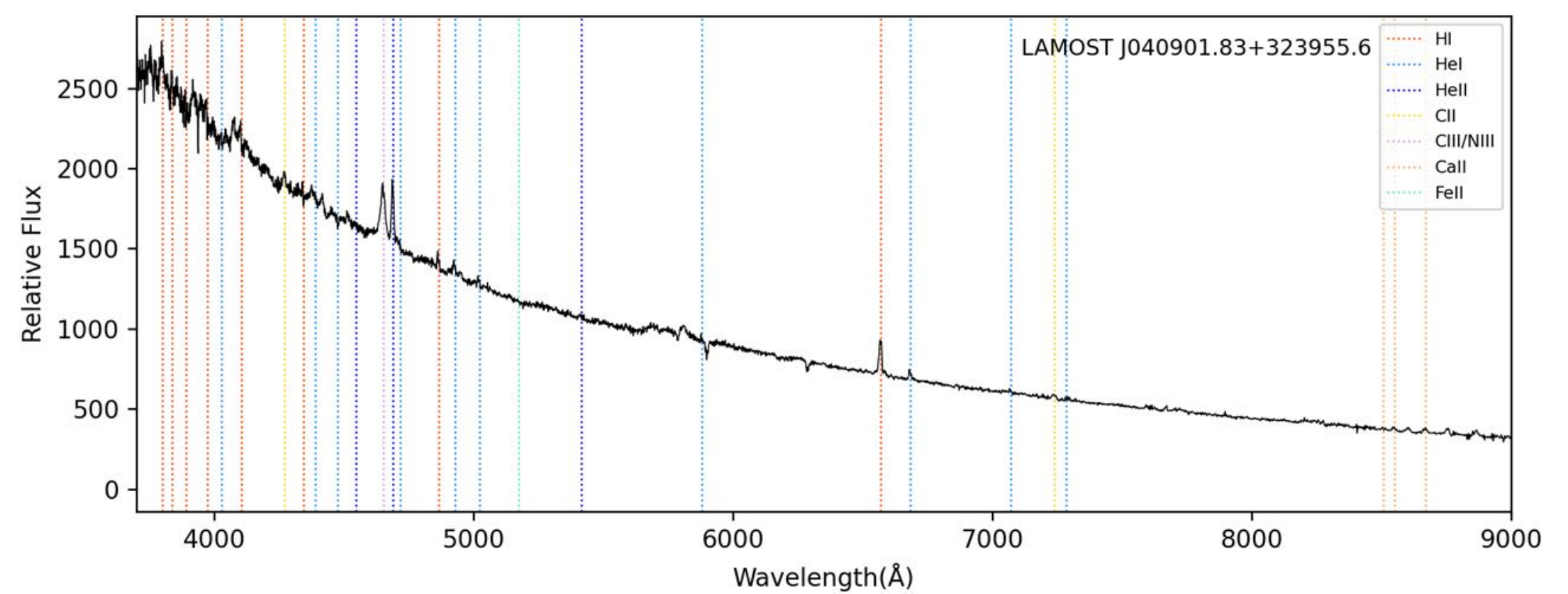}{0.5\textwidth}{}}
	\gridline{\fig{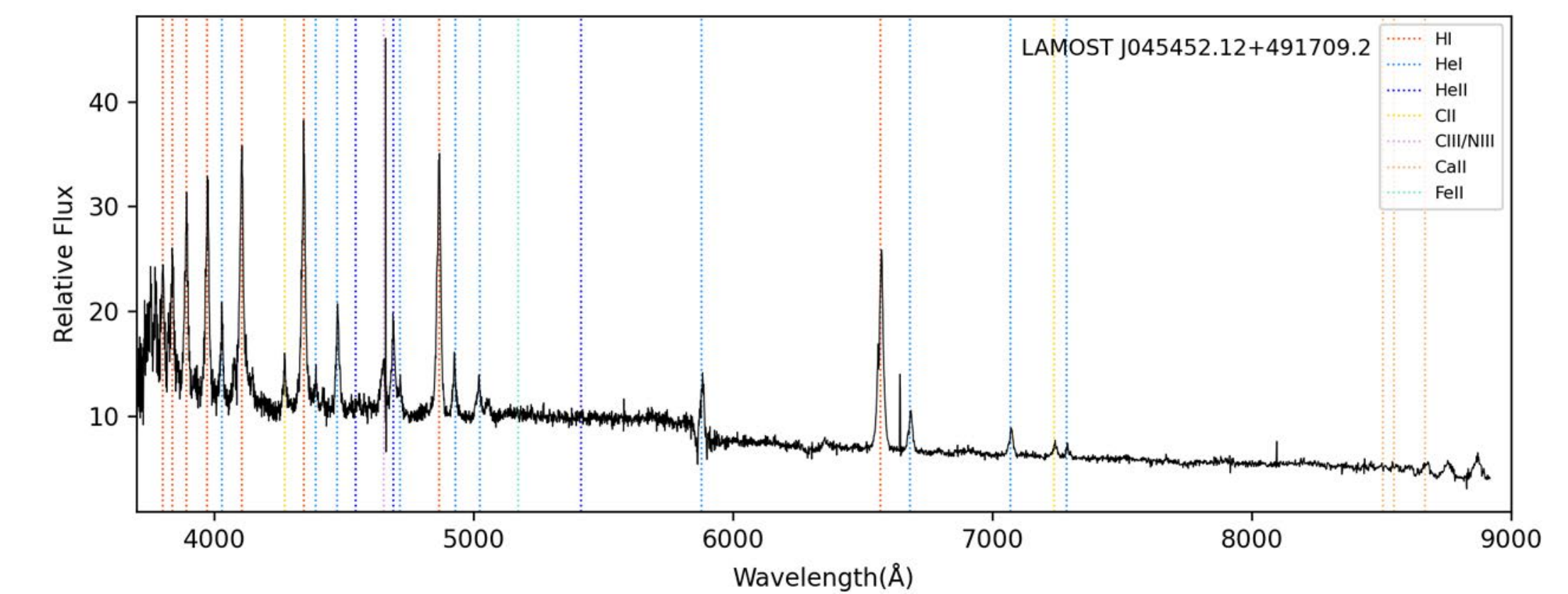}{0.5\textwidth}{}
		\fig{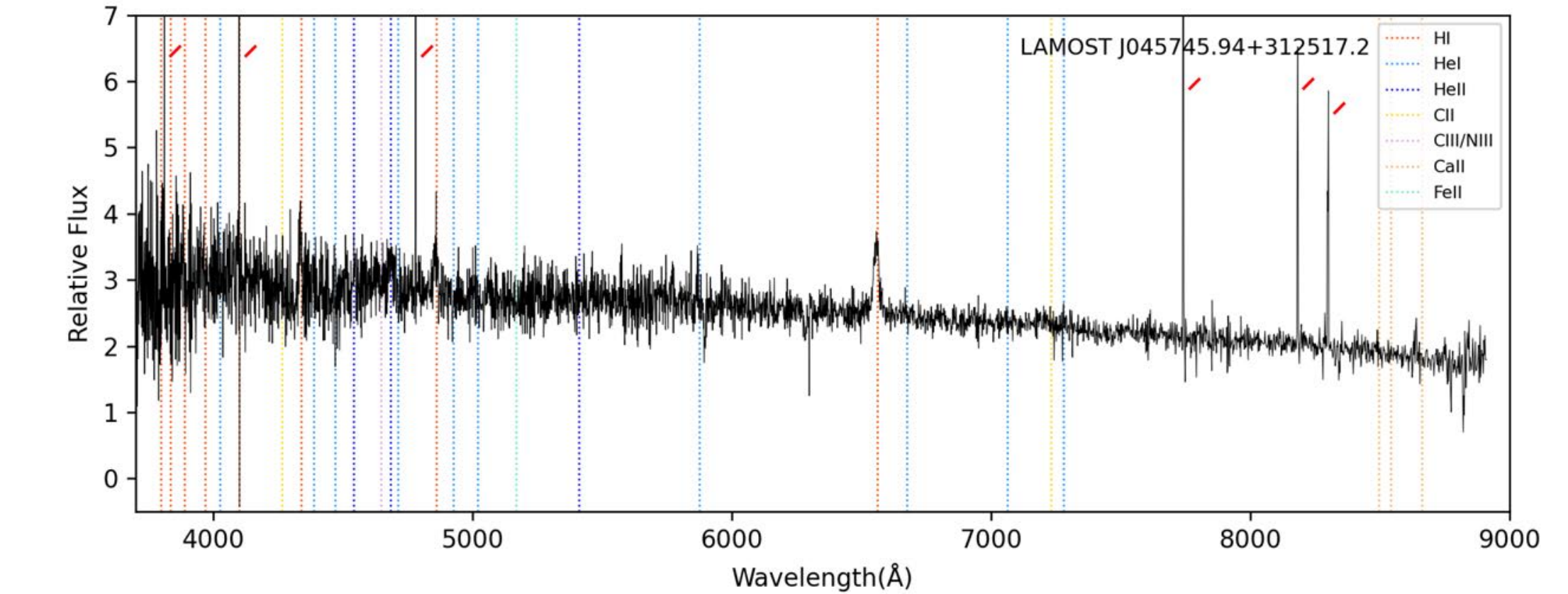}{0.5\textwidth}{}}
\end{figure}
\begin{figure}
	\gridline{\fig{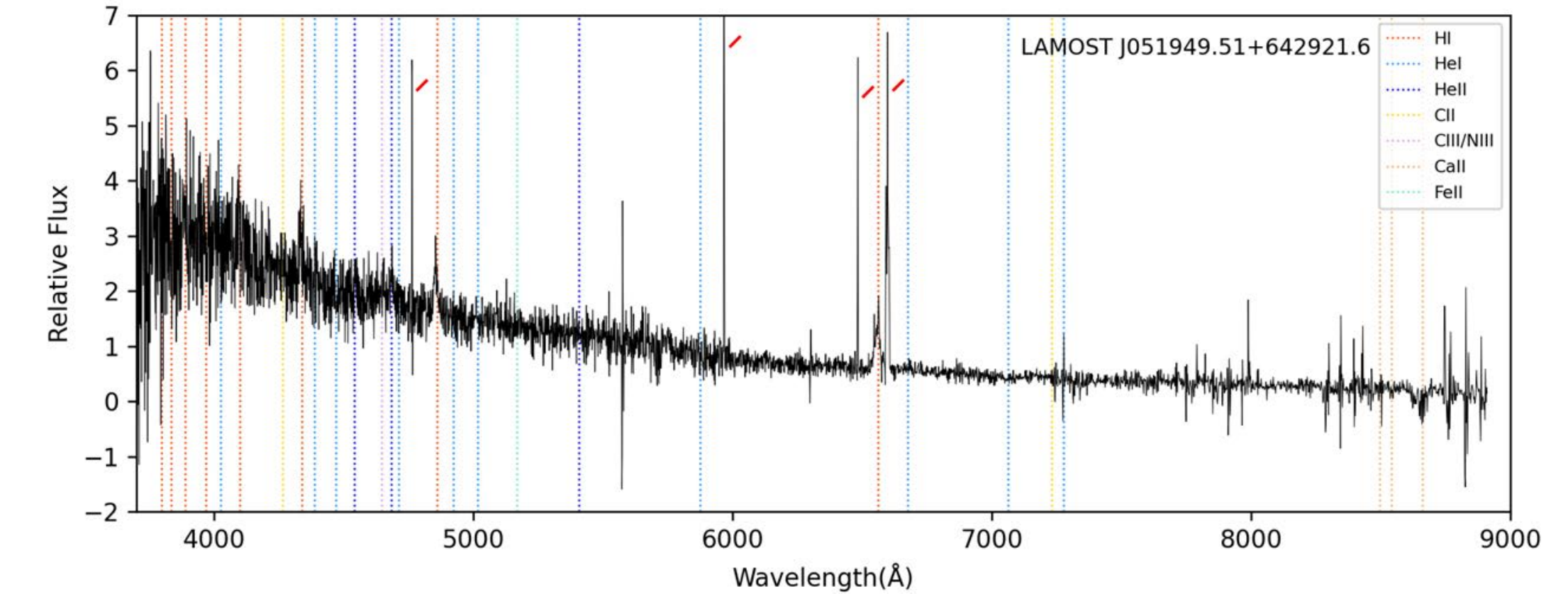}{0.5\textwidth}{}
		\fig{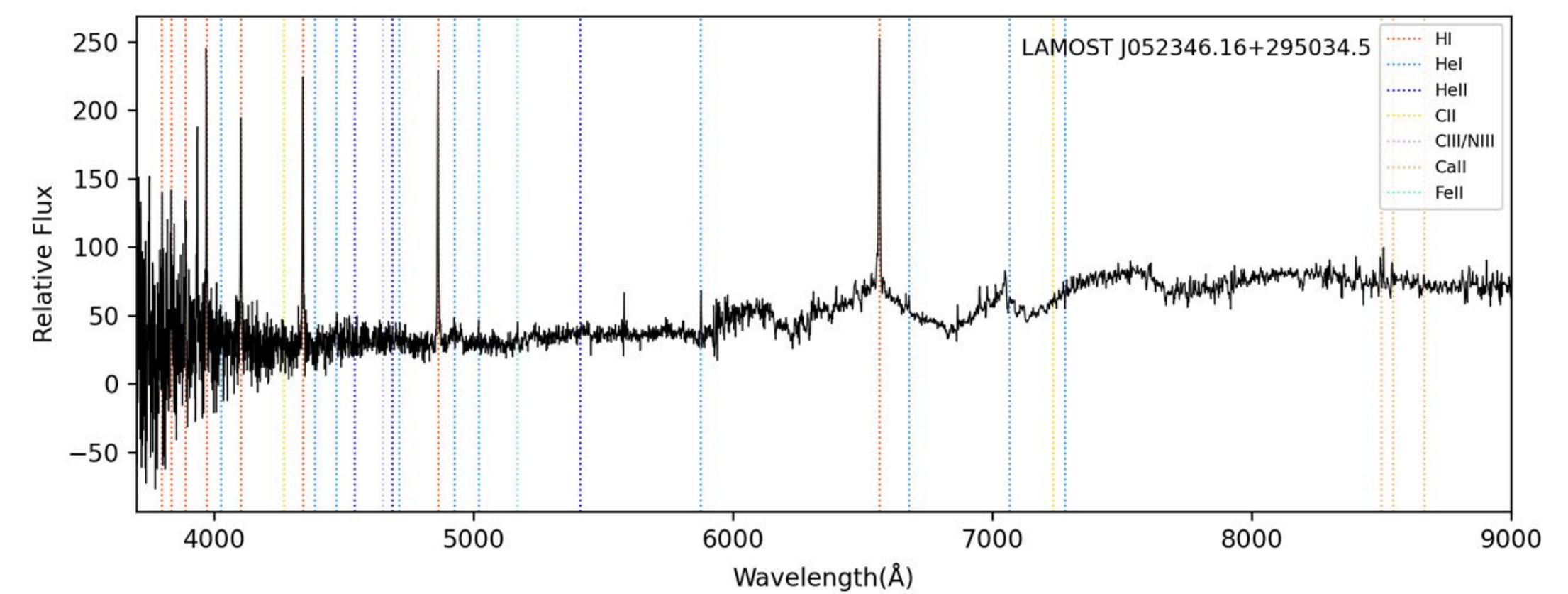}{0.5\textwidth}{}}
	\gridline{\fig{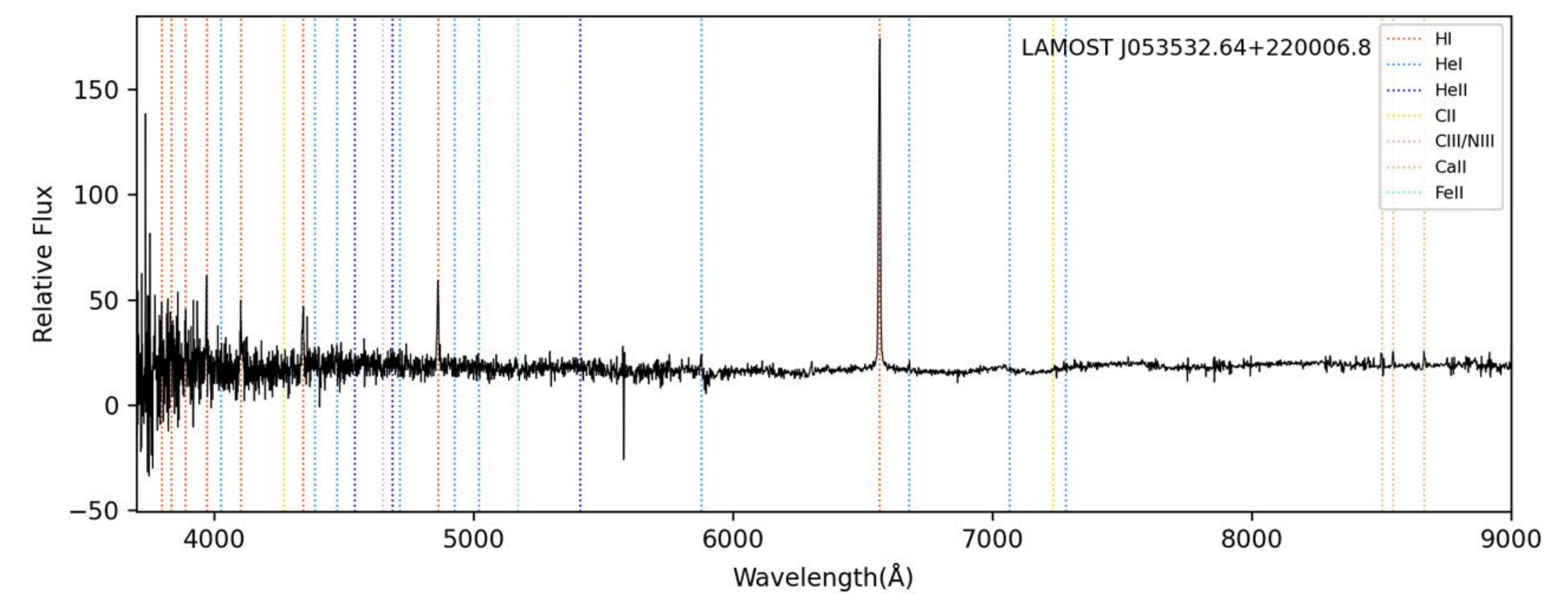}{0.5\textwidth}{}
		\fig{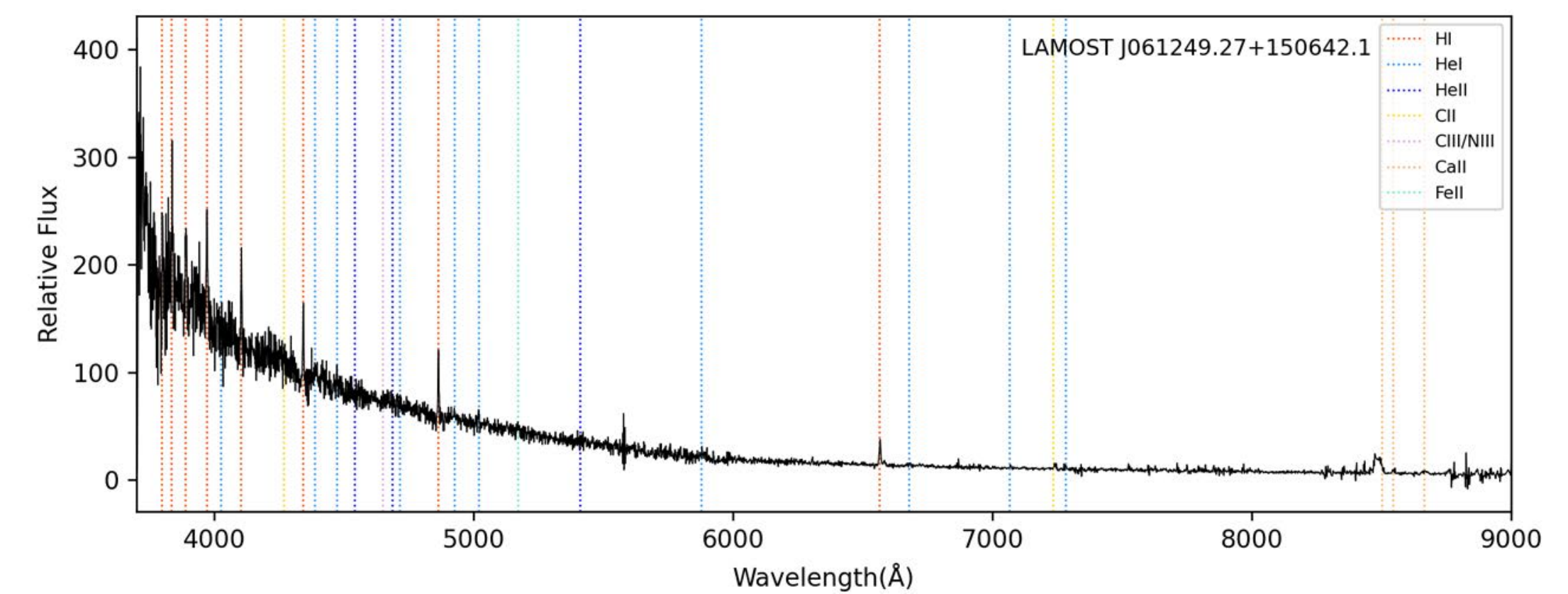}{0.5\textwidth}{}}
	\gridline{\fig{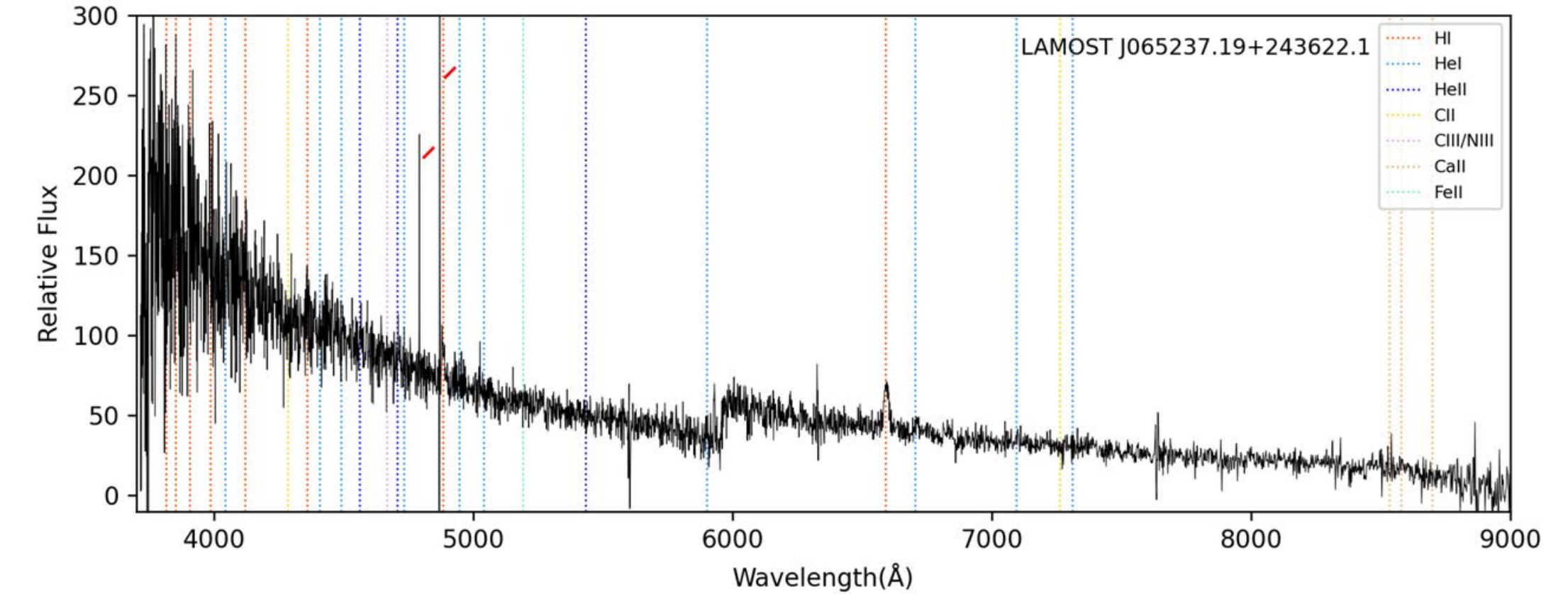}{0.5\textwidth}{}
		\fig{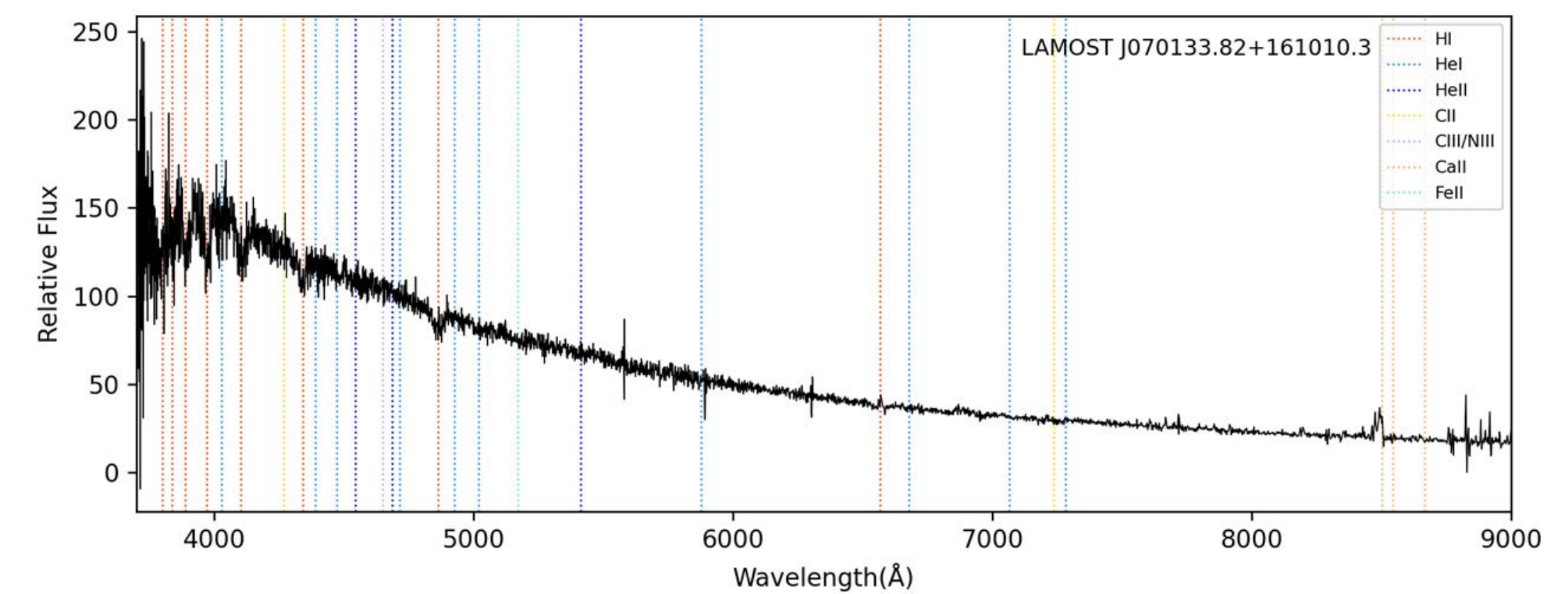}{0.5\textwidth}{}}
	\gridline{\fig{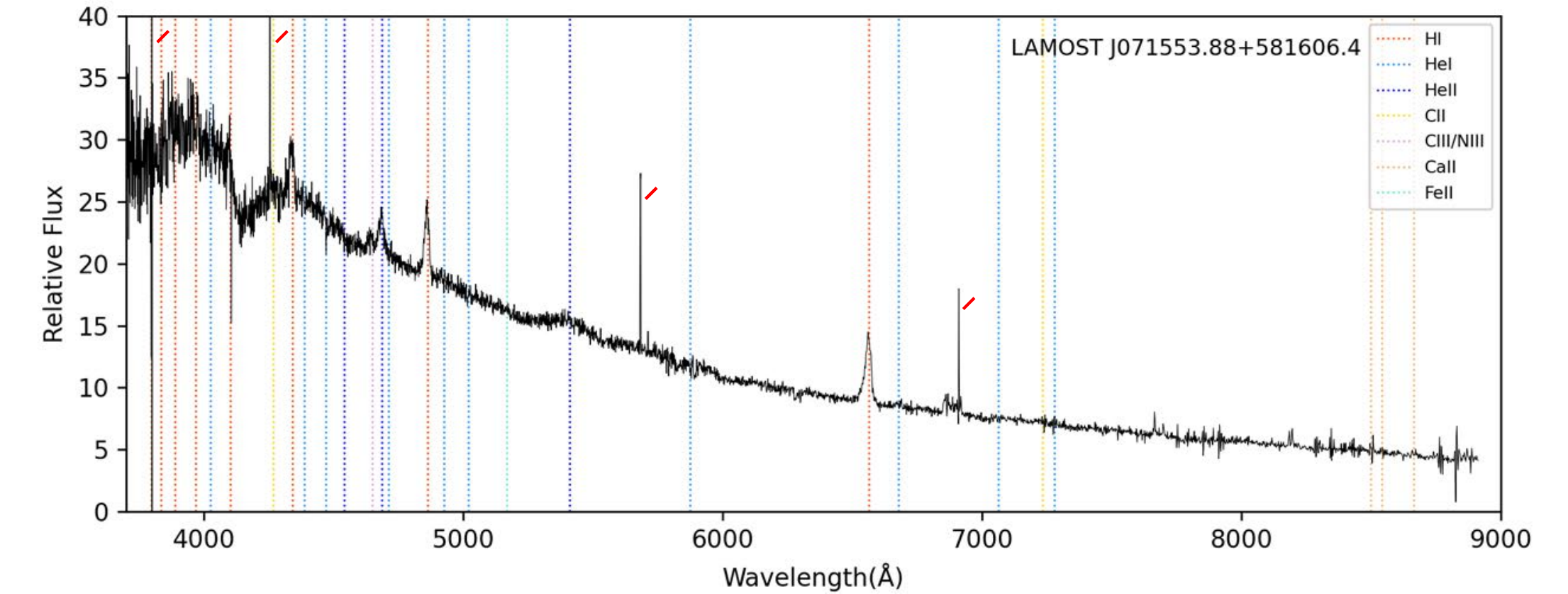}{0.5\textwidth}{}
		\fig{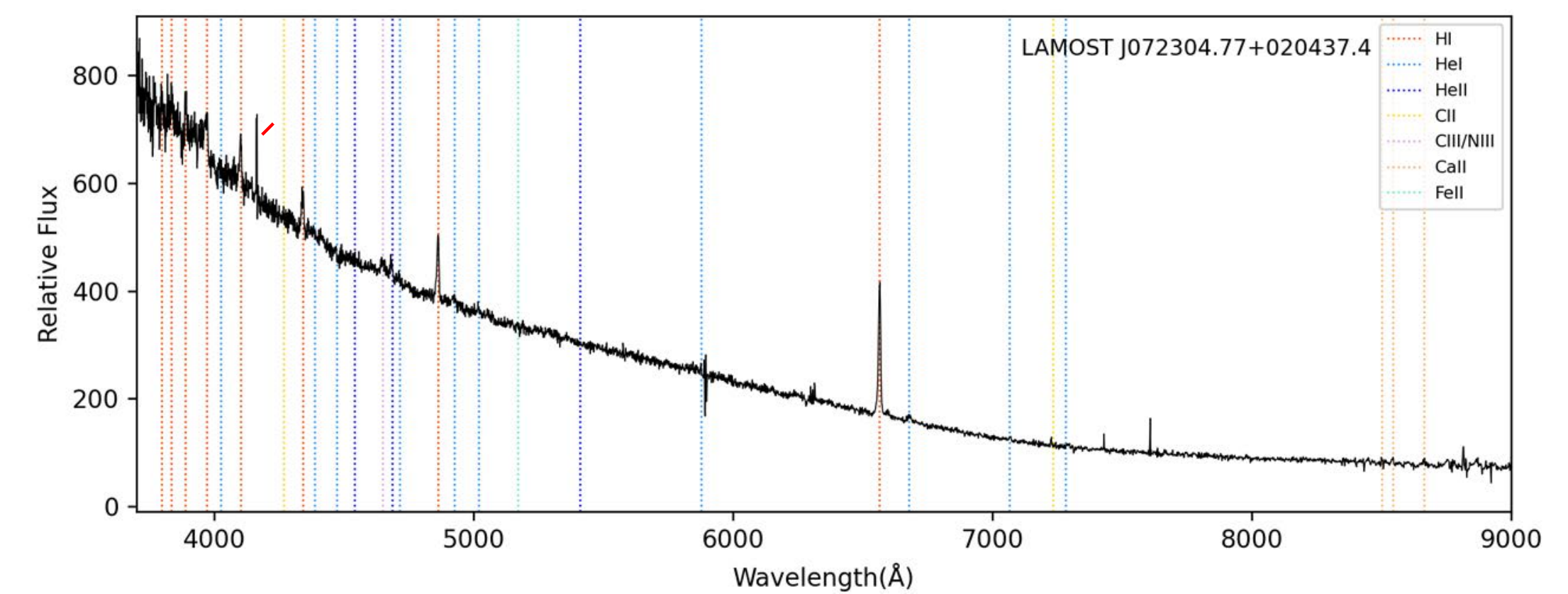}{0.5\textwidth}{}}
	\gridline{\fig{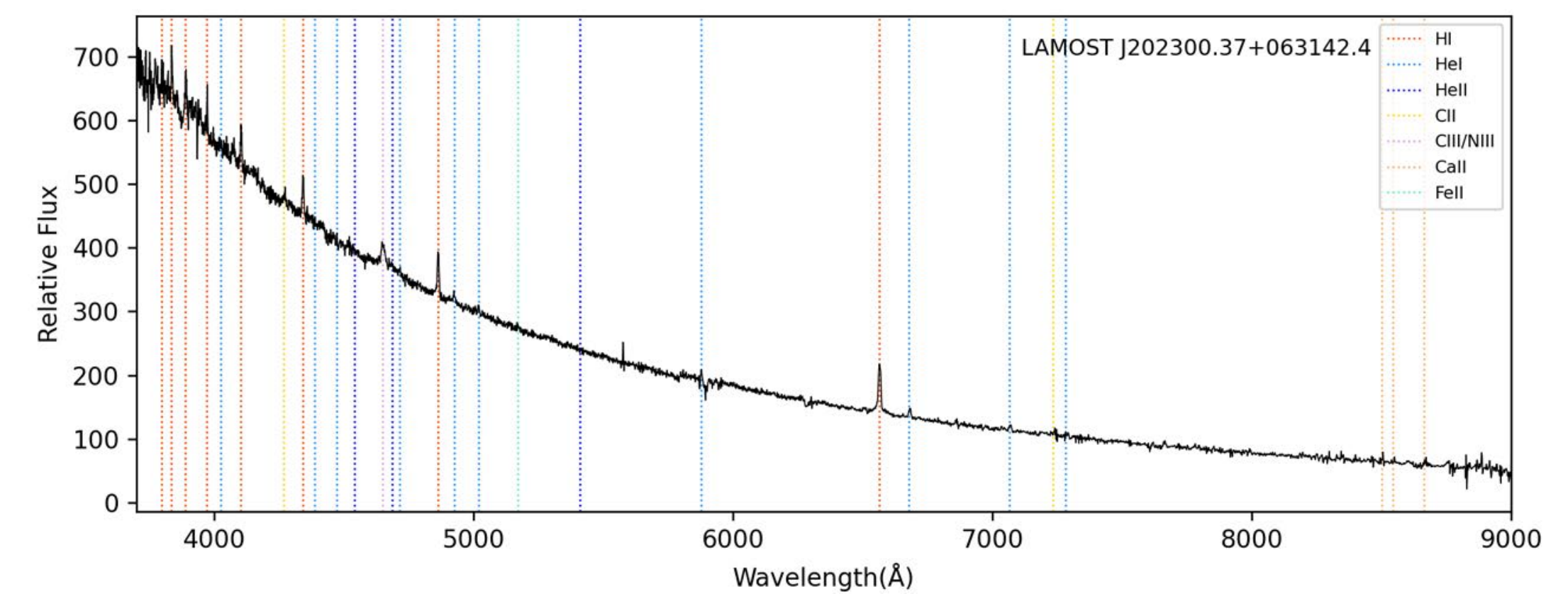}{0.5\textwidth}{}
		\fig{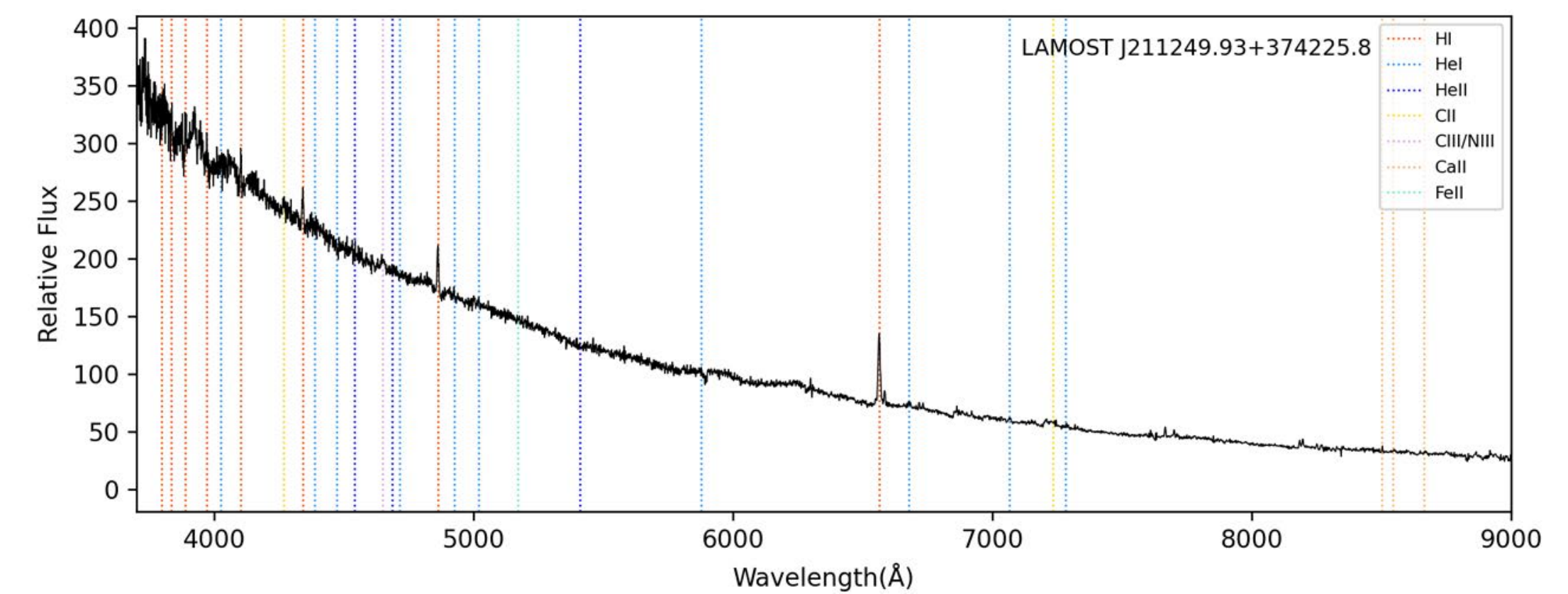}{0.5\textwidth}{}}
\end{figure}

\begin{figure}
	\gridline{\fig{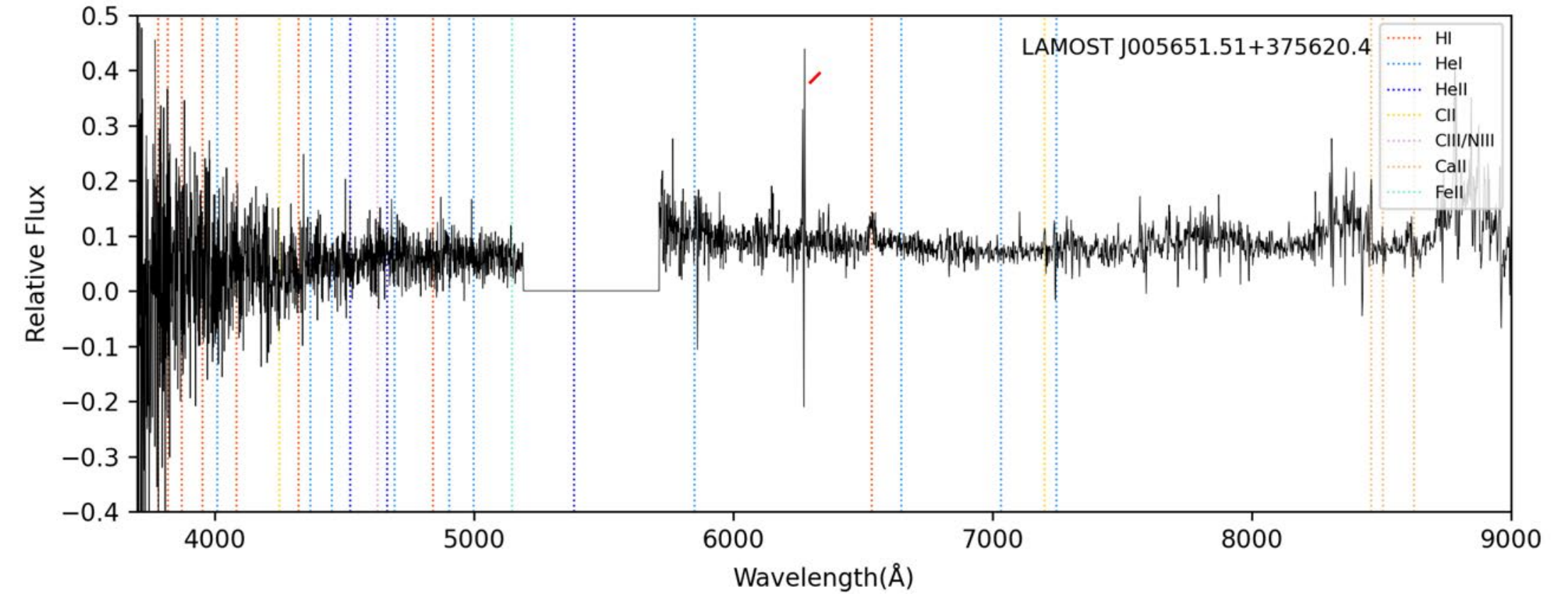}{0.5\textwidth}{}
		\fig{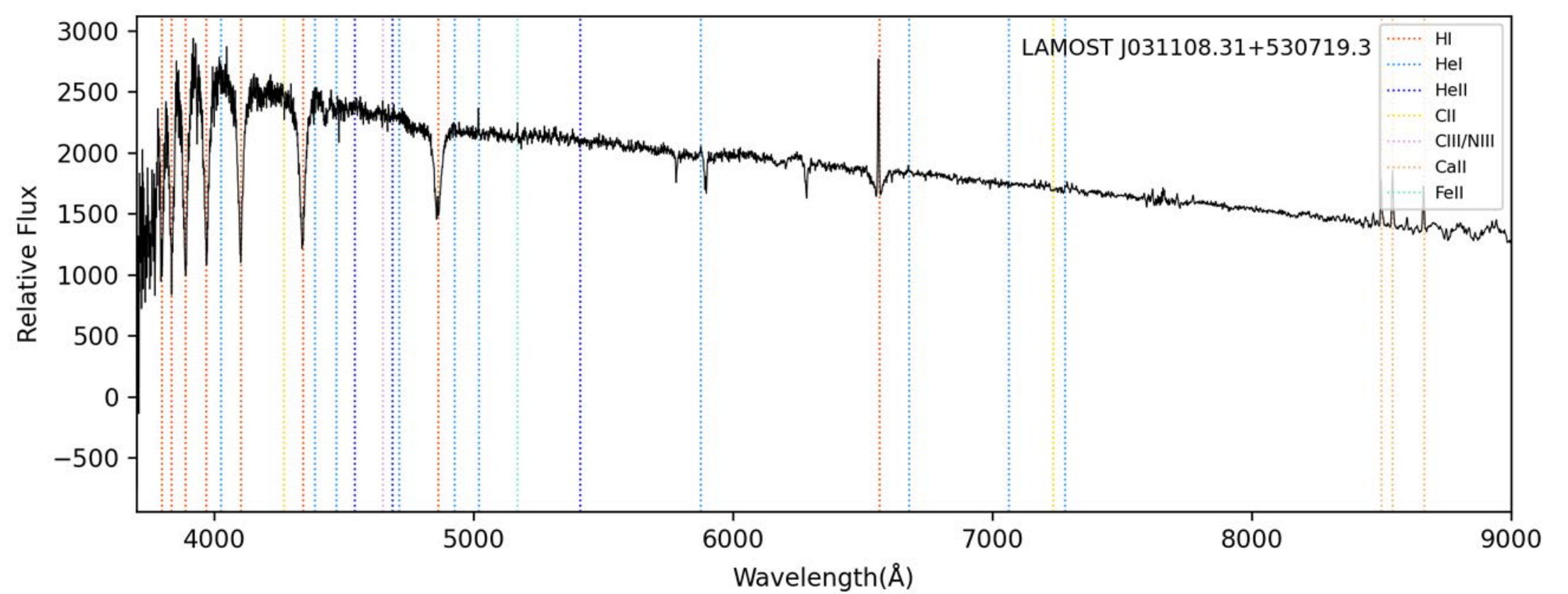}{0.5\textwidth}{}}
	\gridline{\fig{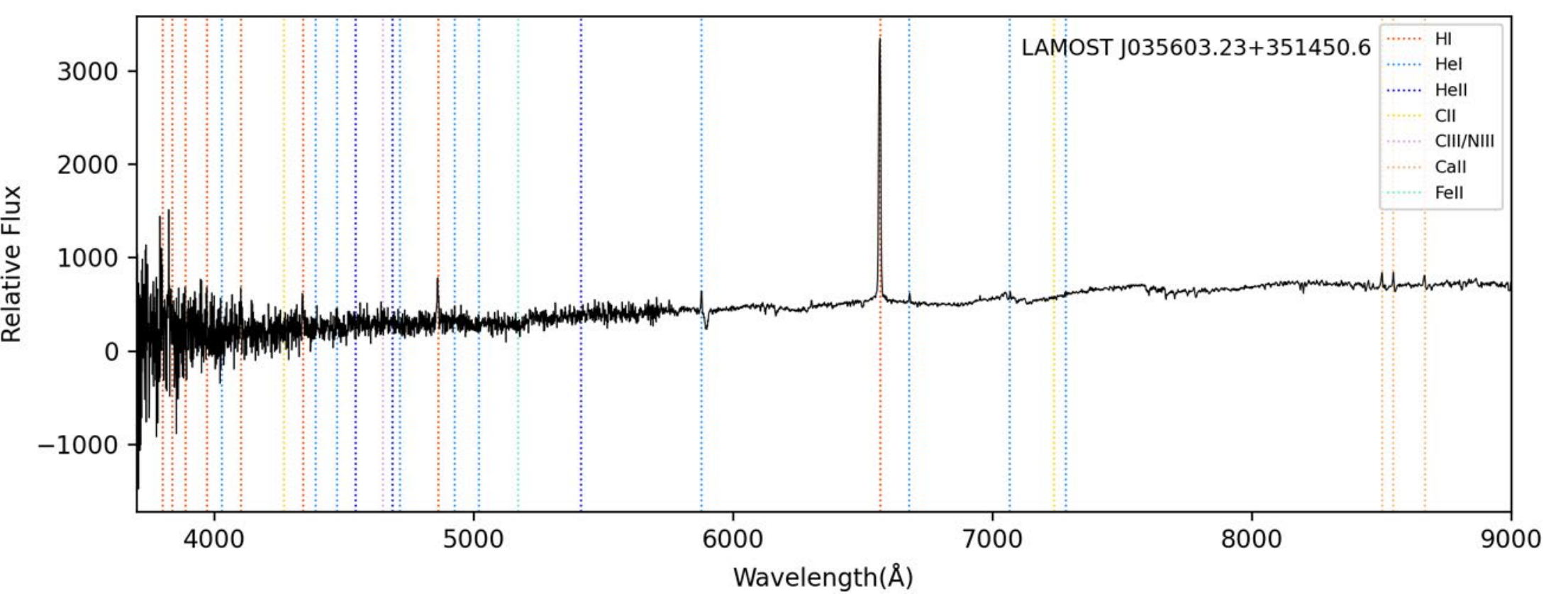}{0.5\textwidth}{}
		\fig{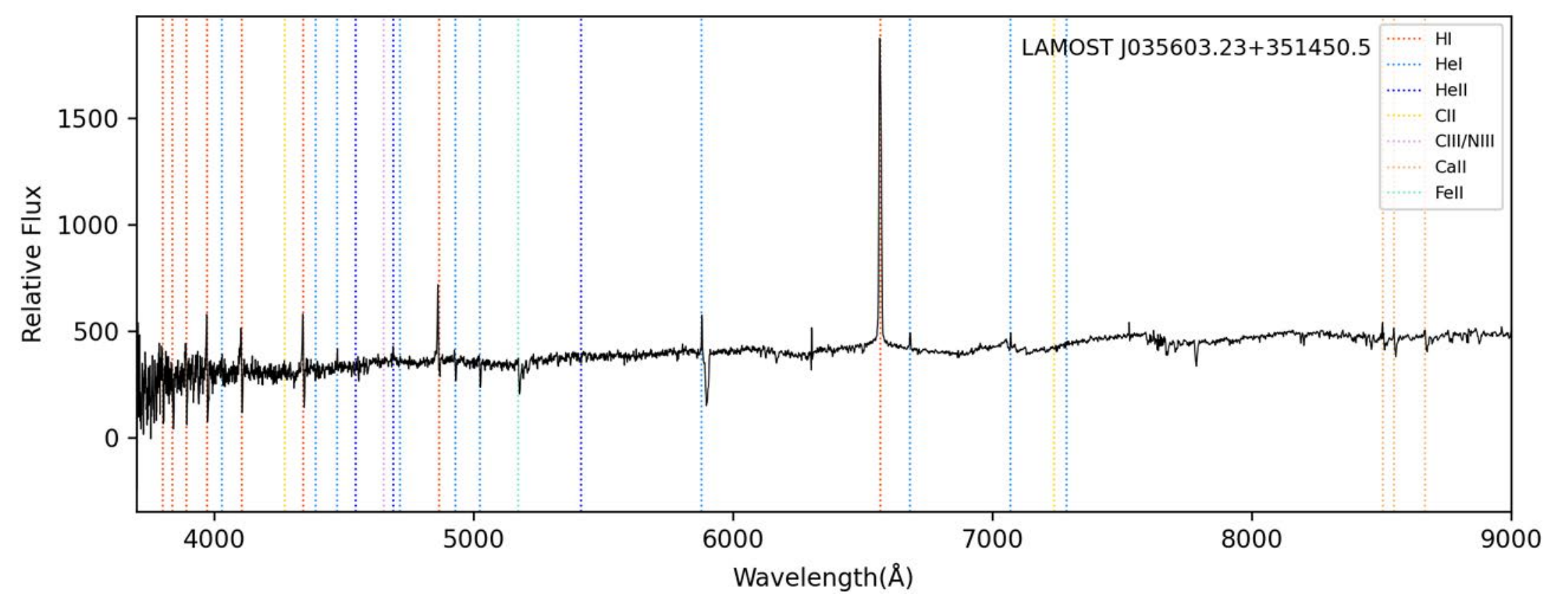}{0.5\textwidth}{}}
	\gridline{\fig{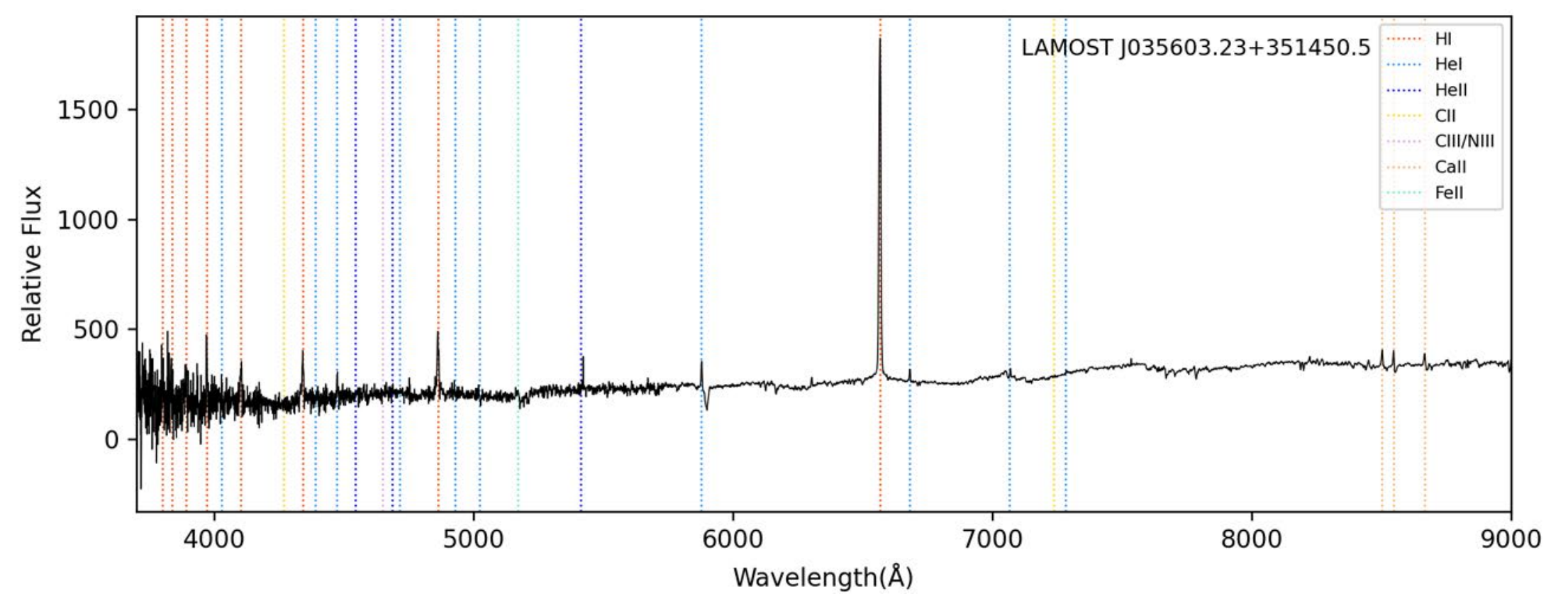}{0.5\textwidth}{}
		\fig{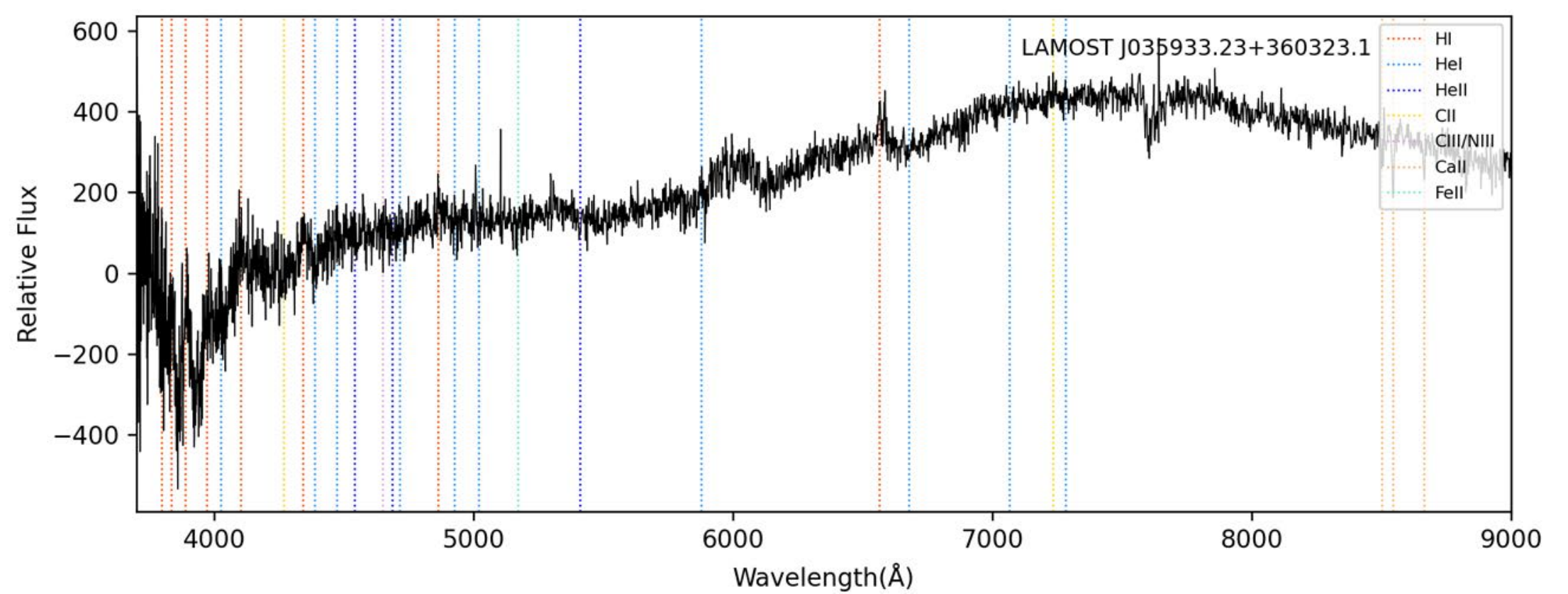}{0.5\textwidth}{}}
	\gridline{\fig{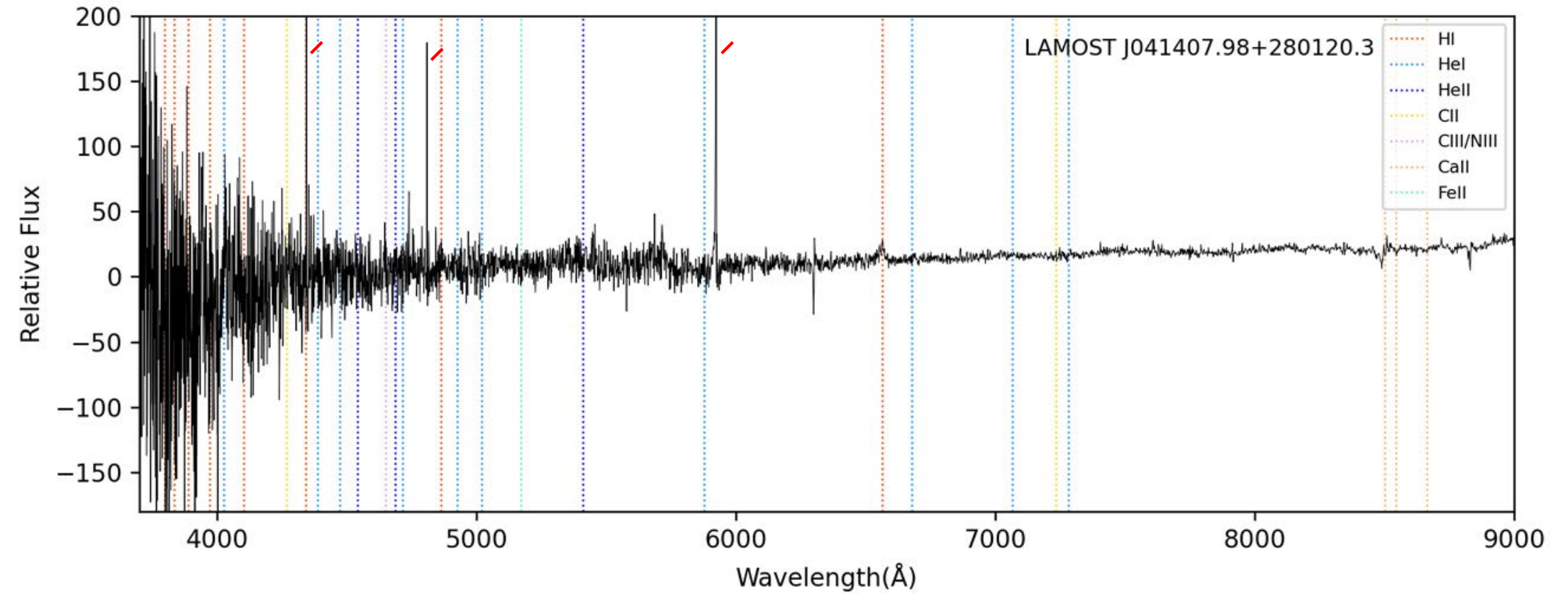}{0.5\textwidth}{}
		\fig{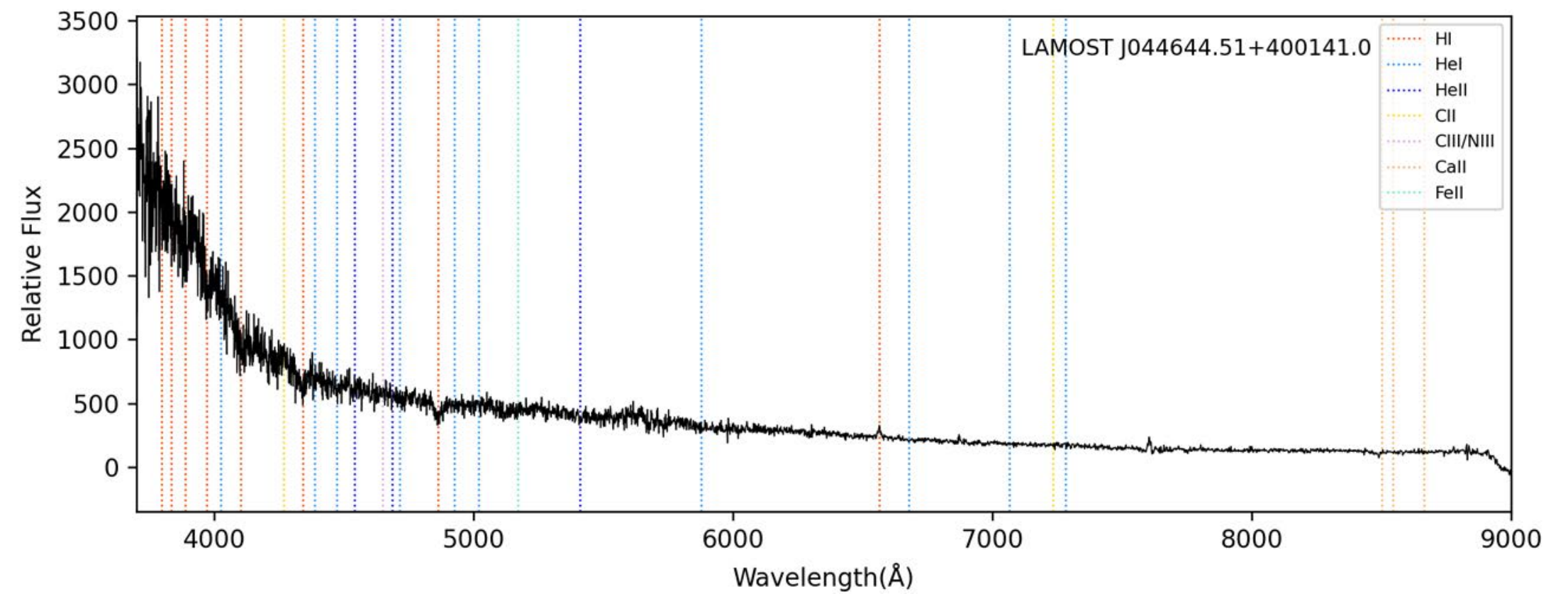}{0.5\textwidth}{}}
	\gridline{\fig{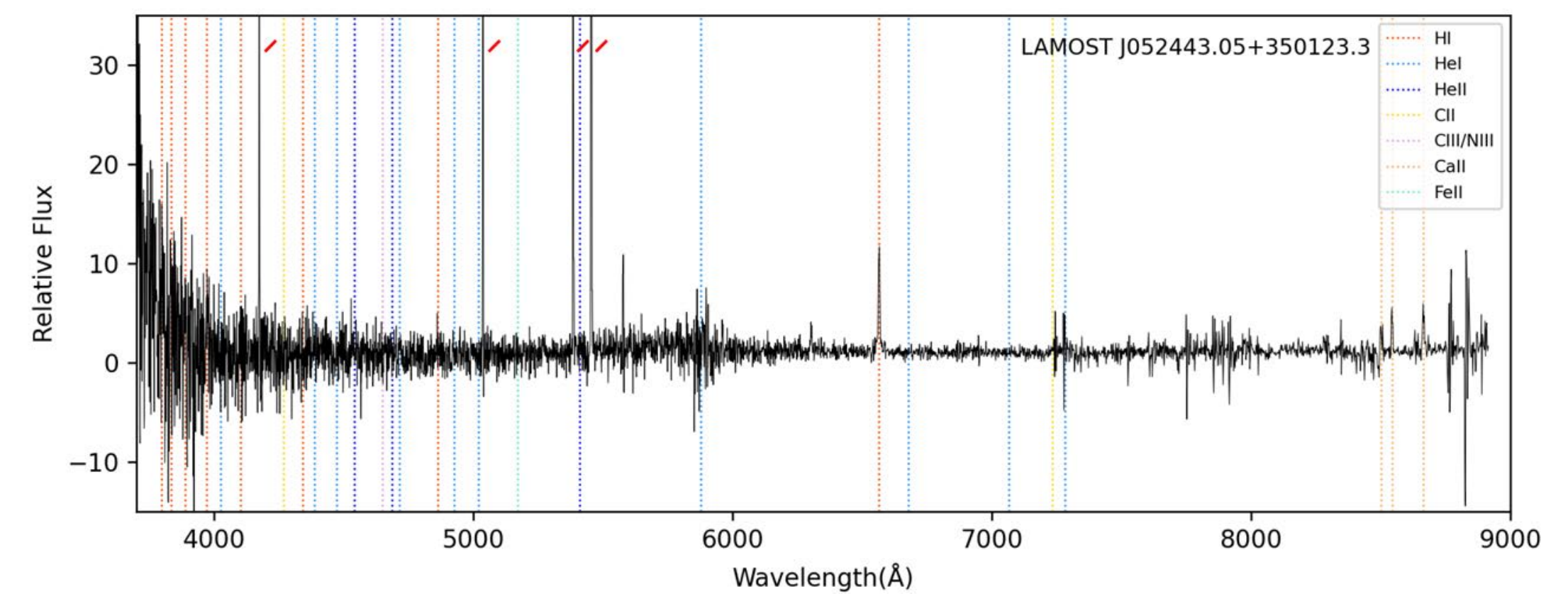}{0.5\textwidth}{}
		\fig{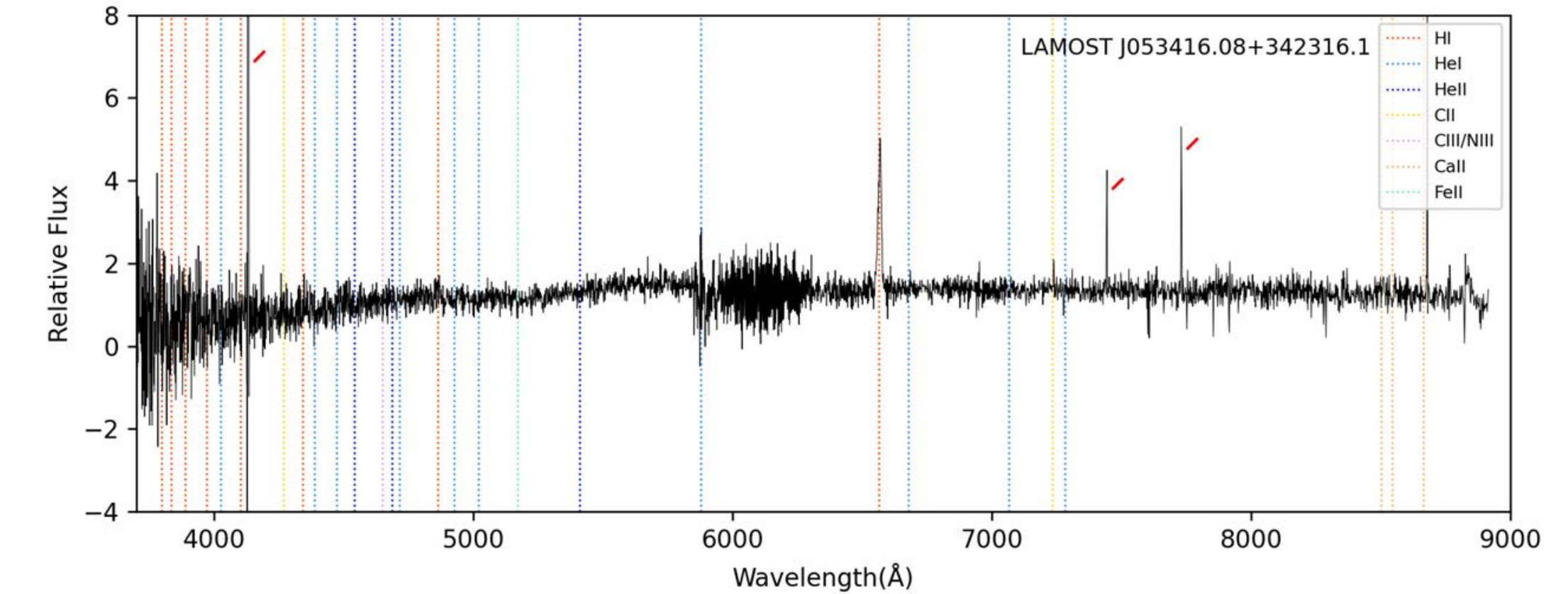}{0.5\textwidth}{}}
\end{figure}
\begin{figure}
	\gridline{\fig{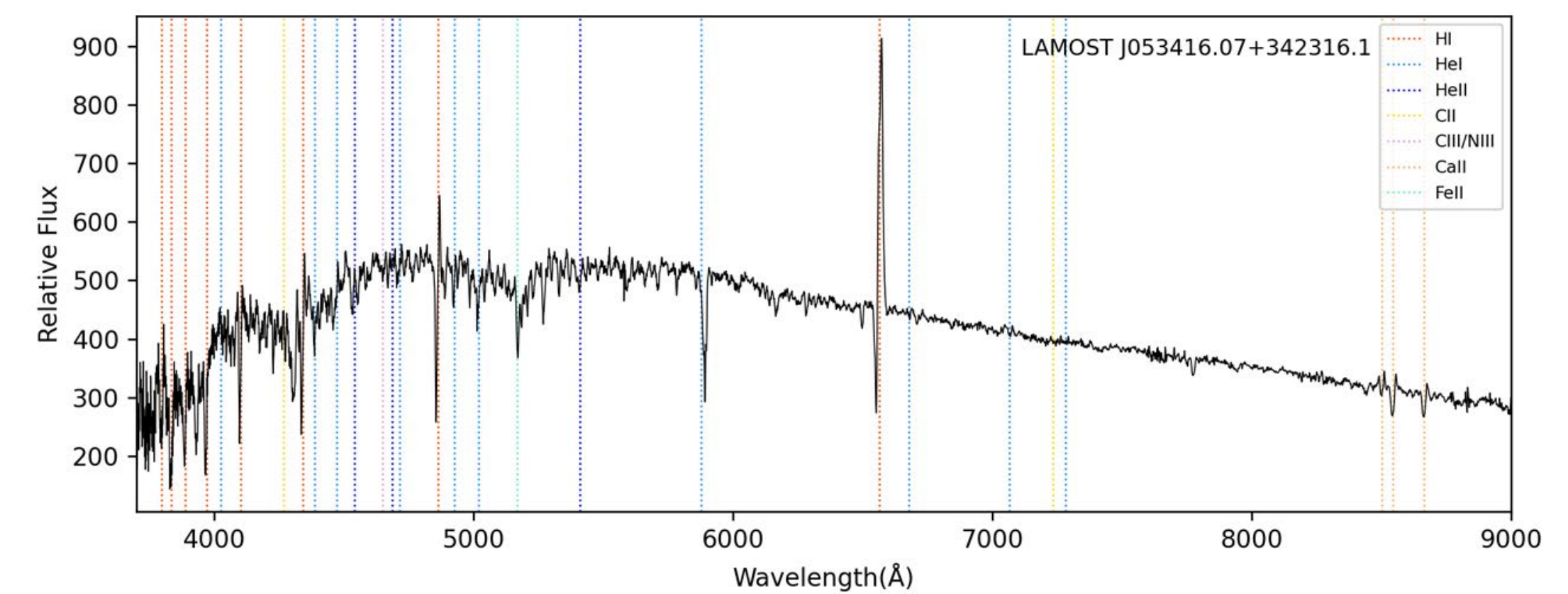}{0.5\textwidth}{}
		\fig{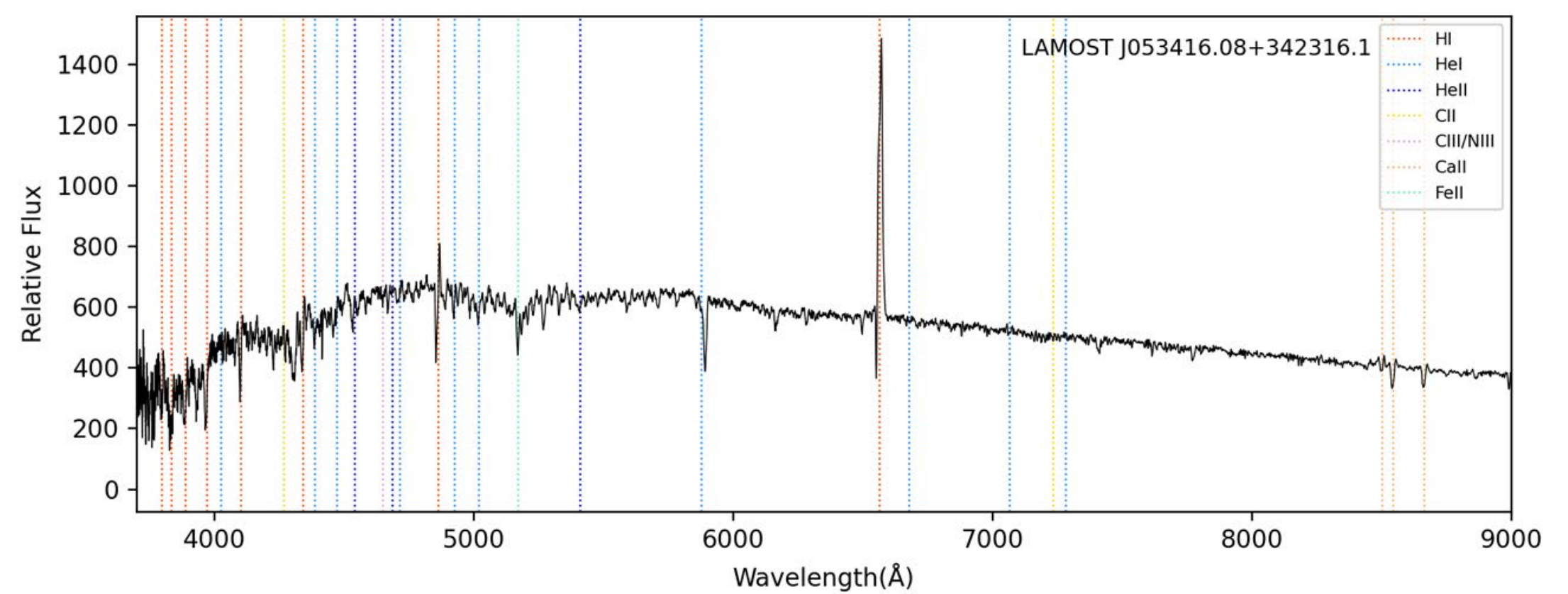}{0.5\textwidth}{}}
	\gridline{\fig{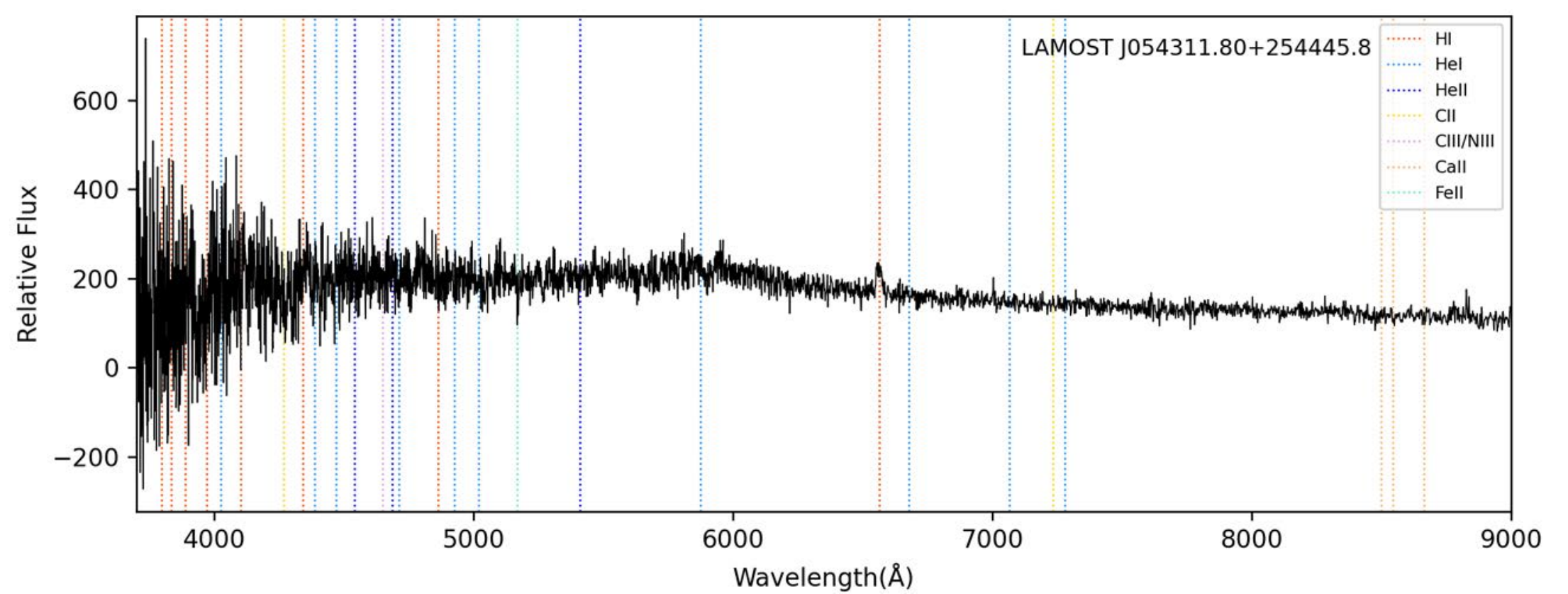}{0.5\textwidth}{}
		\fig{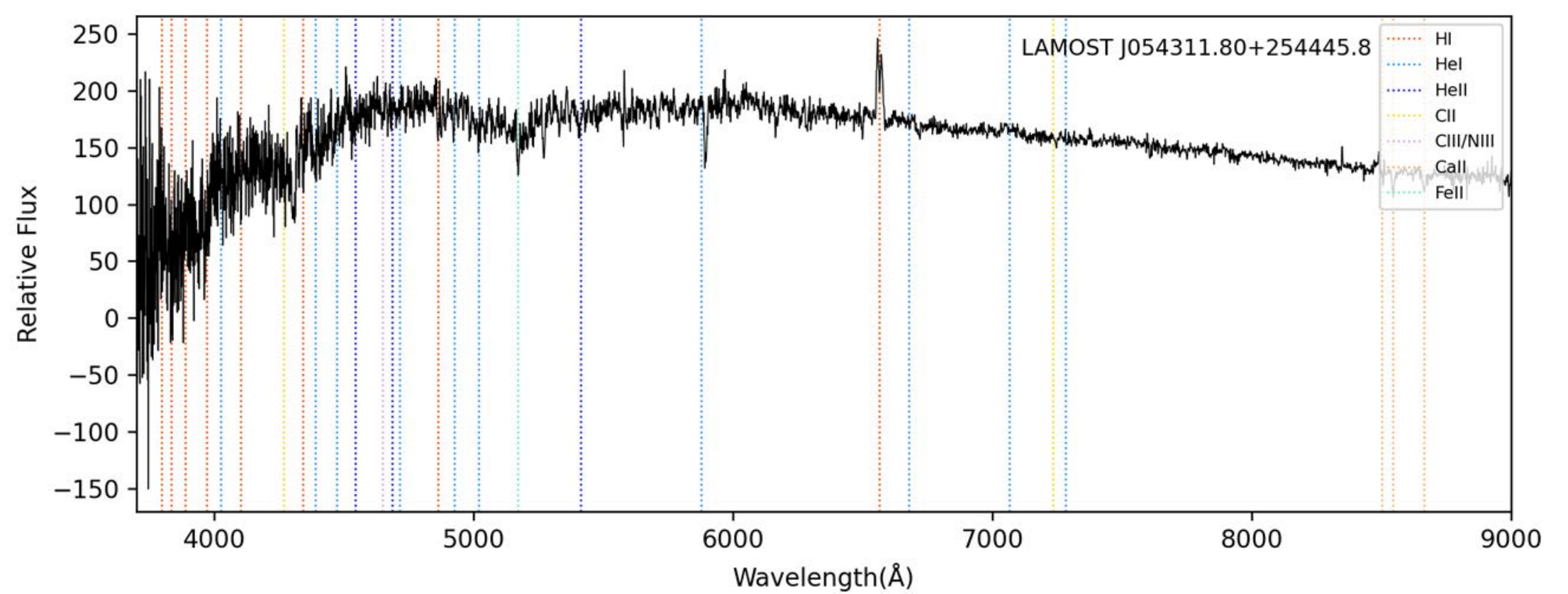}{0.5\textwidth}{}}
	\gridline{\fig{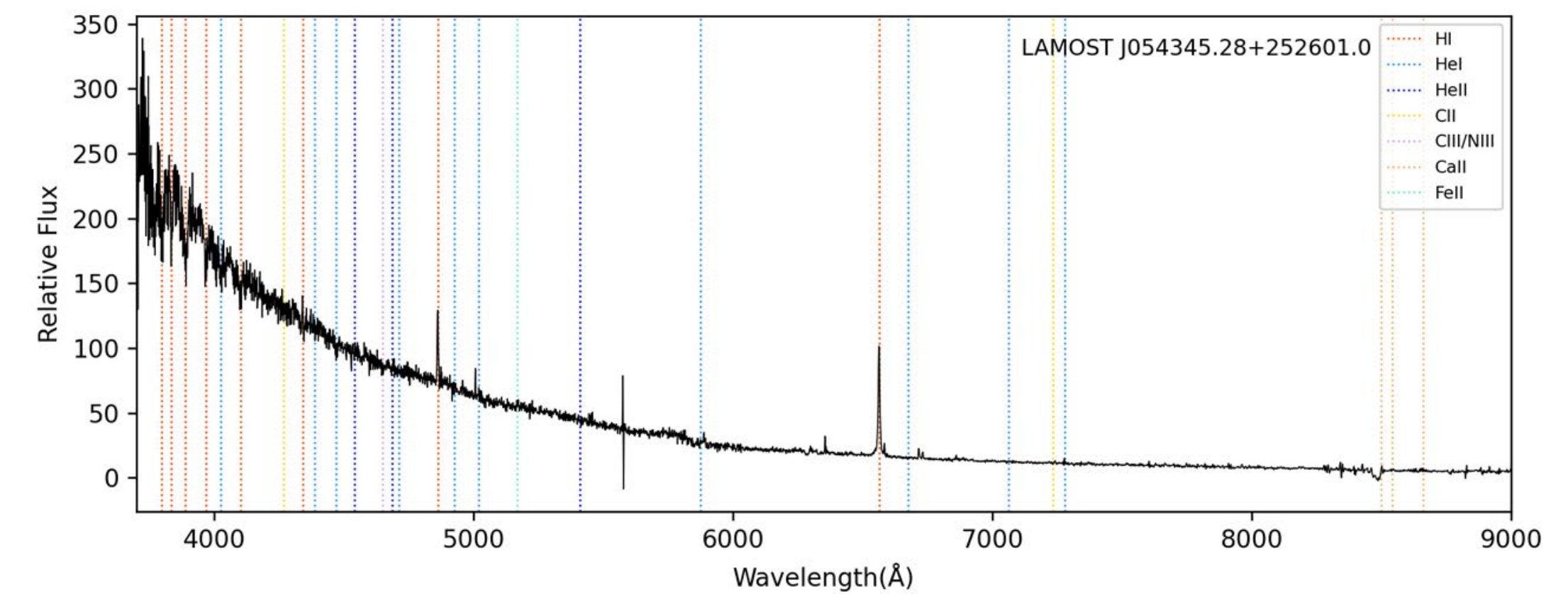}{0.5\textwidth}{}
		\fig{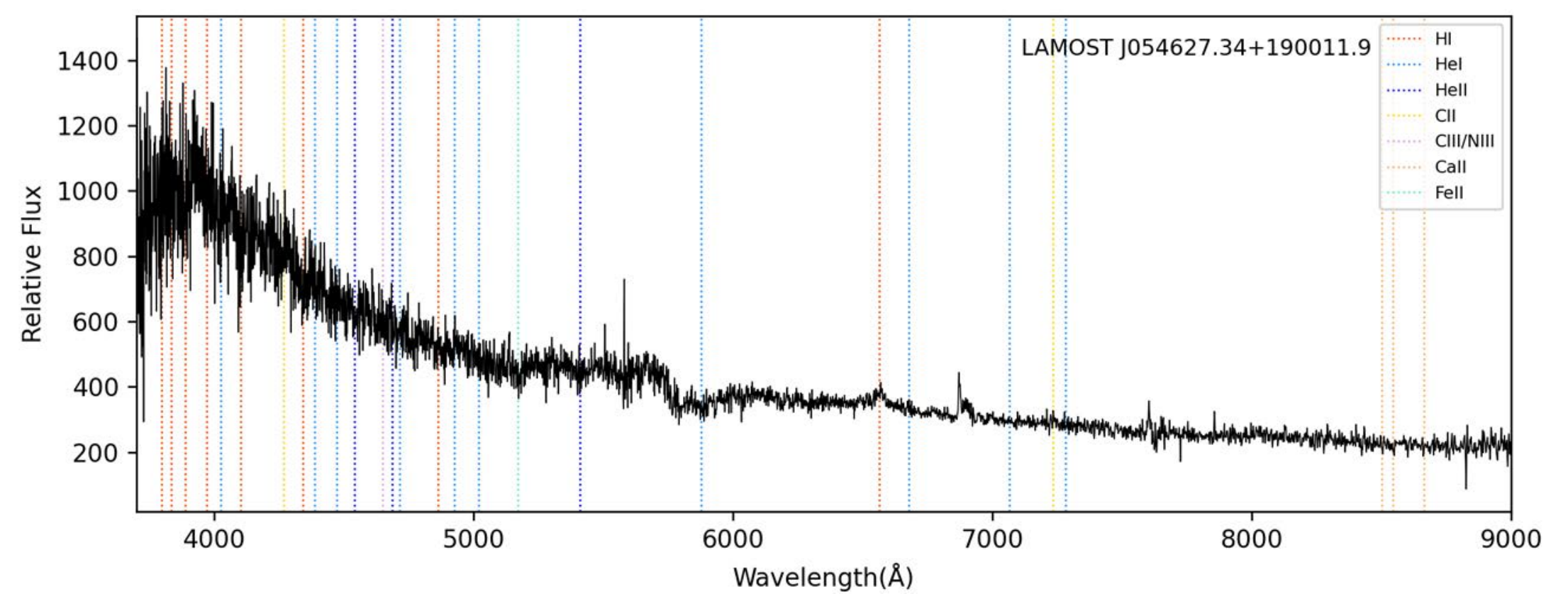}{0.5\textwidth}{}}
	\gridline{\fig{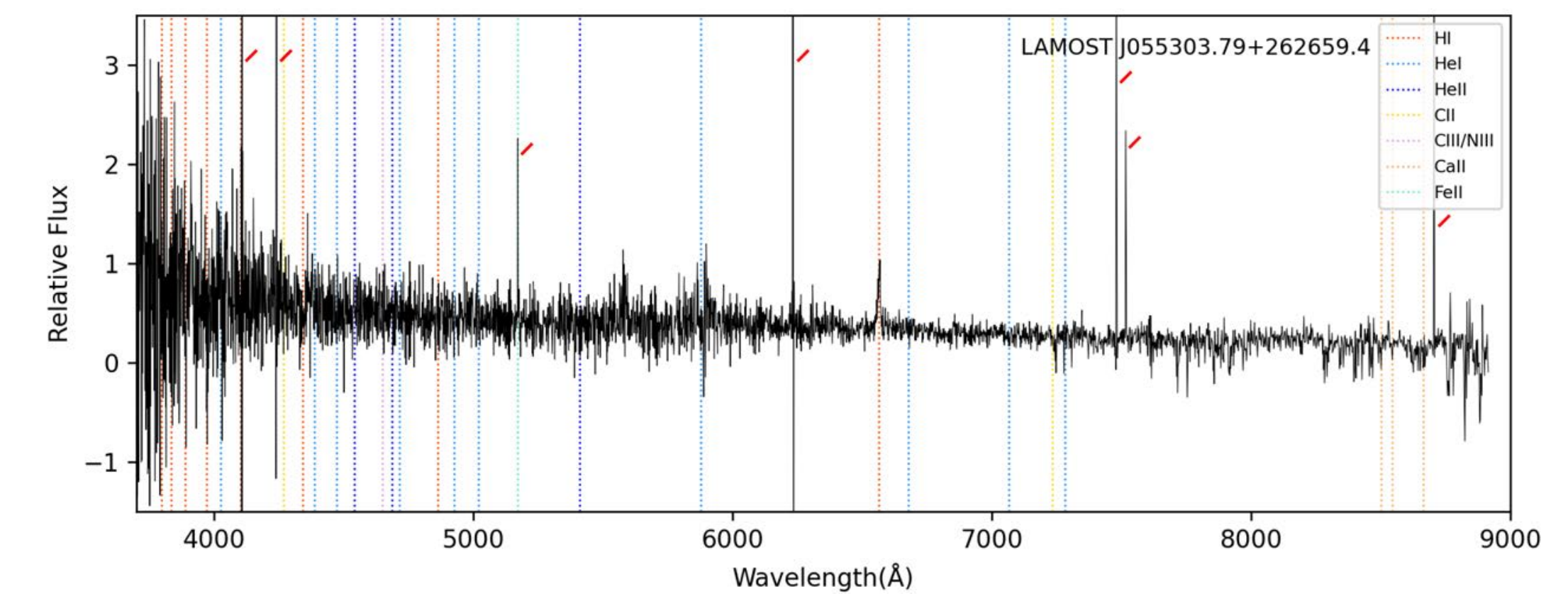}{0.5\textwidth}{}
		\fig{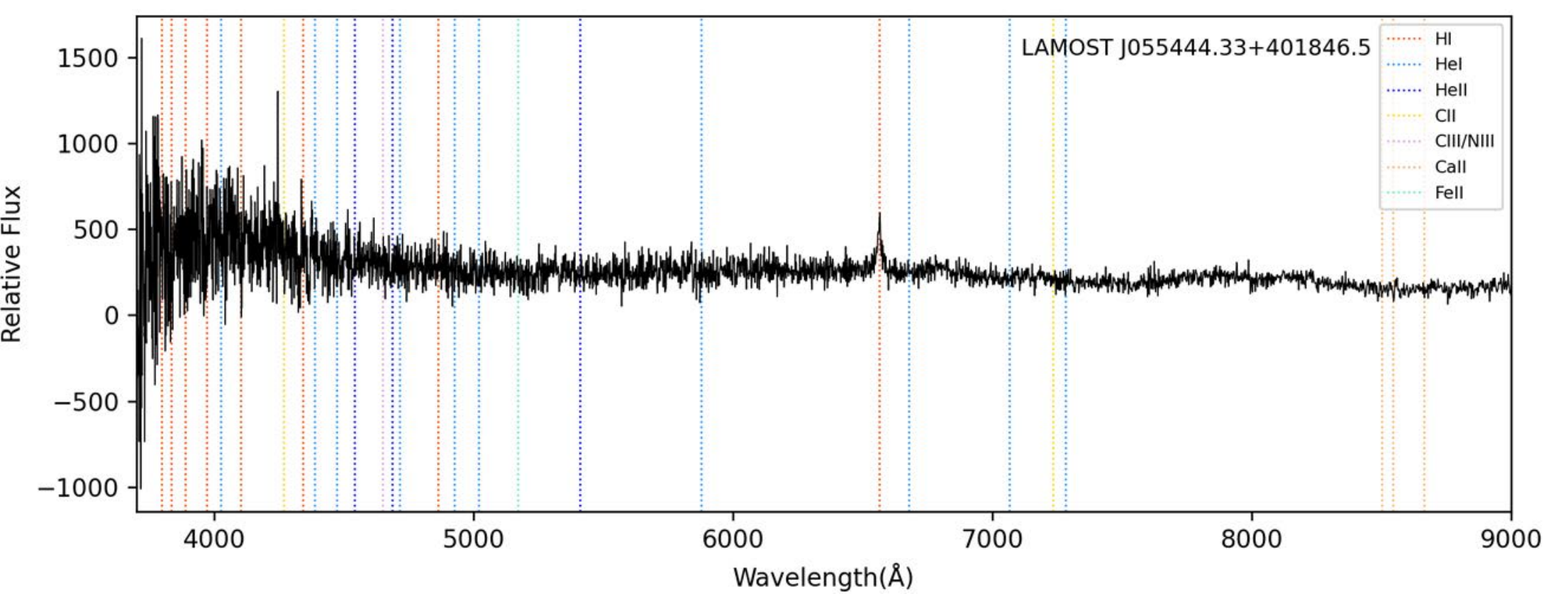}{0.5\textwidth}{}}
	\gridline{\fig{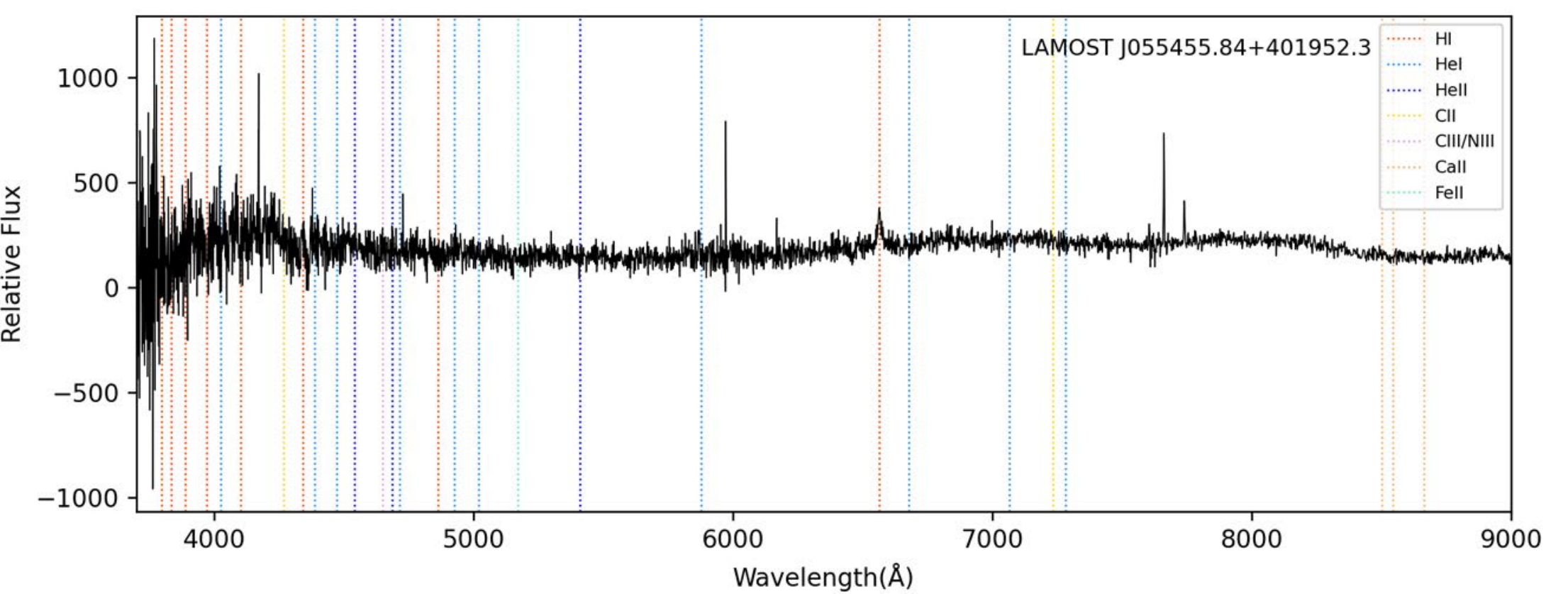}{0.5\textwidth}{}
		\fig{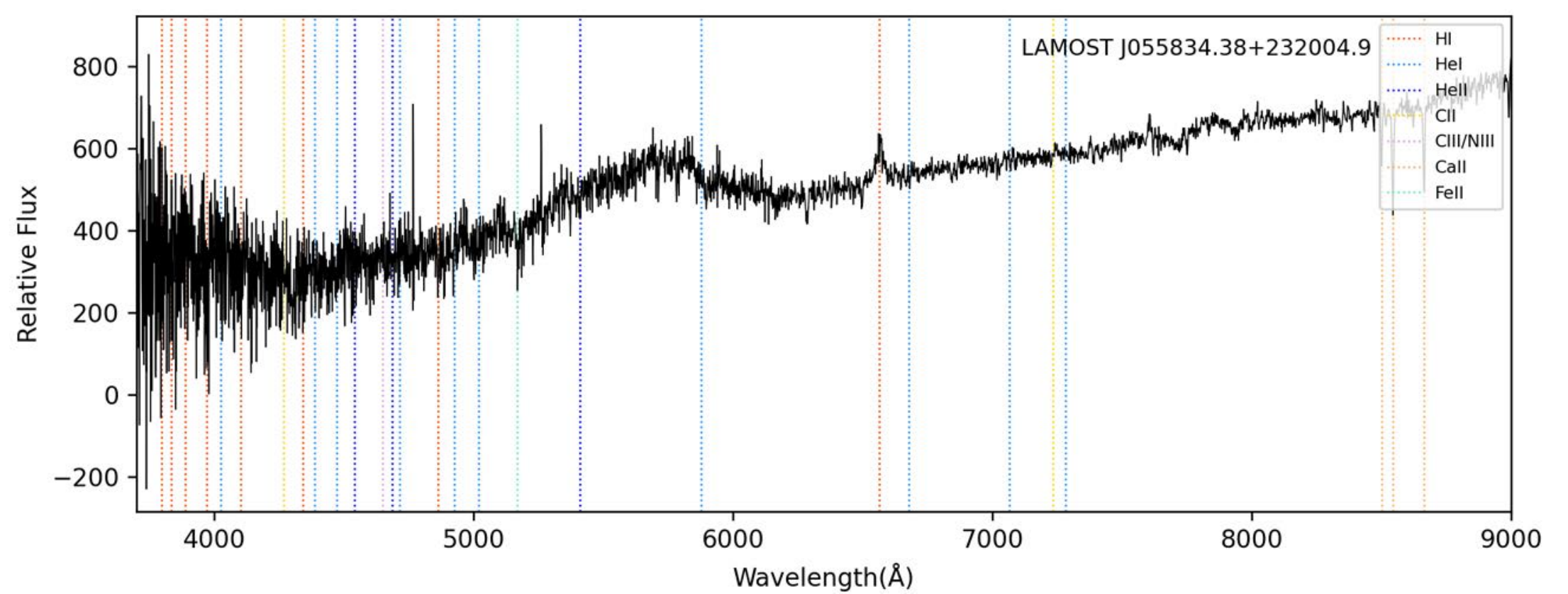}{0.5\textwidth}{}}
\end{figure}
\begin{figure}
	\gridline{\fig{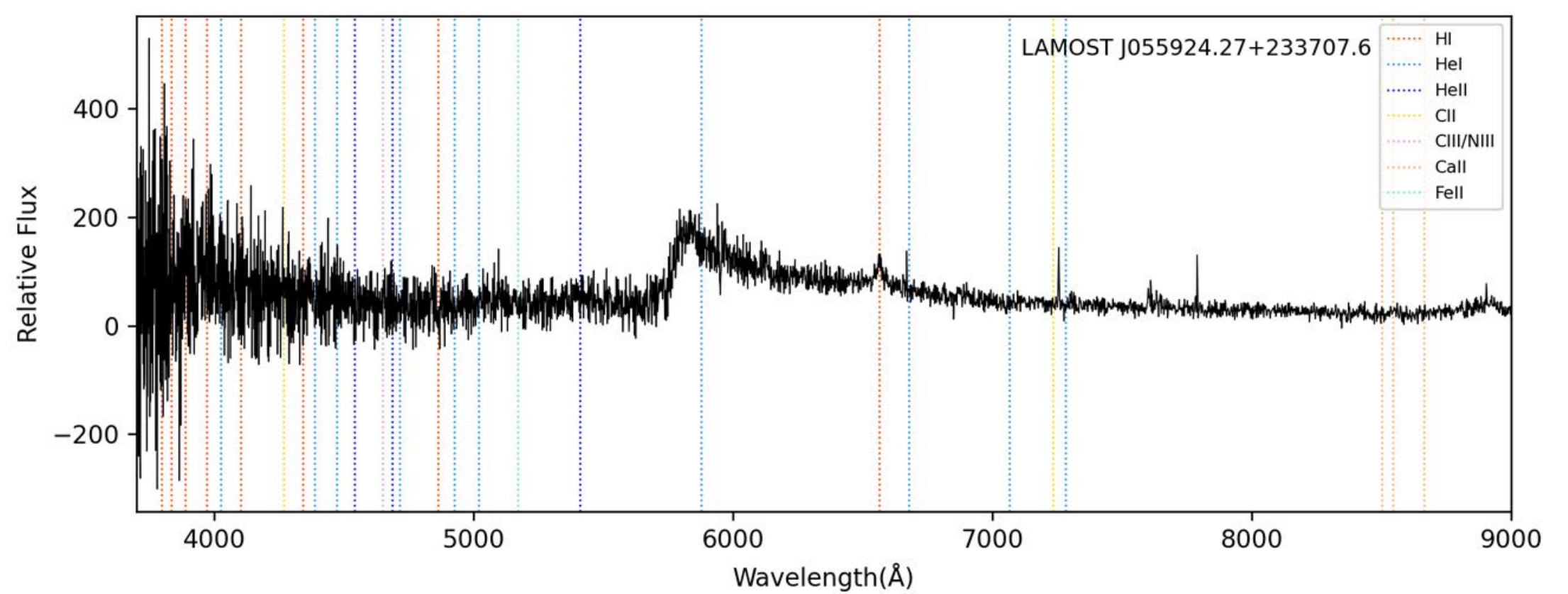}{0.5\textwidth}{}
		\fig{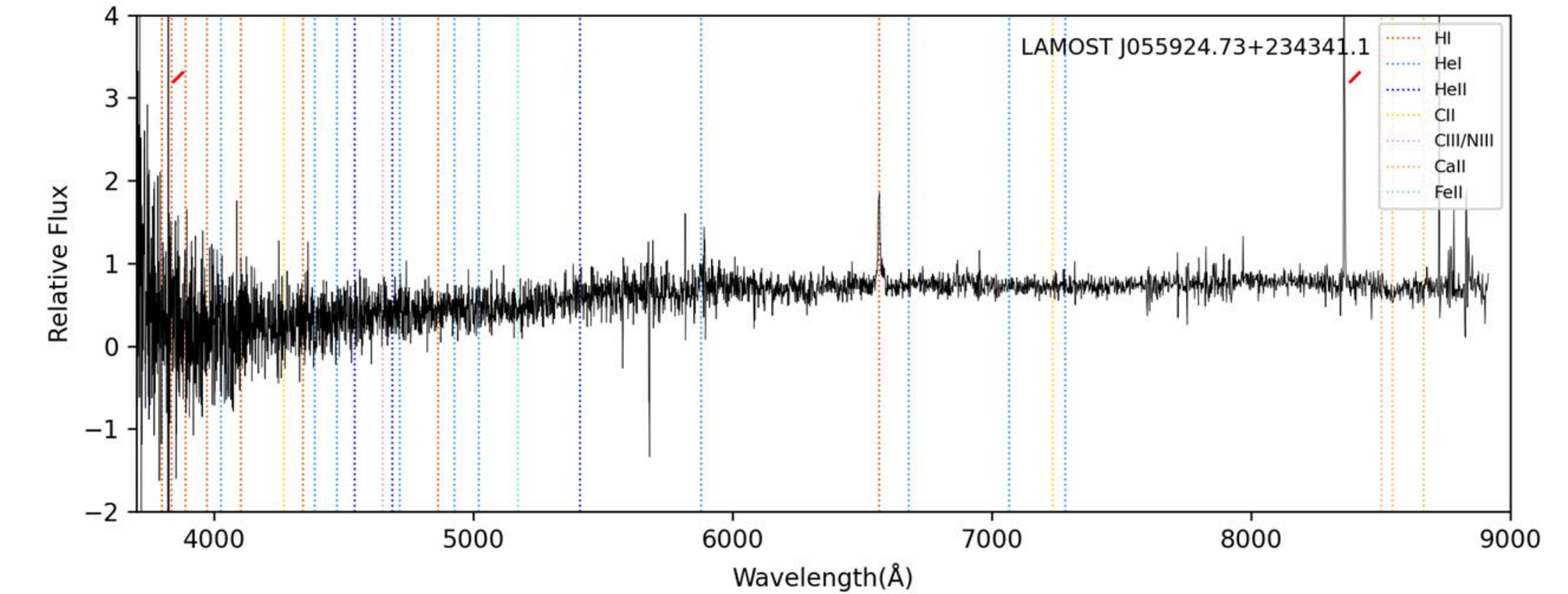}{0.5\textwidth}{}}
	\gridline{\fig{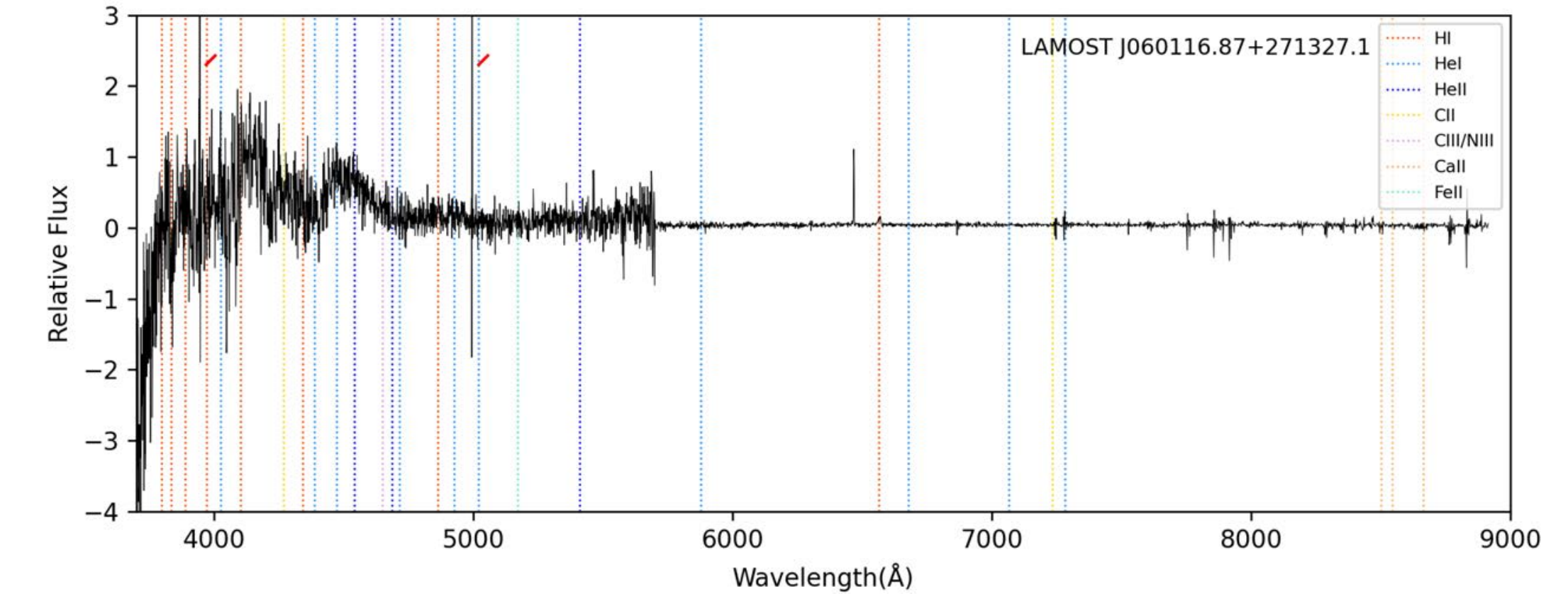}{0.5\textwidth}{}
		\fig{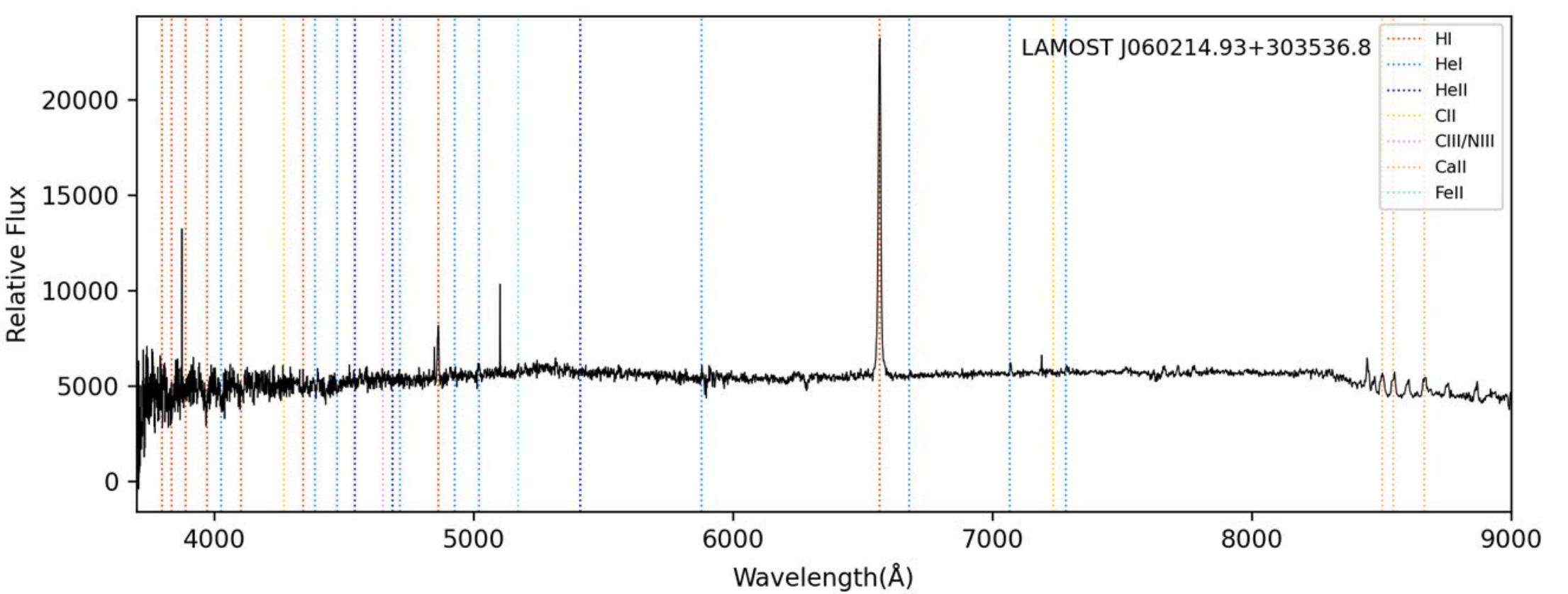}{0.5\textwidth}{}}
	\gridline{\fig{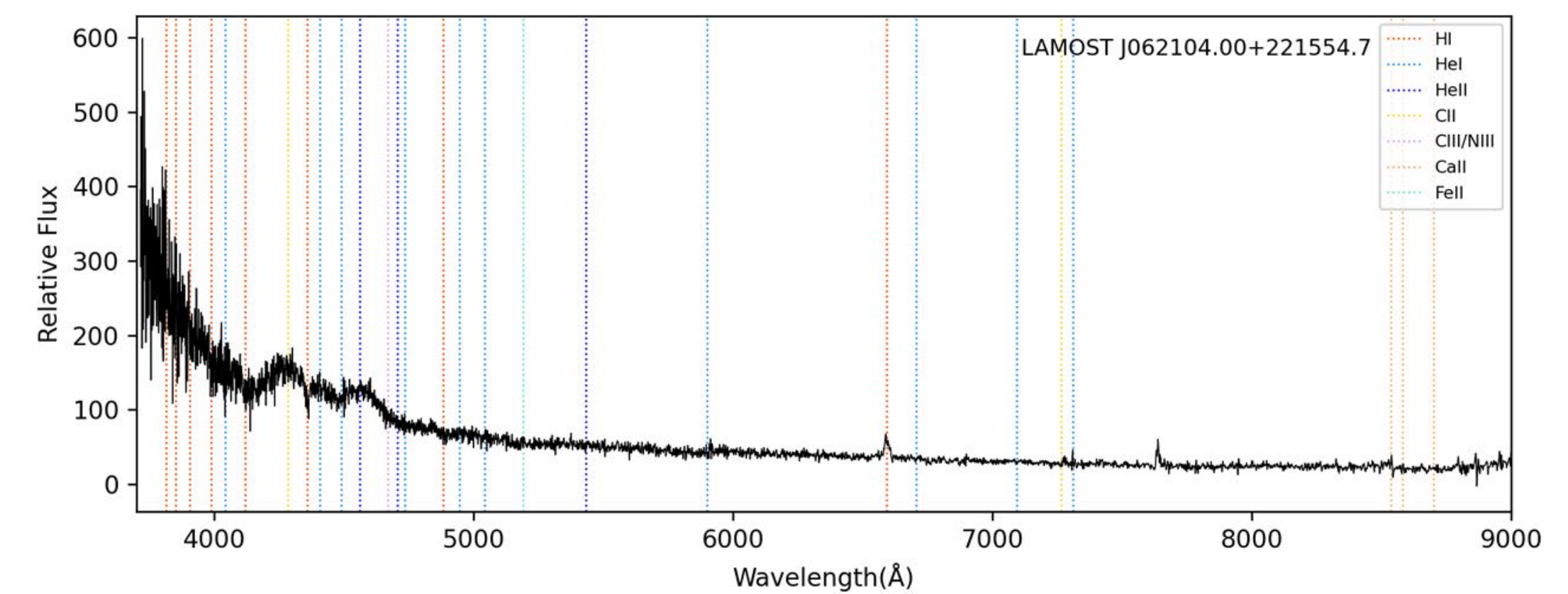}{0.5\textwidth}{}
		\fig{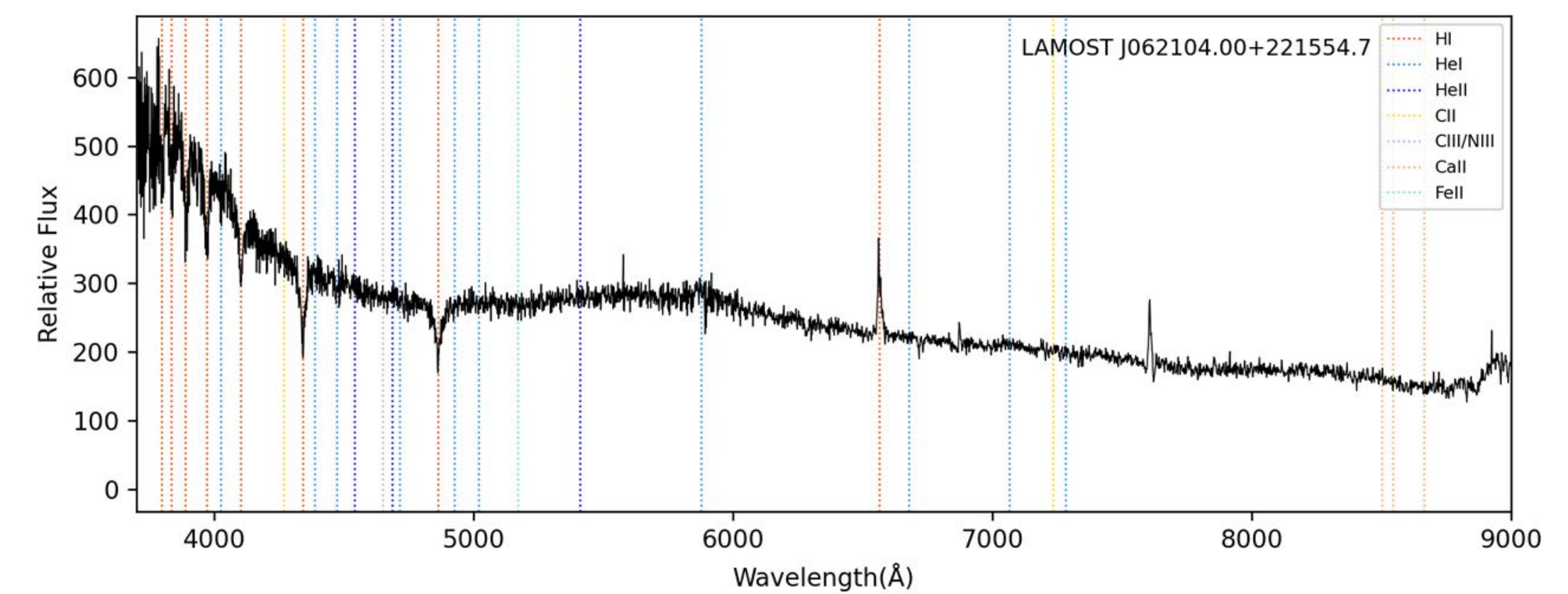}{0.5\textwidth}{}}
	\gridline{\fig{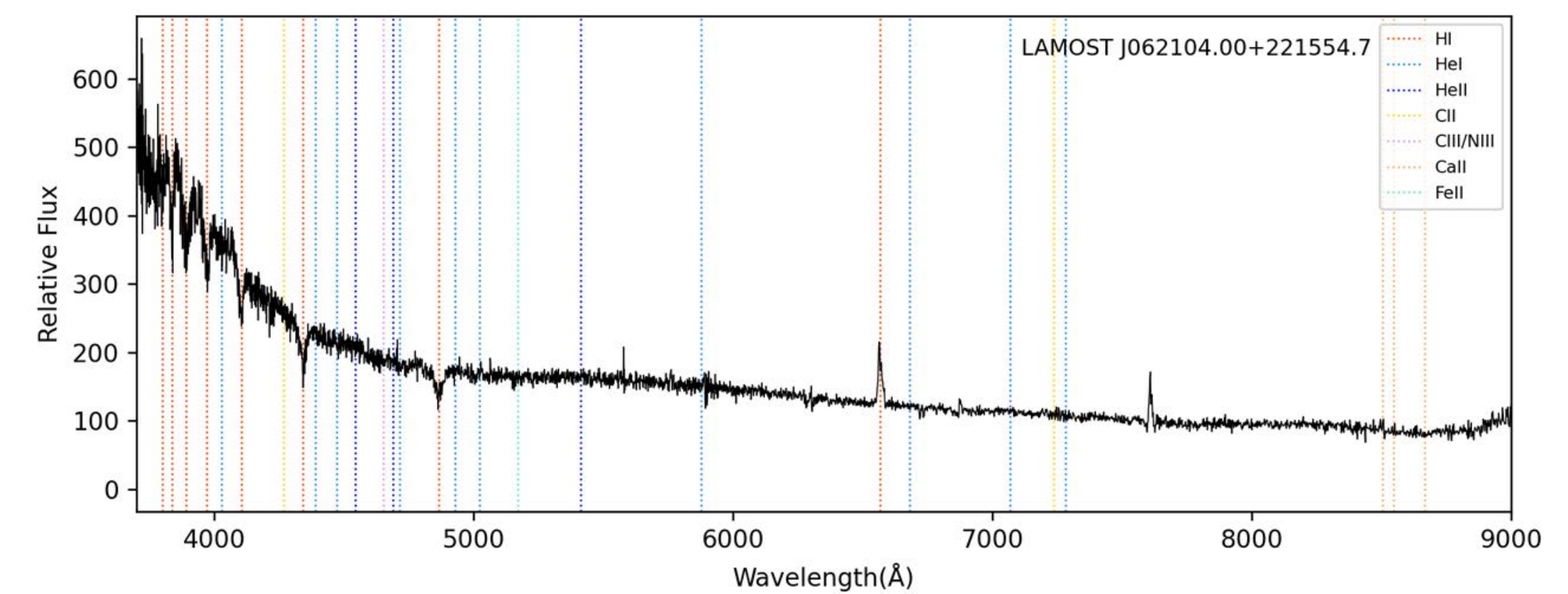}{0.5\textwidth}{}
		\fig{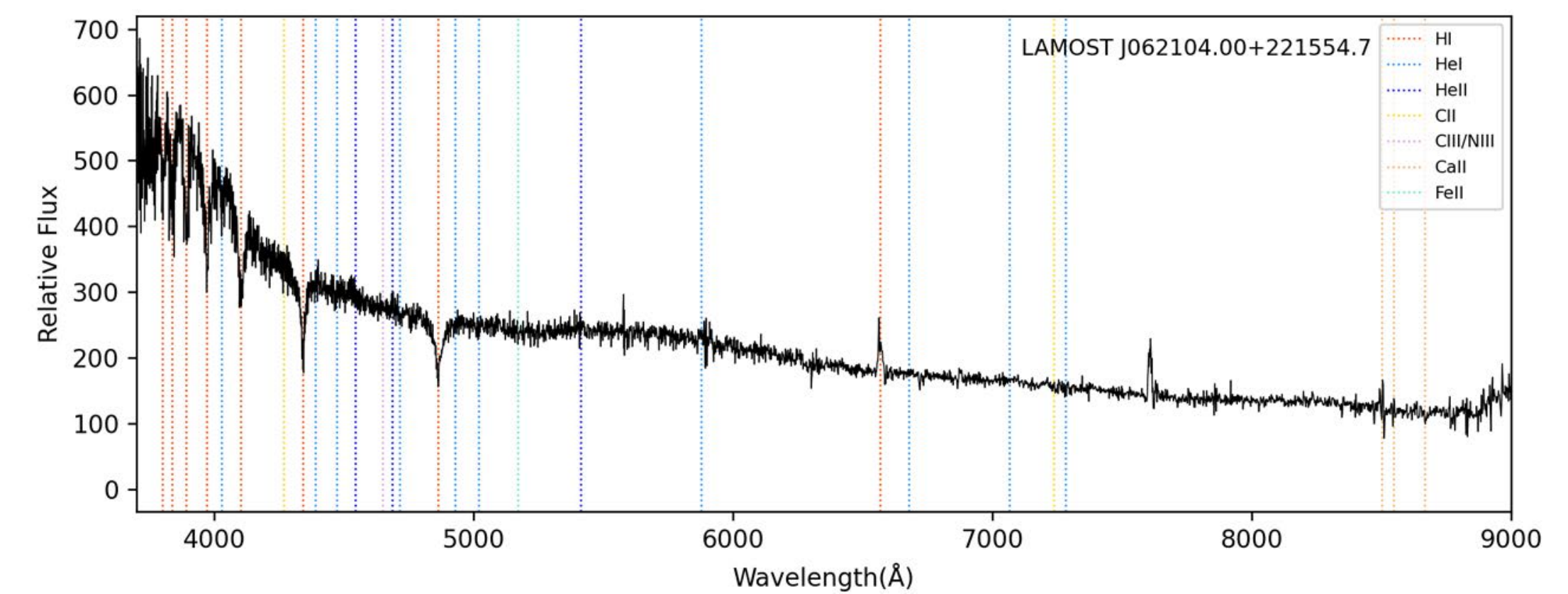}{0.5\textwidth}{}}
	\gridline{\fig{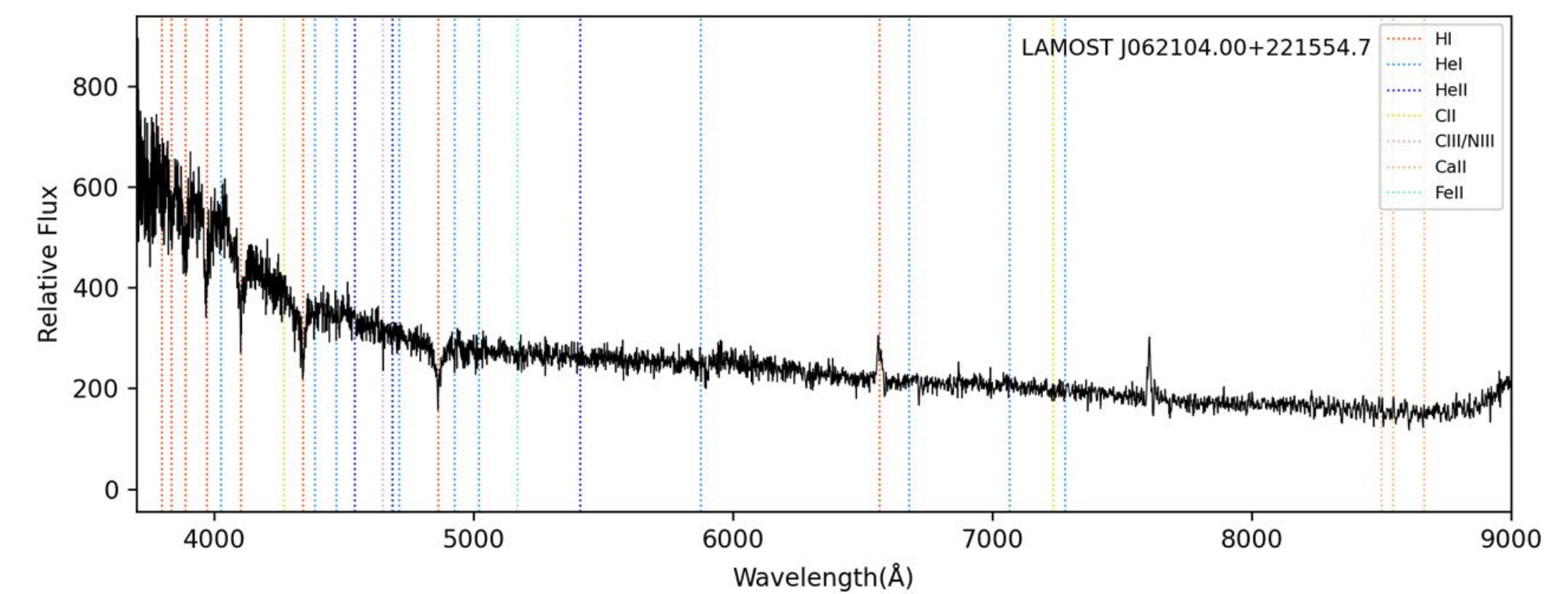}{0.5\textwidth}{}
		\fig{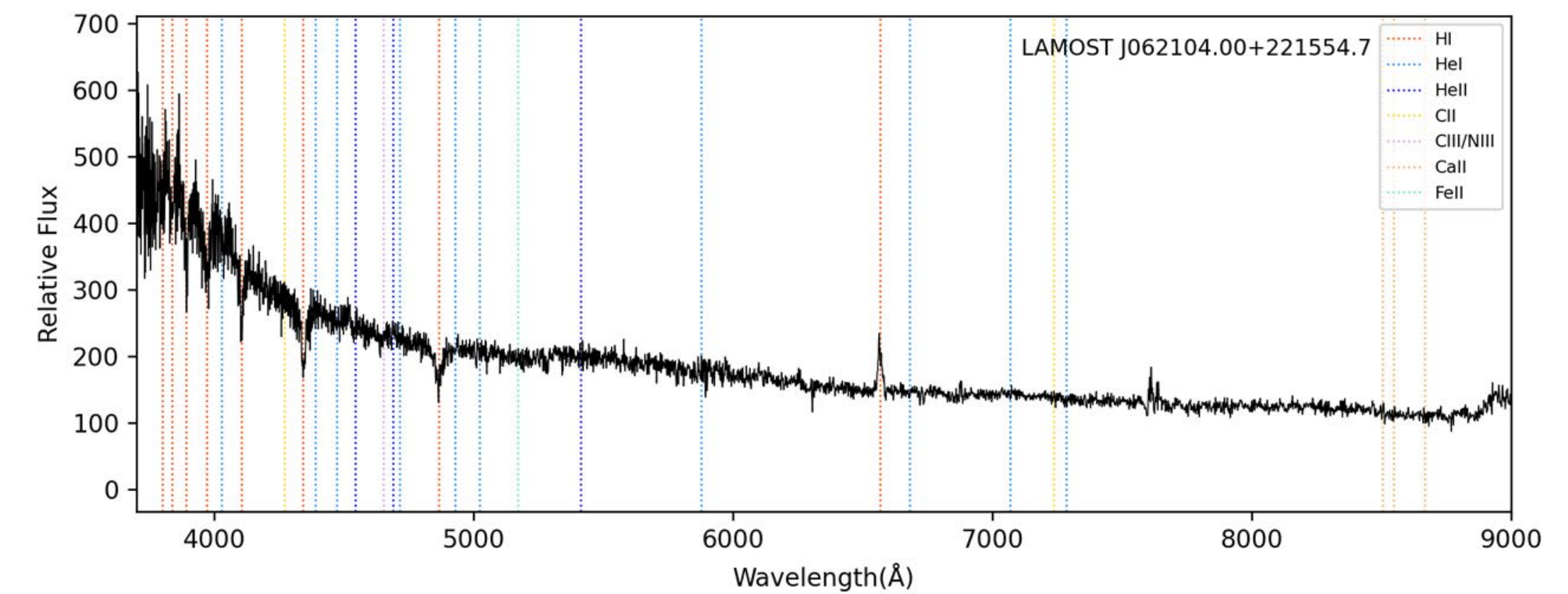}{0.5\textwidth}{}}
\end{figure}
\begin{figure}
	\gridline{\fig{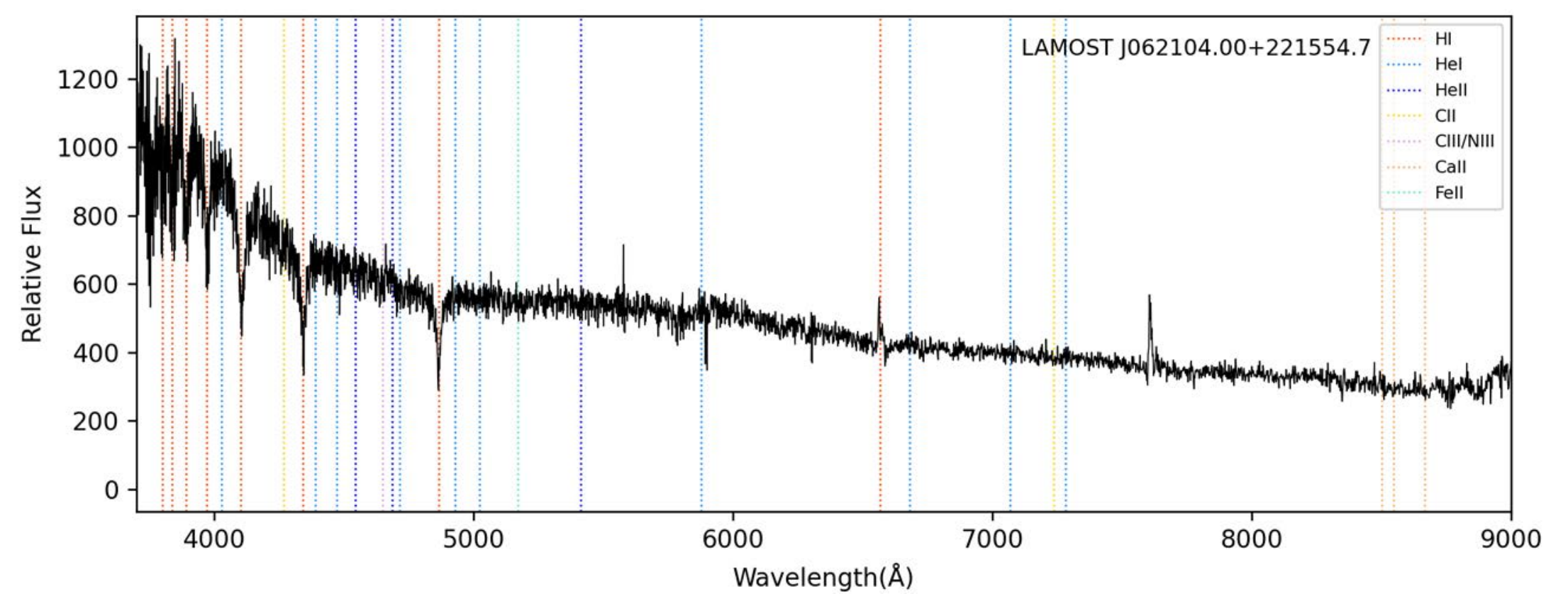}{0.5\textwidth}{}
		\fig{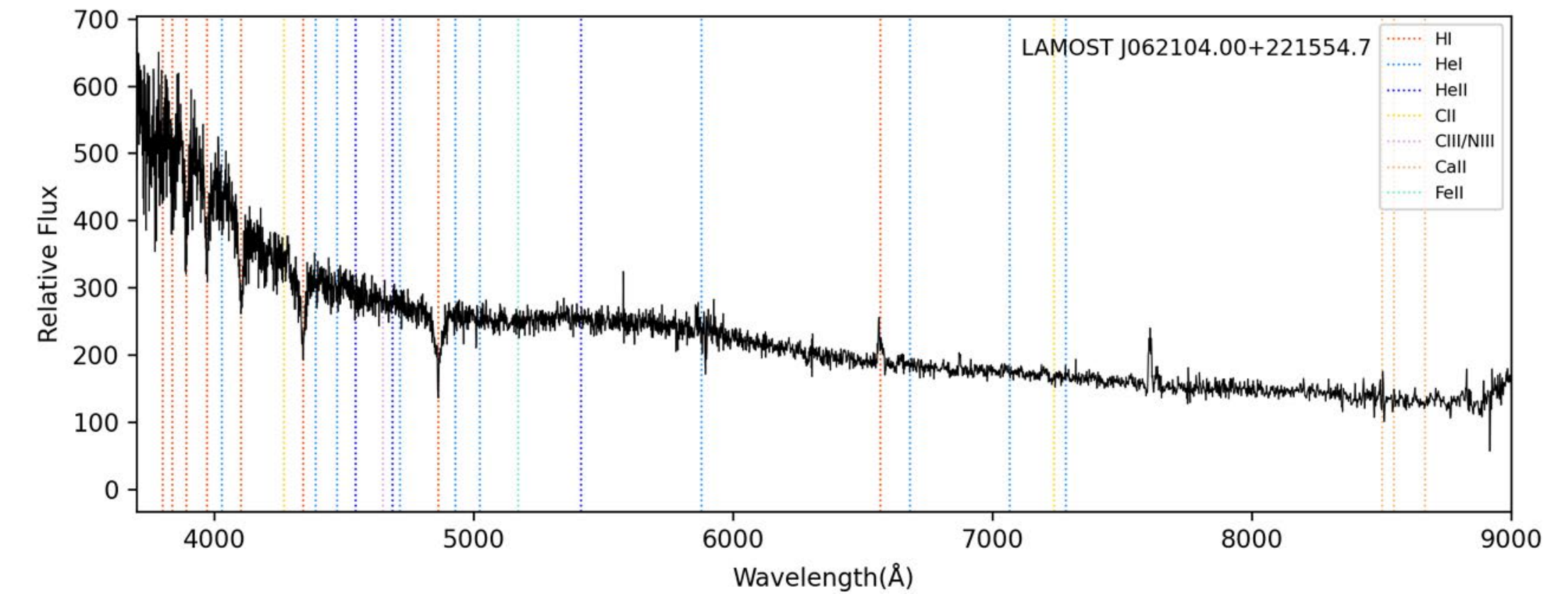}{0.5\textwidth}{}}
	\gridline{\fig{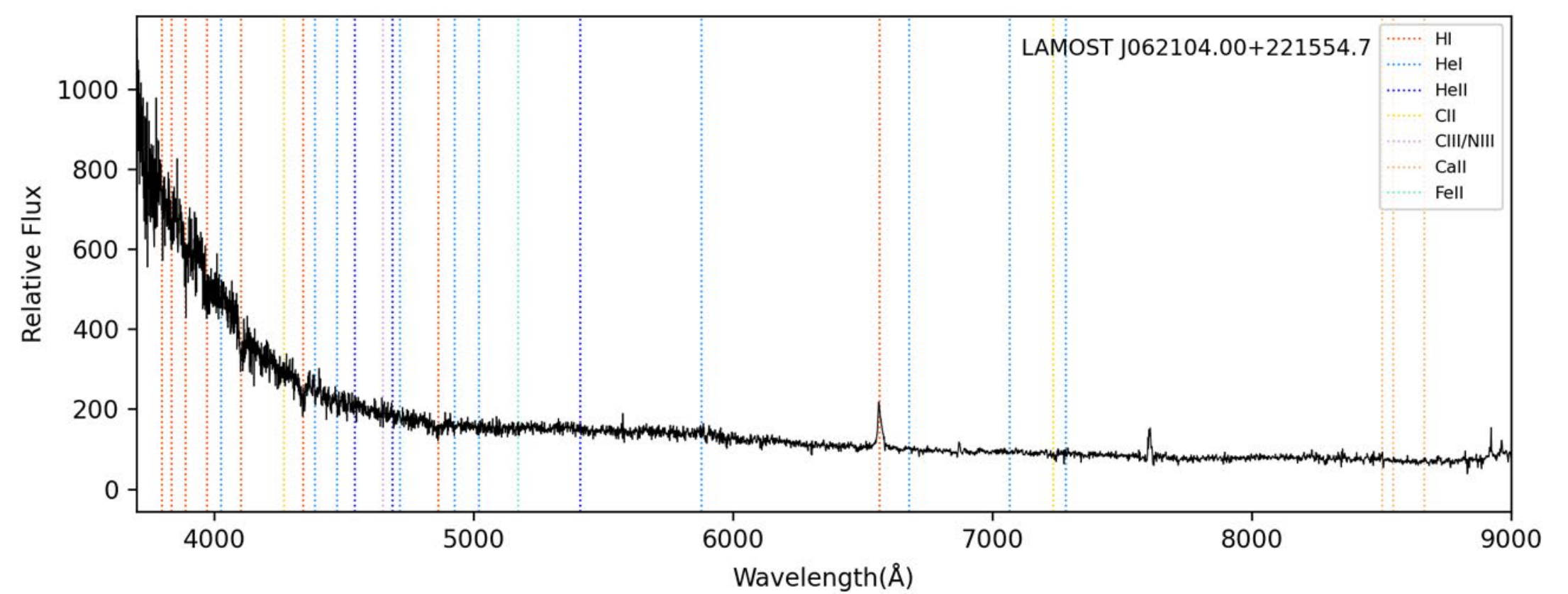}{0.5\textwidth}{}
		\fig{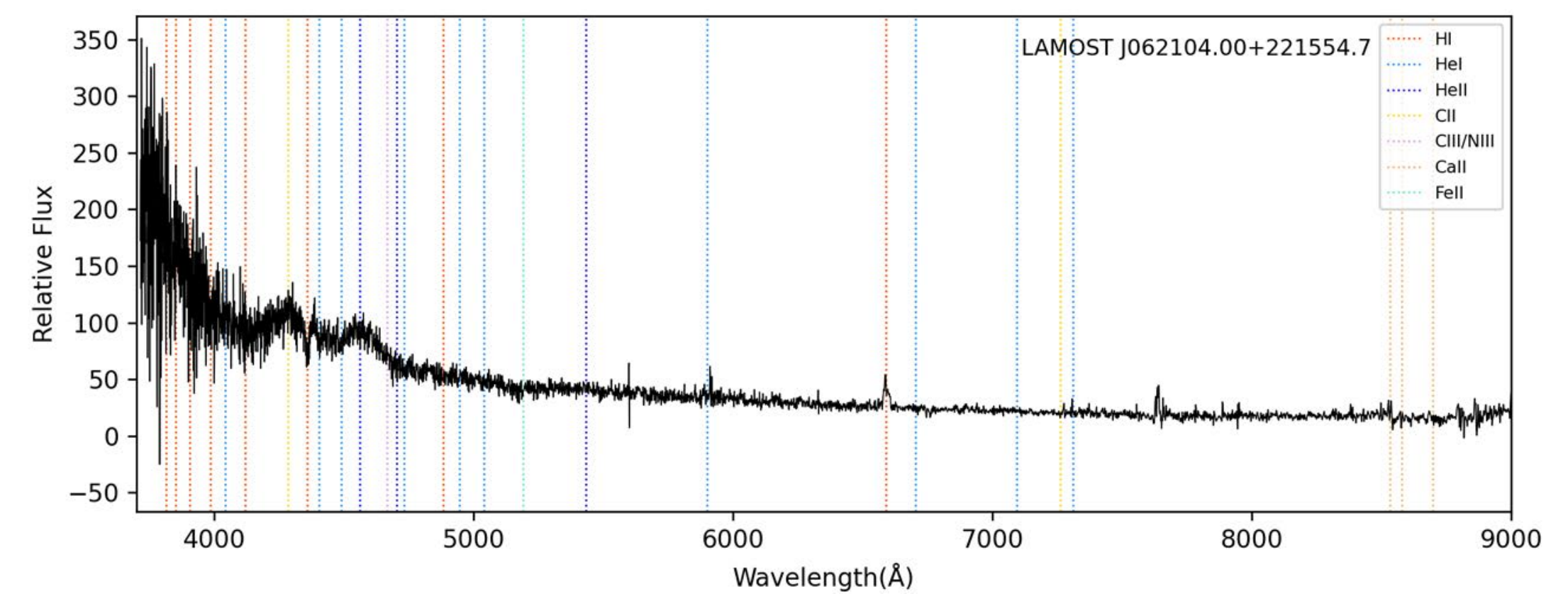}{0.5\textwidth}{}}
	\gridline{\fig{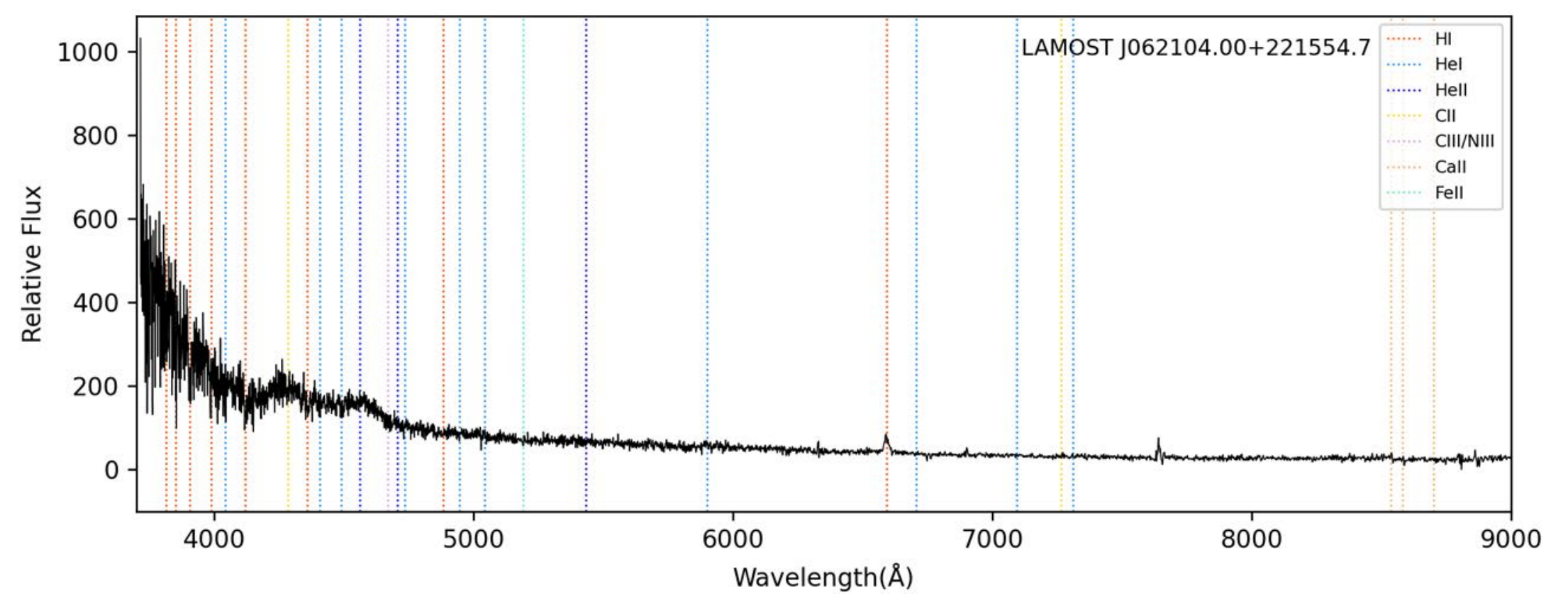}{0.5\textwidth}{}
		\fig{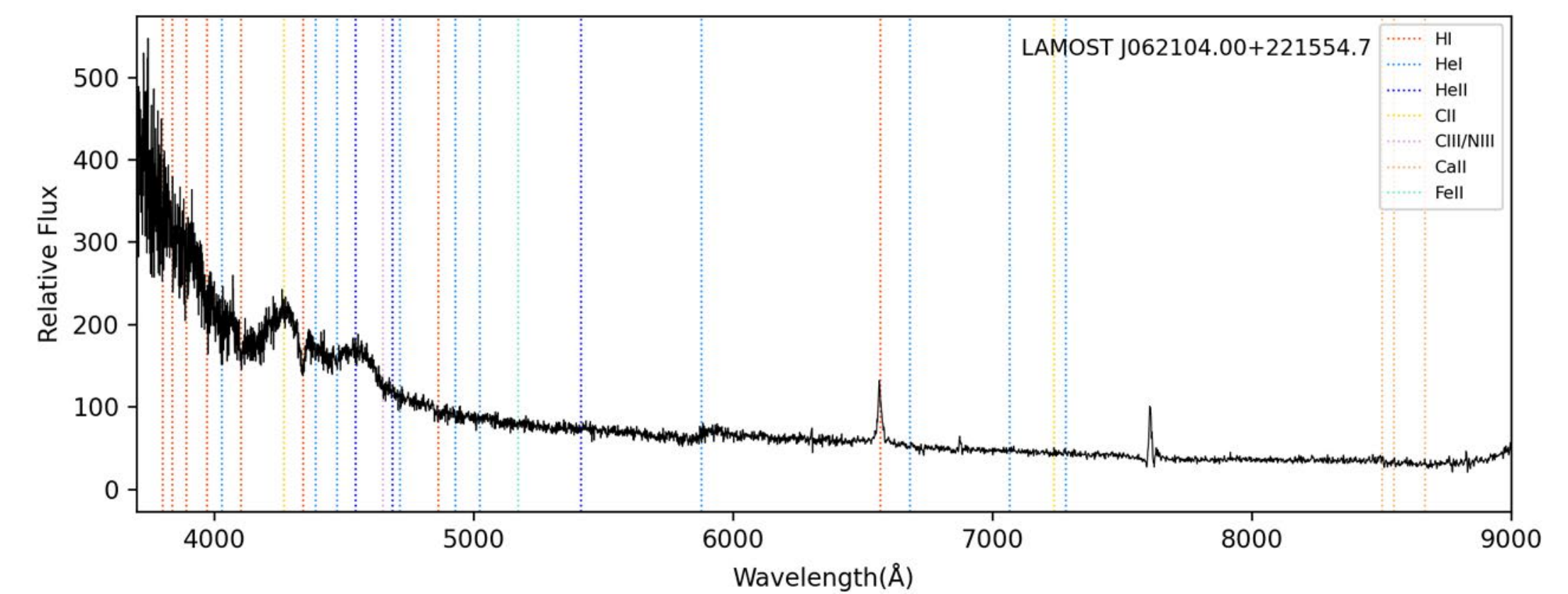}{0.5\textwidth}{}}
	\gridline{\fig{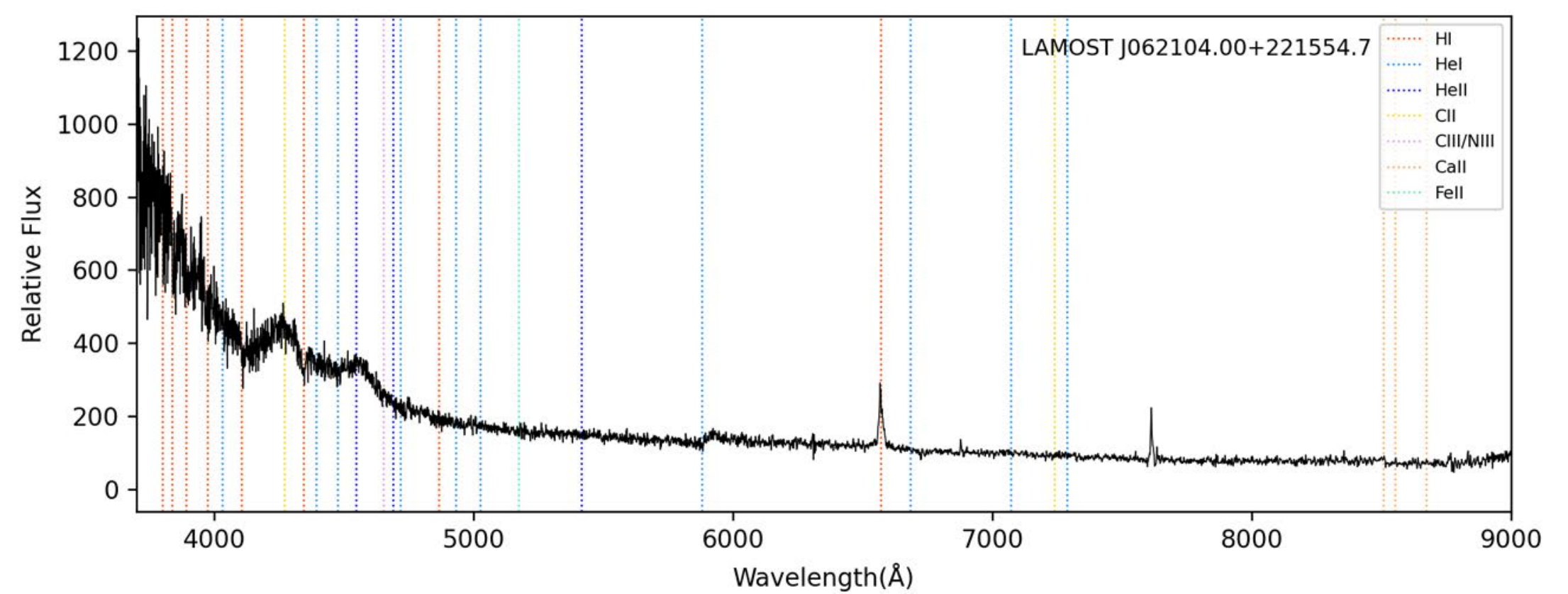}{0.5\textwidth}{}
		\fig{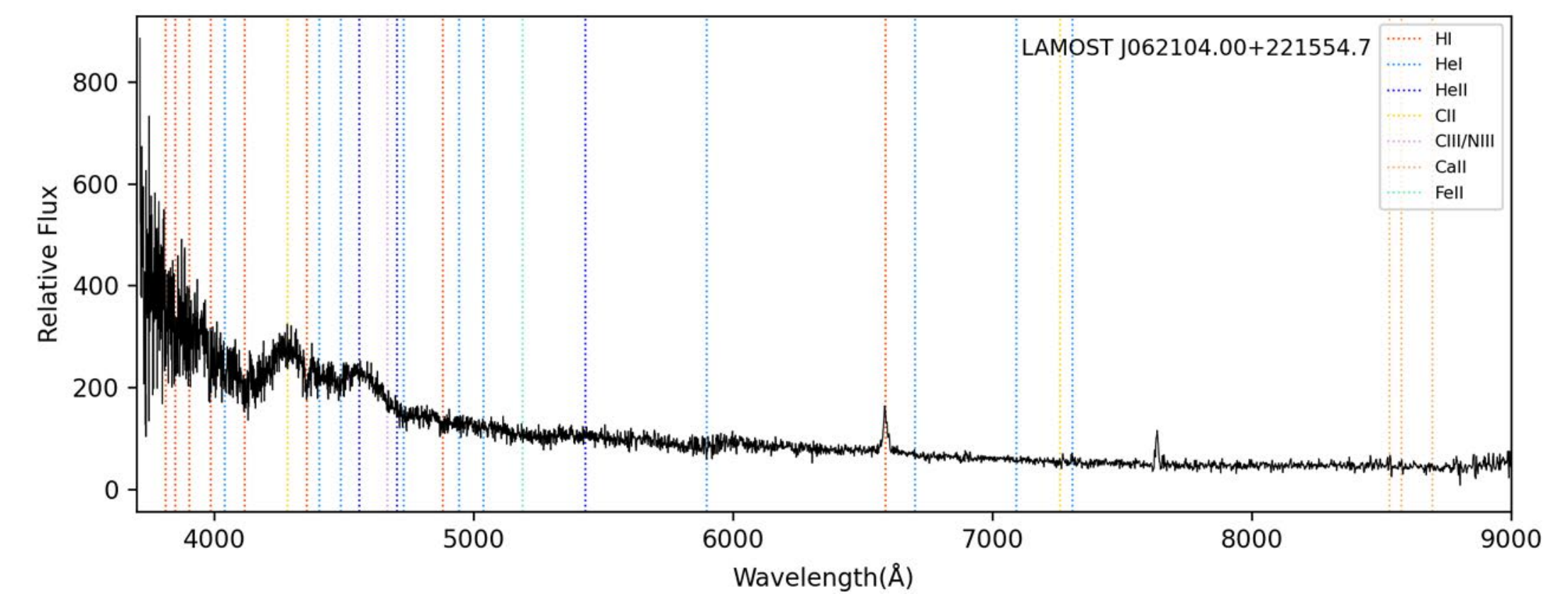}{0.5\textwidth}{}}
	\gridline{\fig{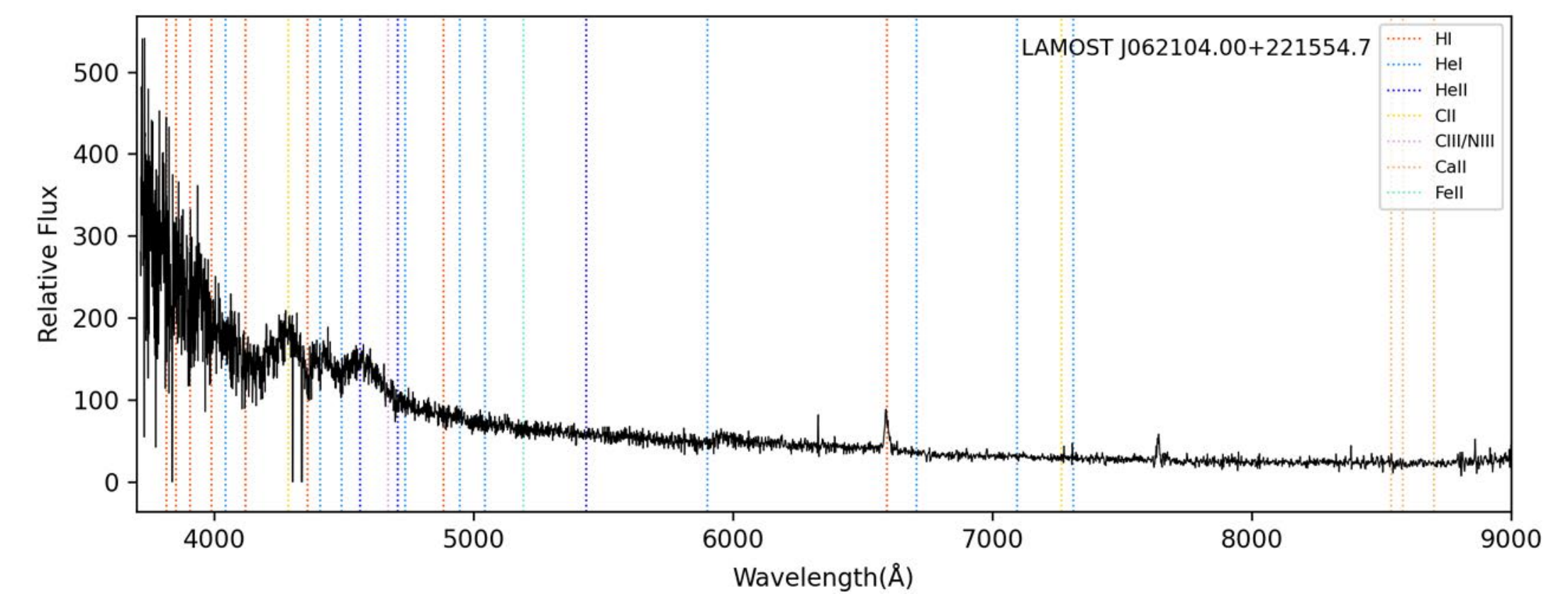}{0.5\textwidth}{}
		\fig{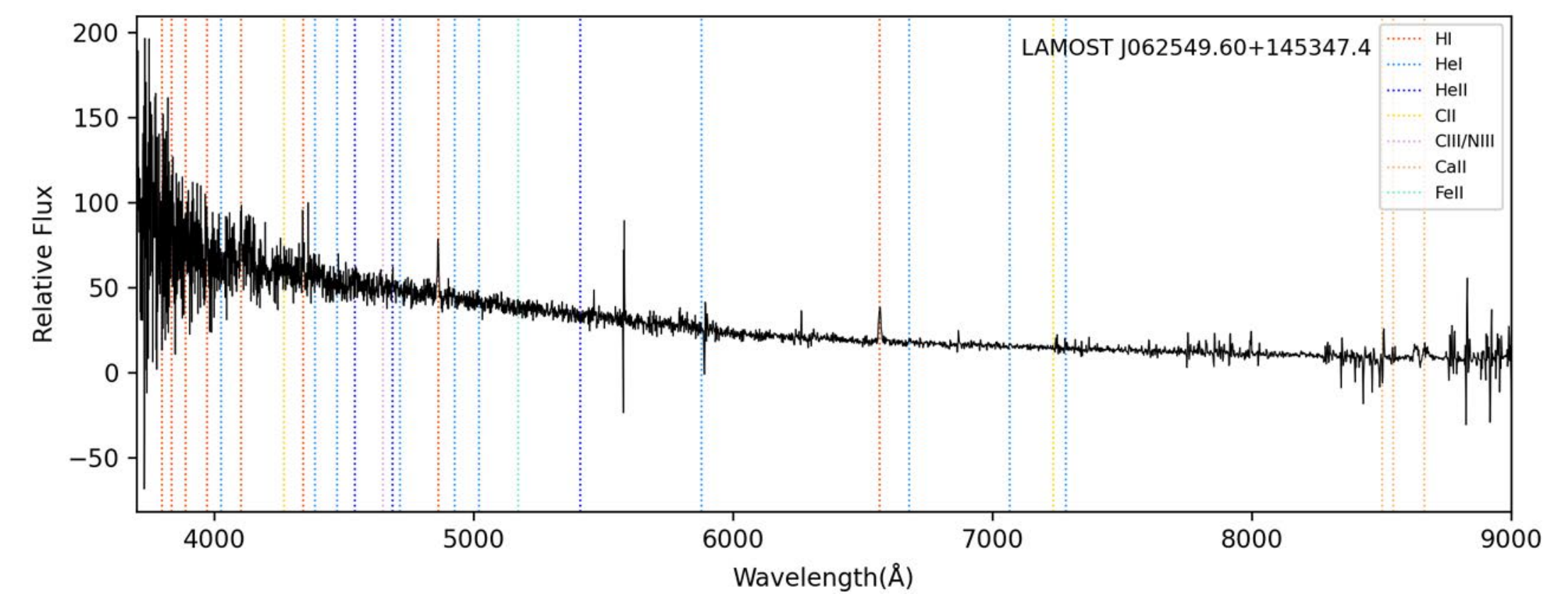}{0.5\textwidth}{}}
\end{figure}
\begin{figure}
	\gridline{\fig{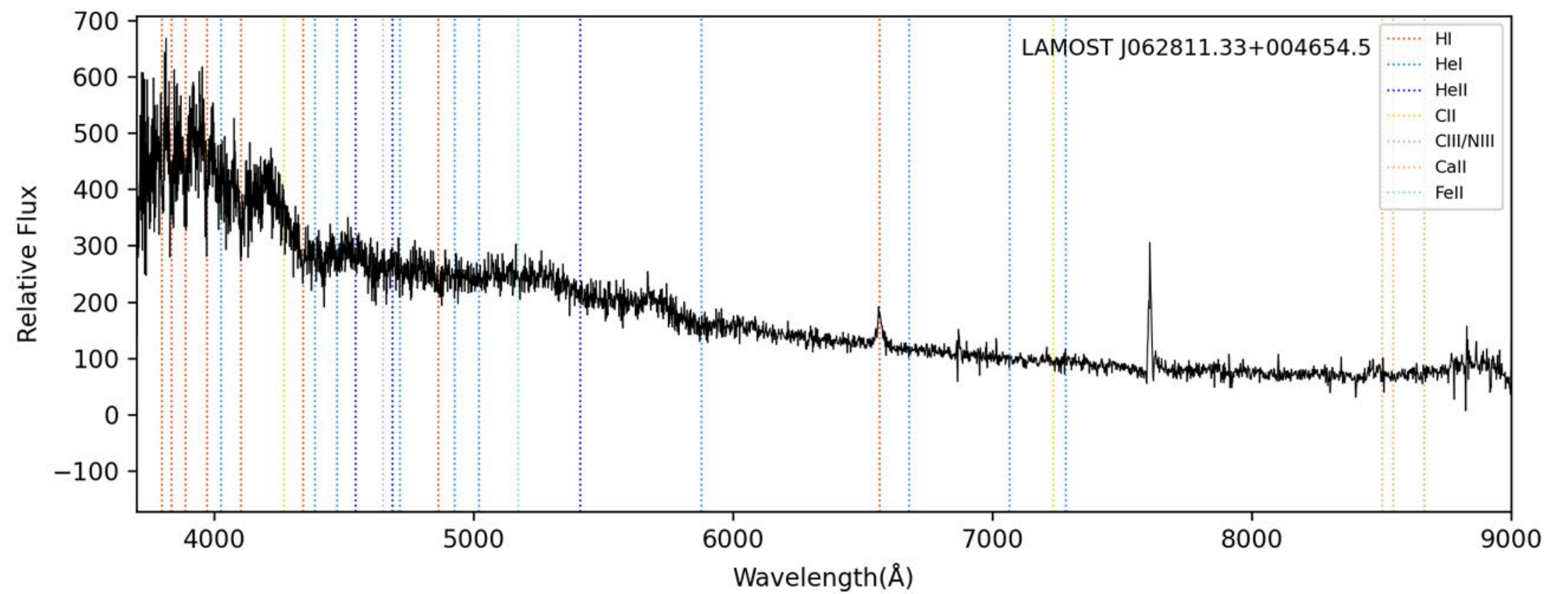}{0.5\textwidth}{}
		\fig{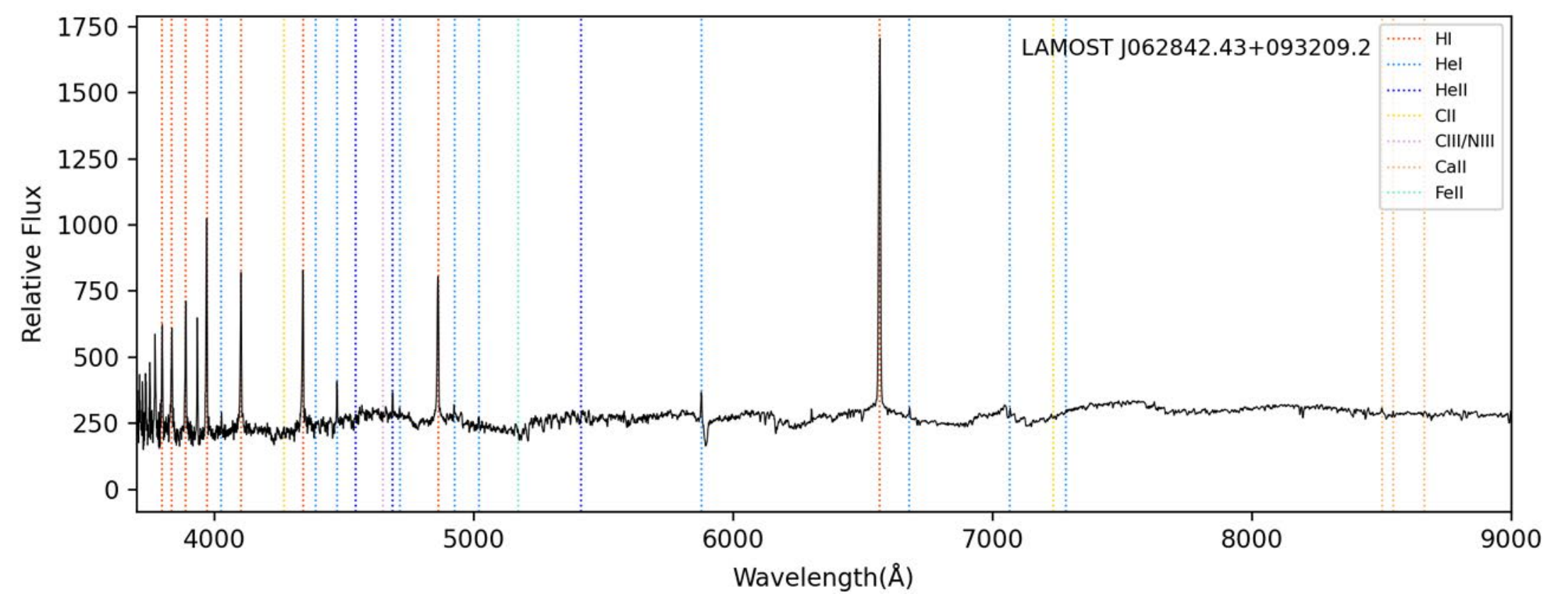}{0.5\textwidth}{}}
	\gridline{\fig{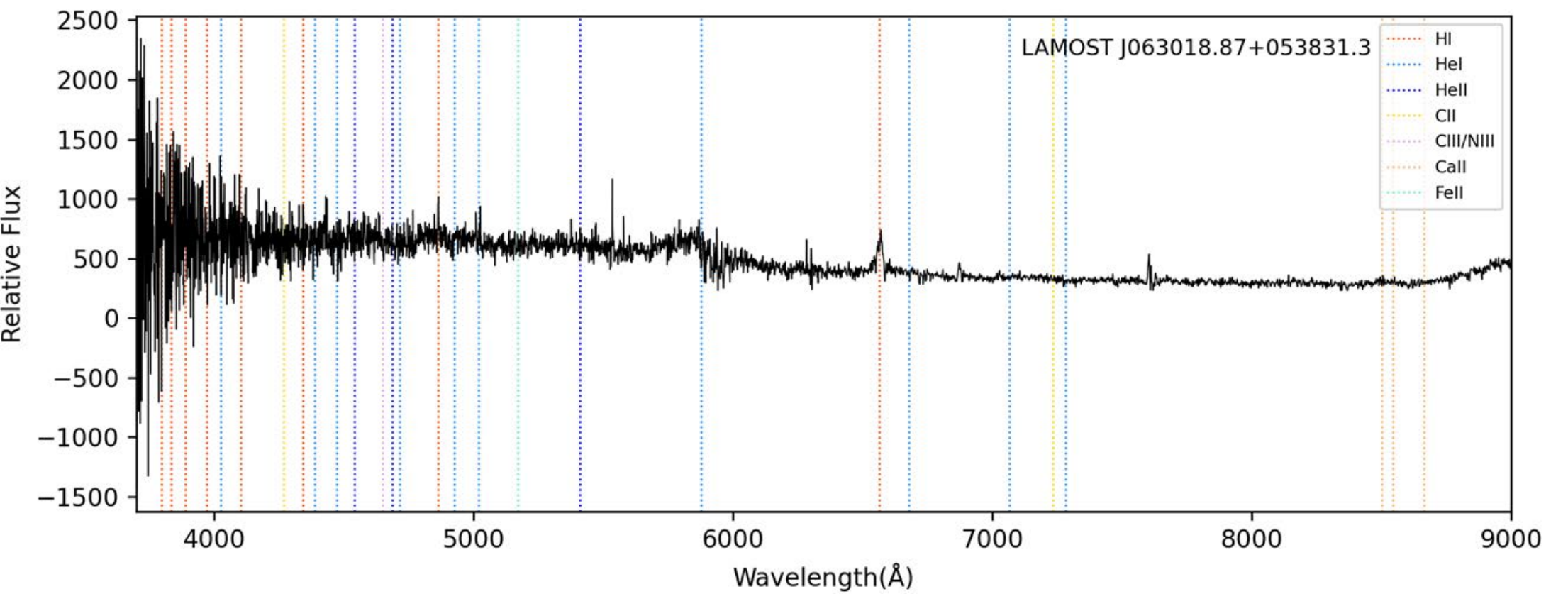}{0.5\textwidth}{}
		\fig{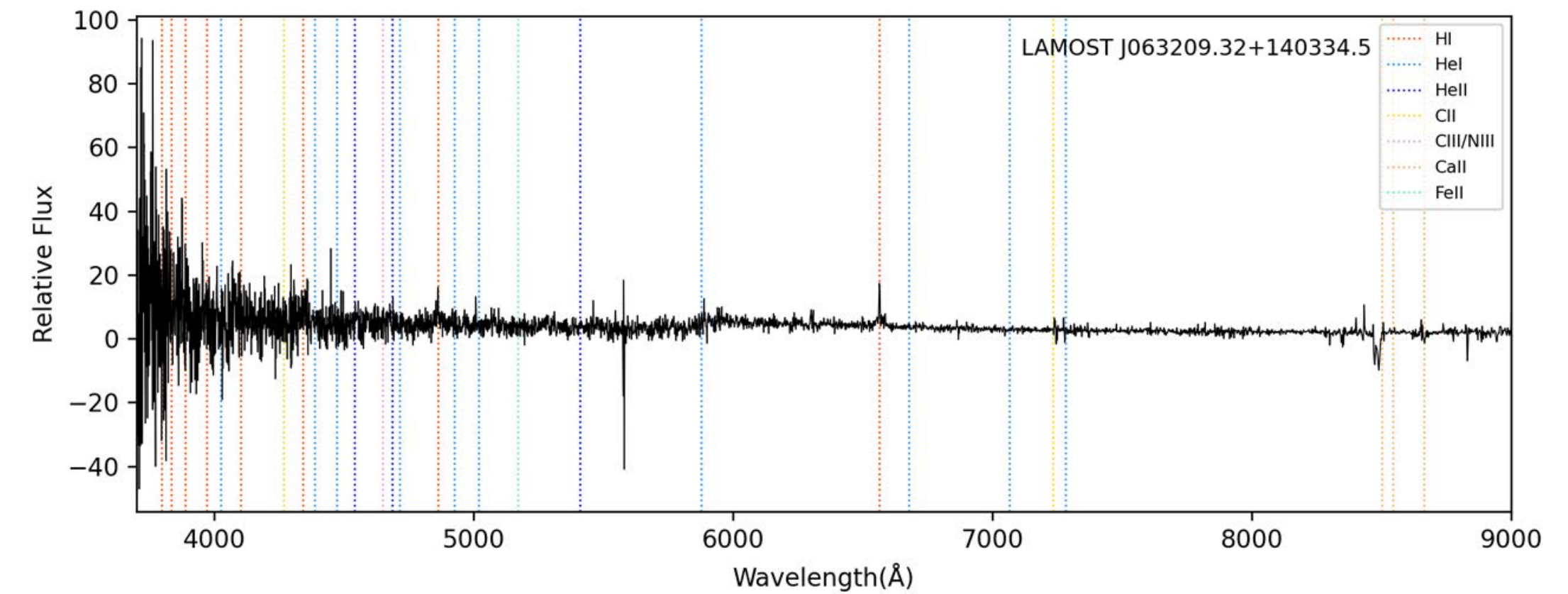}{0.5\textwidth}{}}
	\gridline{\fig{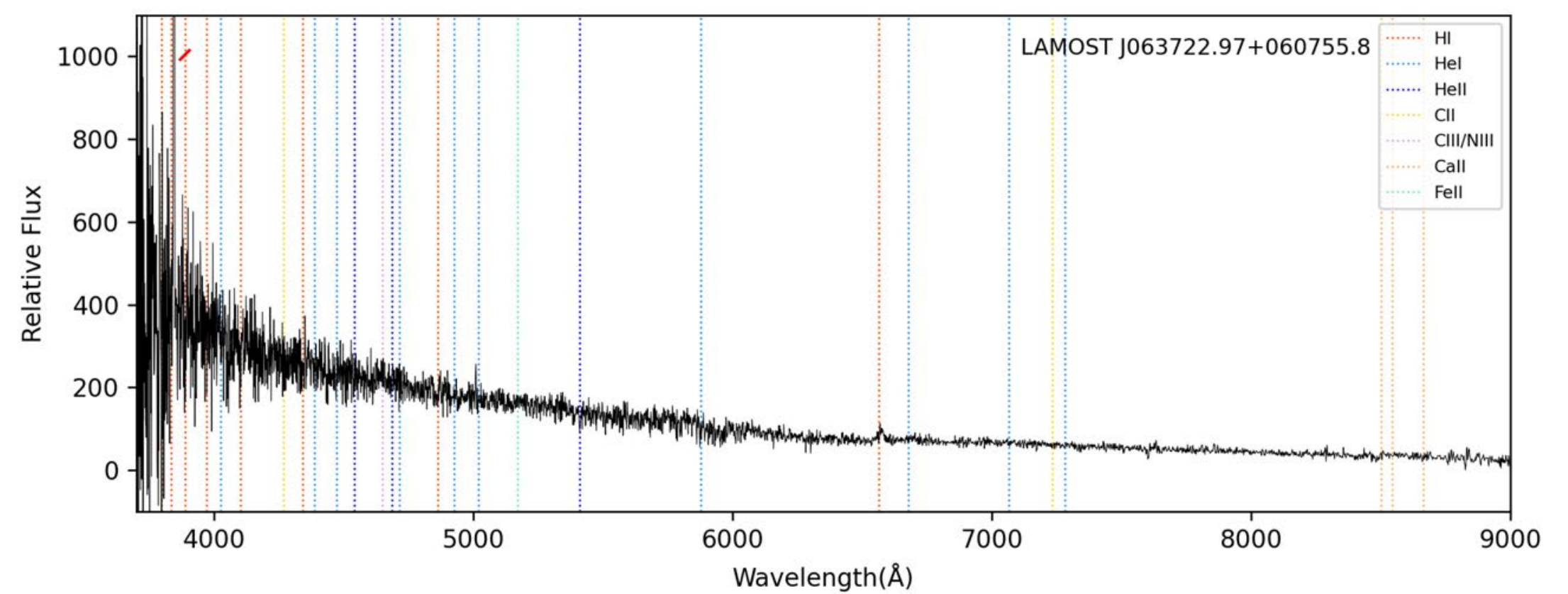}{0.5\textwidth}{}
		\fig{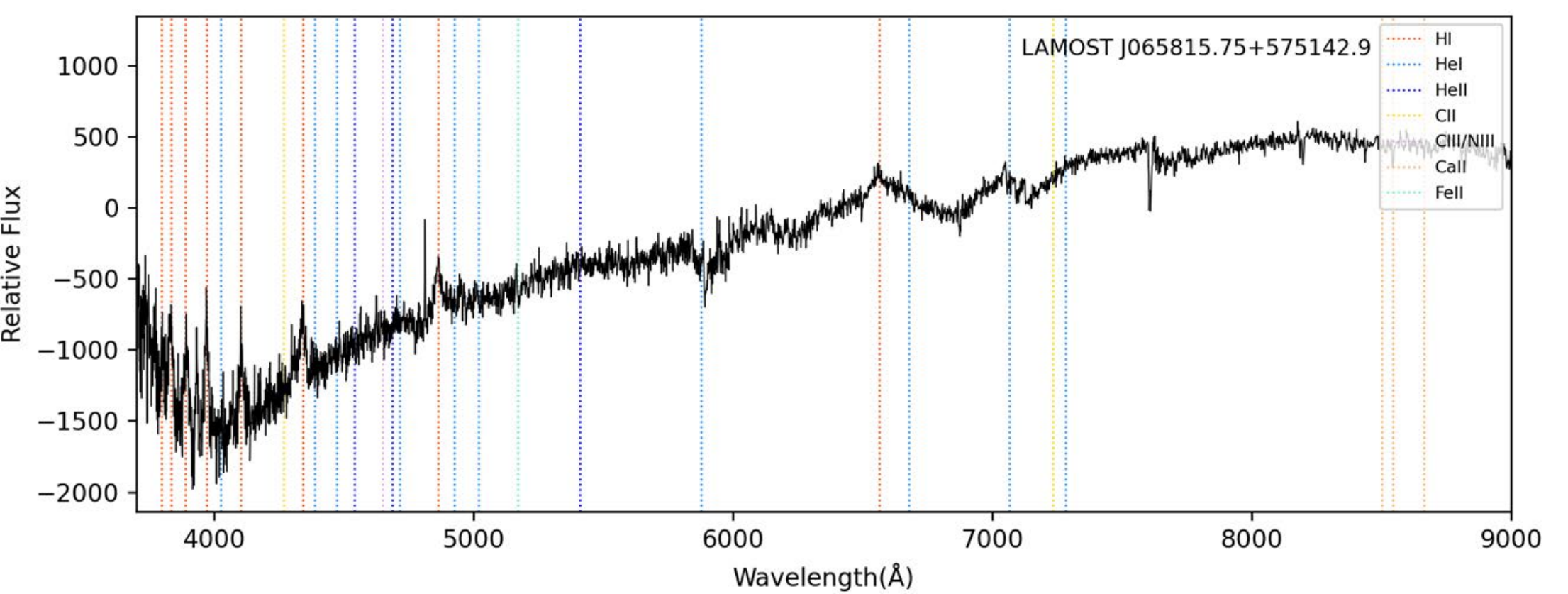}{0.5\textwidth}{}}
	\gridline{\fig{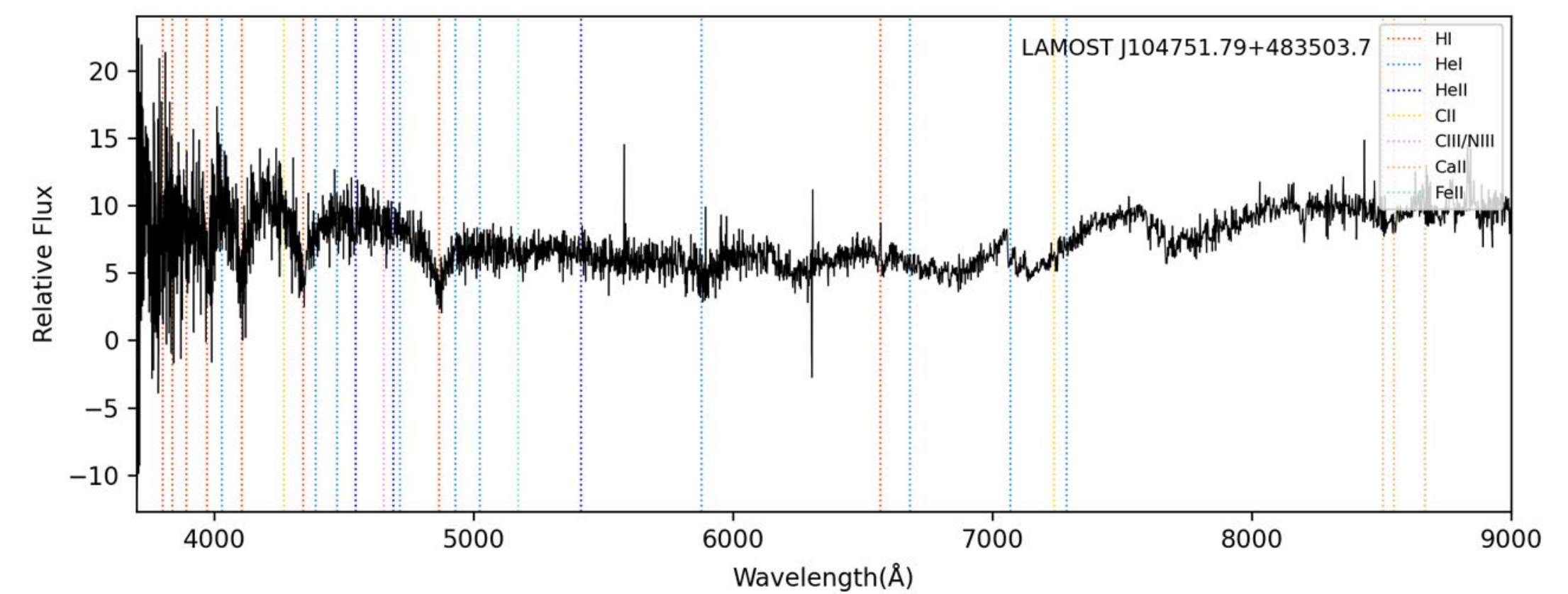}{0.5\textwidth}{}
		\fig{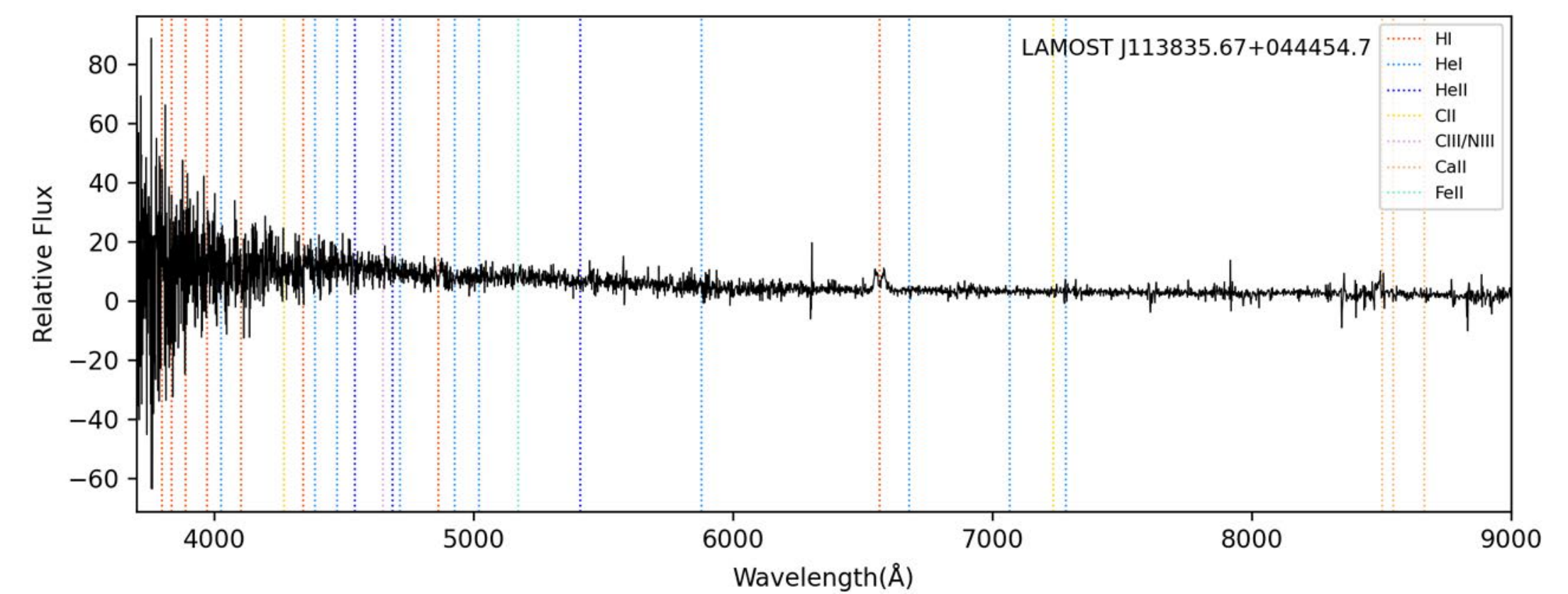}{0.5\textwidth}{}}
	\gridline{\fig{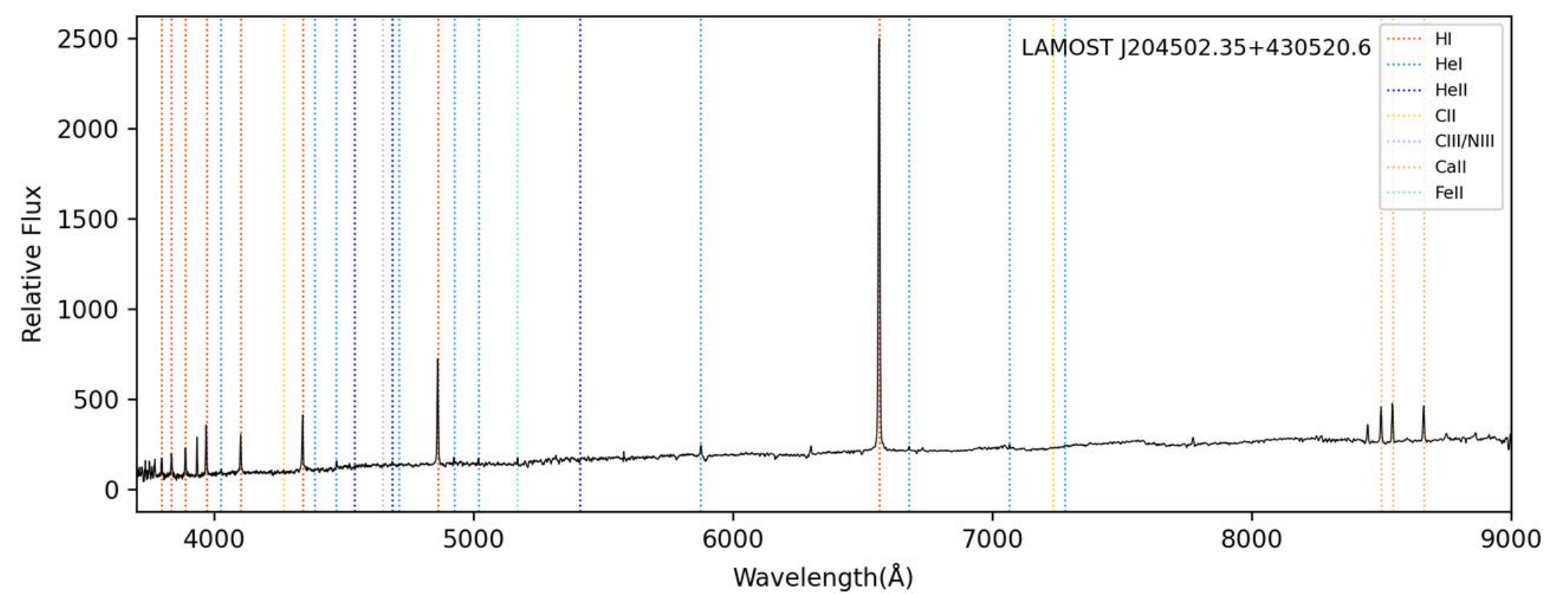}{0.5\textwidth}{}
		\fig{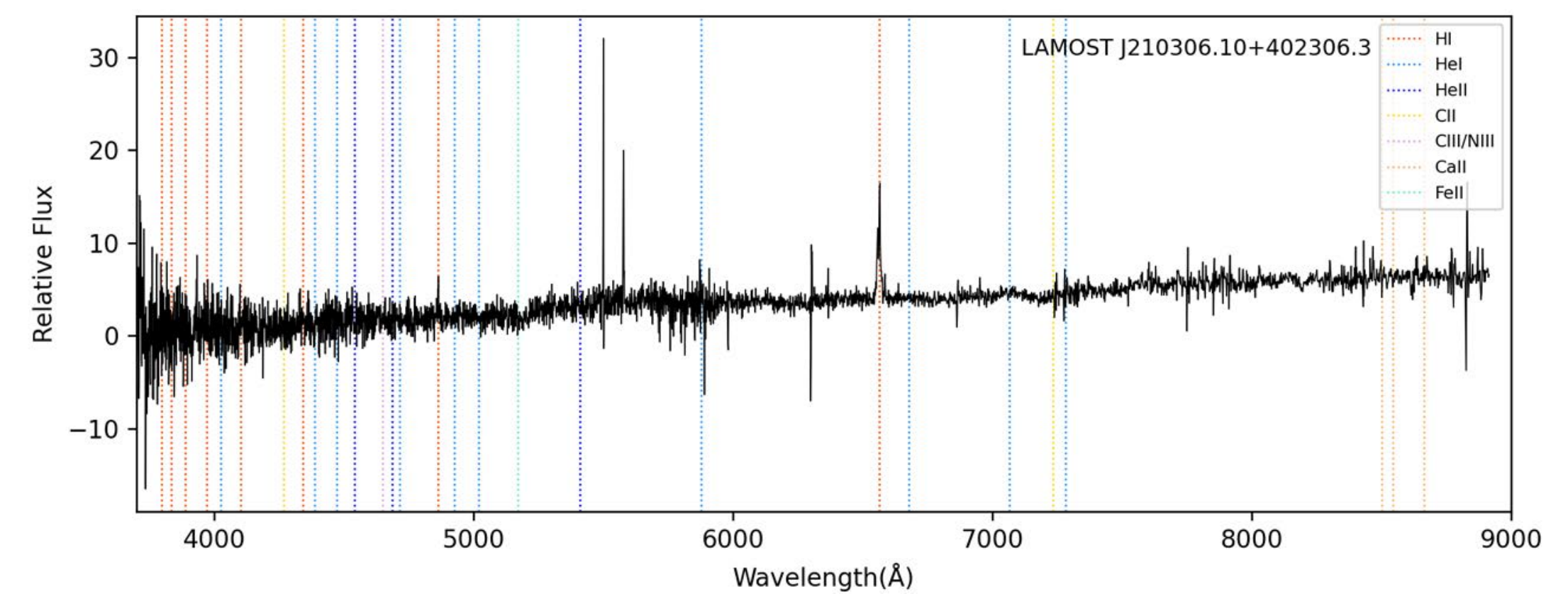}{0.5\textwidth}{}}
\end{figure}

\end{document}